\documentclass[11pt,a4paper]{article}
\pdfoutput=1

\usepackage{graphicx}
\usepackage{float}
\usepackage{afterpage}
\usepackage{epsfig,cite}
\usepackage{amssymb}
\usepackage{amsmath}
\usepackage{dsfont}
\usepackage{multirow}
\usepackage{url}
\usepackage{hyperref}
\usepackage{booktabs}
\usepackage{rotating}
\usepackage{morefloats}
\usepackage{xcolor}
\usepackage{graphicx}
\usepackage{lipsum}

\def\gsim{\mathrel{\rlap{\lower4pt\hbox{\hskip1pt$\sim$}}
    \raise1pt\hbox{$>$}}}         
\def\lsim{\mathrel{\rlap{\lower4pt\hbox{\hskip1pt$\sim$}}
    \raise1pt\hbox{$<$}}} 

\numberwithin{equation}{section}
\numberwithin{figure}{section}
\numberwithin{table}{section}

\begin{document}

\begin{figure}[h]
 \includegraphics[width=0.37\textwidth]{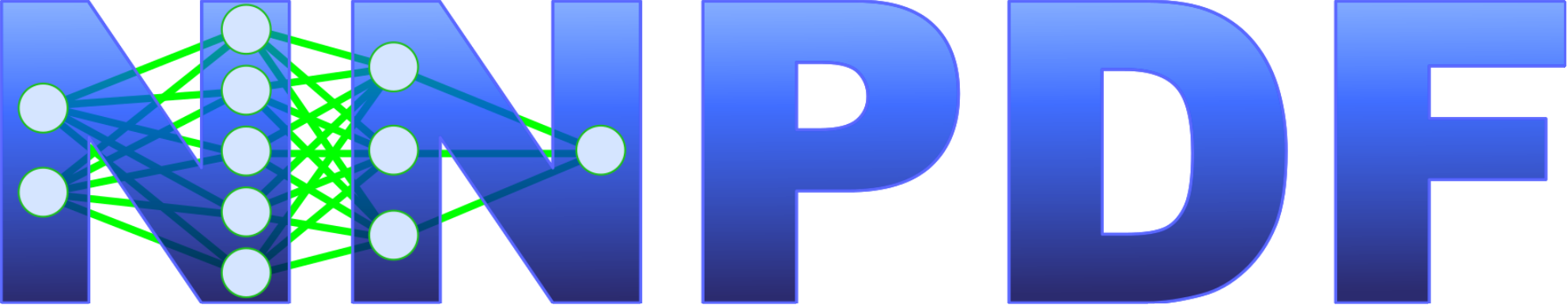}
\end{figure}

\vspace{-2.0cm}
\begin{flushright}
CERN-TH-2017-122\\
OUTP-16-15P \\
Nikhef/2016-047 \\
\end{flushright}

\vspace{.9cm}
\begin{center}
  {\Large \bf A determination of the fragmentation functions\\[3.0pt] 
  of pions, kaons, and protons with faithful uncertainties}
\vspace{1cm}

{\bf  The NNPDF Collaboration:} \\
Valerio~Bertone$^{1}$,
Stefano~Carrazza$^{2}$,
Nathan~P.~Hartland$^{1}$,\\[0.2pt] 
Emanuele~R.~Nocera$^{3}$ and Juan~Rojo$^{1}$

\vspace{.3cm}
{\it
~$^{1}$ Department of Physics and Astronomy, 
       VU University, NL-1081 HV Amsterdam,\\
       and Nikhef Theory Group, 
       Science Park 105, 1098 XG Amsterdam, The Netherlands\\
~$^{2}$ Theoretical Physics Department, CERN, CH-1211 Geneva, Switzerland\\
~$^{3}$ Rudolf Peierls Centre for Theoretical Physics, University of Oxford,\\
       1 Keble Road, OX1 3NP Oxford, United Kingdom\\
}

\vspace{1.2cm}

{\bf \large Abstract}

\end{center}

We present NNFF1.0, a new determination of the fragmentation functions (FFs) 
of charged pions, charged kaons, and protons/antiprotons from an analysis of 
single-inclusive hadron production data in electron-positron annihilation.
This determination, performed at leading, next-to-leading, and 
next-to-next-to-leading order in perturbative QCD, is based on the NNPDF 
methodology, a fitting framework designed to provide a statistically sound 
representation of FF uncertainties and to minimise any procedural bias.
We discuss novel aspects of the methodology used in this analysis, namely
an optimised parametrisation of FFs and a more efficient $\chi^2$ minimisation 
strategy, and validate the FF fitting procedure by means of closure tests.
We then present the NNFF1.0 sets, and discuss their fit
quality, their perturbative convergence, and their stability upon
variations of the kinematic cuts and the fitted dataset.
We find that the systematic inclusion of higher-order QCD corrections
significantly improves the description of the data, especially in the
small-$z$ region.
We compare the NNFF1.0 sets to other recent sets of FFs, finding in general a 
reasonable agreement, but also important differences.
Together with existing sets of unpolarised and polarised parton 
distribution functions (PDFs), FFs and PDFs are now available
from a common fitting framework for the first time.

\clearpage

\tableofcontents

\section{Introduction}
\label{sec:introduction}

In the framework of Quantum Chromodynamics (QCD), Fragmentation Functions 
(FFs)~\cite{Field:1977fa} encode the long-distance dynamics of the interactions 
among quarks and gluons which lead to their hadronisation 
in a hard-scattering process~\cite{Collins:1981uk,Collins:1981uw}.
In order to obtain theoretical predictions for the observables involving 
identified hadrons in the final state, FFs have to be 
convoluted~\cite{Collins:1989gx} with partonic cross-sections encoding instead 
the short-distance dynamics of the interaction.
If the hard-scattering process is initiated by nucleons, additional 
convolutions with the Parton Distribution Functions 
(PDFs)~\cite{Forte:2010dt,Forte:2013wc,Rojo:2015acz}, 
the space-like counterparts of FFs, are required.

The knowledge of FFs is an important ingredient in our understanding
of non-perturbative QCD dynamics, as well as an essential tool in the 
description of a number of processes used to examine the internal
structure of nucleons. 
For example, processes probing nucleon momentum, spin, flavour and spatial 
distributions~\cite{Albino:2008aa}, as well as the dynamics
of cold~\cite{Sassot:2009sh} and hot~\cite{Aamodt:2010jd} nuclear matter.
While partonic cross-sections can be computed perturbatively in QCD,
FFs cannot, although their dependence on the factorisation scale results
in the perturbatively computable DGLAP evolution equations~\cite{Gribov:1972ri,
Lipatov:1974qm,Altarelli:1977zs,Dokshitzer:1977sg}.
In this respect, FFs and PDFs are on the same footing.
Therefore, as PDFs, FFs need to be determined from a global analysis of a
suitable set of experimental measurements, possibly from a variety of
hard-scattering processes (see {\it e.g.} 
Refs.~\cite{Albino:2008gy,Metz:2016swz} for a review).

These processes include hadron production in electron-positron
Single-Inclusive Annihilation (SIA), in lepton-nucleon Semi-Inclusive
Deep-Inelastic Scattering (SIDIS), and in proton-proton ($pp$) collisions.
Information from SIDIS multiplicities and from $pp$ collisions is particularly 
useful in order to achieve a complete flavour decomposition into quark and 
antiquark FFs alongside a direct determination of the gluon FF.
However, SIA remains the theoretically cleanest process among the three, since 
its interpretation does not require the simultaneous knowledge of PDFs.

Recent progress in the determination of FFs has been focused on
charged pions and kaons, for which data are more abundant as they
dominate the identified hadron yields.
In the last few years, at least three groups have determined sets of
FFs with uncertainties for these two hadronic species:
DEHSS~\cite{deFlorian:2014xna,deFlorian:2017lwf},
HKKS~\cite{Hirai:2016loo}, and JAM~\cite{Sato:2016wqj}.
All these determinations were performed at next-to-leading order (NLO) 
accuracy in perturbative QCD. 
Their primary focus was put on quantifying the effects of the
inclusion of new measurements, although in the HKKS and JAM fits these
were limited to SIA.
These FF analyses also introduced some methodological and theoretical 
improvements over previous determinations.
Specifically, in order to achieve a more reliable estimate of the
uncertainties of FFs, various techniques widely used in PDF
determinations have been adopted.
For example, the iterative Hessian approach developed in
Refs.~\cite{Pumplin:2000vx,Pumplin:2001ct} has been used in the DEHSS
analyses, while the iterative Monte Carlo procedure developed in
Ref.~\cite{Sato:2016tuz} has been used in the JAM analysis.
Separately, theoretical investigations of the effect of 
next-to-next-to-leading order (NNLO) QCD
corrections~\cite{Anderle:2015lqa}, of a general treatment of
heavy-quark mass effects~\cite{Epele:2016gup}, and of all-order
small-$z$ resummation~\cite{Anderle:2016czy} were performed in
the framework of the DEHSS analyses, although only for the SIA production
of charged pions.

Despite this progress, available determinations of FFs are still
potentially affected by some sources of procedural bias, the size and
effect of which are difficult to quantify.
First, the parametrisation of FFs in terms of a simple functional
form, customary in current analyses, may not encapsulate all the
possible functional behaviours of the FFs.
Second, the Hessian method supplemented with a tolerance parameter,
used to determine FF uncertainties for instance in the HKKS
analysis, lacks a robust statistical interpretation.
Third, if a global analysis of FFs including PDF-dependent processes
is carried out, PDFs and FFs should be determined within a consistent
methodology.
This is not the case in current global analyses like DEHSS.

This work represents a first step in overcoming some of these limitations.
Building on the preliminary results of Refs.~\cite{Nocera:2017qgb,
Bertone:2017xsf}, here we present NNFF1.0, a new determination of the
FFs of charged pions, charged kaons, and protons/antiprotons from a
comprehensive set of SIA measurements.
This analysis is performed at leading order (LO), NLO, and NNLO
accuracy in perturbative QCD.
This analysis is based on the NNDPF methodology, a fitting framework
designed to provide a statistically sound representation of FF
uncertainties and to reduce potential procedural biases as much as possible.
This is achieved by representing FFs as a Monte Carlo sample, from
which central values and uncertainties can be computed respectively as
a mean and a standard deviation, and by parametrising FFs with a
flexible function provided by a neural network.

The NNPDF methodology for the determination of PDFs was originally applied to
the analysis of inclusive Deep-Inelastic Scattering (DIS) structure
functions~\cite{Forte:2002fg,DelDebbio:2004xtd}, and then extended to a
determination of the PDFs of the proton, first from DIS data
only~\cite{DelDebbio:2007ee,Ball:2008by,Ball:2009mk} and then from a
wider dataset including hadron collider
data~\cite{Ball:2010de,Ball:2011mu,Ball:2011uy}.
Several developments have been achieved since then.
These include determinations of PDFs with LHC
data~\cite{Ball:2012cx,Ball:2014uwa,Ball:2017nwa}, of PDFs with effects of
threshold resummation~\cite{Bonvini:2015ira}, of PDFs with QED
corrections~\cite{Ball:2013hta}, of a fitted charm
PDF~\cite{Ball:2016neh}, and of longitudinally polarised
PDFs~\cite{Ball:2013lla,Nocera:2014gqa}.
Applying the NNPDF framework to a determination of FFs is therefore a
natural extension of the NNPDF fits.
It is also a first step towards a simultaneous determination of
polarised and unpolarised PDFs and FFs, as recently attempted by the
JAM collaboration~\cite{Ethier:2017zbq}.

This paper is organised as follows.
In Sect.~\ref{sec:expdata} we present the dataset used in our analysis,
along with the corresponding observables and kinematic cuts.
In Sect.~\ref{sec:theory} we discuss the theoretical details of the NNFF1.0 
determination, including the computation of the observables, the evolution of 
FFs, and our choice of physical parameters.
In Sect.~\ref{sec:methodology} we revisit the NNPDF fitting
methodology and its application to our current determination of FFs.
Specifically, we focus on the aspects of the parametrisation and
minimisation strategy which are introduced here for the first time.
We validate the fitting methodology by means of closure tests.
In Sect.~\ref{sec:results} we present the NNFF1.0 sets, their
fit quality, their perturbative convergence, and their comparison
with other available FF sets.
In Sect.~\ref{sec:stability} we study their stability upon variations in
the kinematic cuts and the fitted dataset.
Finally, in Sect.~\ref{sec:conclusions} we conclude
and outline possible future developments.
The delivery of the NNFF1.0 sets is discussed in
Appendix~\ref{sec:delivery}.

\section{Experimental data}
\label{sec:expdata}

The determination of FFs presented in this work is based on a comprehensive 
dataset from SIA, {\it i.e.} electron-positron annihilation into a single 
identified hadron $h$, inclusive over the rest of the final state $X$,
\begin{equation}
e^+(k_1) + e^-(k_2) \xrightarrow{\gamma,Z^0}h(P_h) + X
\,\mbox{.}
\label{eq:SIAprocess}
\end{equation}
This process is the time-like counterpart of inclusive DIS, to which
it is related by crossing symmetry.
Similarly to the Bjorken-$x$ variable in DIS, one usually defines the
scaling variable
\begin{equation}
z=2P_h \cdot q/Q^2 
\,\mbox{,}
\label{eq:zscaling}
\end{equation}
with $P_h$ the four-momentum of the outgoing identified hadron, $q=k_1+k_2$
the four-momentum of the exchanged virtual gauge boson, $\sqrt{q^2}=Q$. 
The center-of-mass energy of the electron-positron collision is given by
$\sqrt{s}=Q$.
In this section we provide the details of the dataset included
in this analysis.
We describe first the experiments considered, then the corresponding
physical observables and finally the kinematic cuts that we apply to
the data.

\subsection{The NNFF1.0 dataset}
\label{sec:dataset}

The dataset entering the NNFF1.0 analysis is based on electron-positron
SIA cross-sections for the sum of charged pion, charged kaon, and
proton/antiproton production, {\it i.e.}
$h=\pi^++\pi^-,K^++K^-,p+\bar{p}$ in Eq.~(\ref{eq:SIAprocess}).
These cross-sections are differential with respect to either the scaling
variable $z$, Eq.~(\ref{eq:zscaling}), or a closely related
quantity (see Sect.~\ref{sec:observables}).
We include measurements performed by experiments at
CERN (ALEPH~\cite{Buskulic:1994ft},
DELPHI~\cite{Abreu:1998vq}, and OPAL~\cite{Akers:1994ez}), DESY
(TASSO~\cite{Brandelik:1980iy,Althoff:1982dh,Braunschweig:1988hv}),
KEK (BELLE~\cite{Leitgab:2013qh,Seidl:2015lla} and
TOPAZ~\cite{Itoh:1994kb}), and SLAC (BABAR~\cite{Lees:2013rqd},
TPC~\cite{Aihara:1988su}, and SLD~\cite{Abe:2003iy}).

In addition to inclusive measurements, we also
include flavour-tagged measurements from TPC~\cite{Lu:1986mc},
DELPHI~\cite{Abreu:1998vq} and SLD~\cite{Abe:2003iy} experiments.
The tagged quark flavour refers to the primary quark-antiquark pair produced 
in the $Z/\gamma^*$ decay. 
For these measurements, differential cross-sections corresponding to either 
the sum of light quarks ($u$, $d$, $s$) or to individual charm and bottom 
quarks ($c$, $b$) are provided, with the former obtained by subtracting the 
latter from the inclusive untagged cross-sections.
Unlike inclusive untagged data, heavy-flavour tagged data cannot be measured
directly, but are instead unfolded from flavour enriched samples based
on Monte Carlo simulations. 
These are therefore affected by additional model uncertainties.
The OPAL experiment has also measured fully separated flavour-tagged 
probabilities for a quark flavour to produce a jet containing the hadron 
$h$~\cite{Abbiendi:1999ry}.
We do not include these data because they do not allow for an unambiguous
interpretation in perturbative QCD beyond LO.

We now discuss specific features of some of the datasets
included in the NNFF1.0 analysis.
In the case of the BABAR experiment, two sets of data are available,
based on {\it prompt} and {\it conventional} yields, respectively.
The former includes primary hadrons or decay products from particles
with lifetime shorter than $\tau=10^{-11}$ s.
The latter includes all decay products with lifetime up to $3\times 10^{-1}$ s.
The conventional cross-sections are about $5\% - 15\%$ larger than the prompt 
ones for charged pions and about $10\% - 30\%$ for protons/antiprotons. 
They are almost the same for charged kaons.
Although the conventional dataset was derived by means of an analysis closer
to that adopted by other experiments, we include the prompt dataset
in our baseline fit of charged pions and proton/antiproton FFs.
The motivation for this choice is that the prompt measurements are more
consistent with other SIA data than the conventional measurements.
A similar choice based on similar considerations was adopted in
previous analyses of charged pion
FFs~\cite{deFlorian:2014xna,Sato:2016wqj}.

In the case of the BELLE experiment, various sets of data are also available.
In a first analysis~\cite{Leitgab:2013qh}, based on an integrated
luminosity $\mathcal{L}=68$ fb$^{-1}$, differential cross-sections
were extracted only for charged pions and charged kaons.
A second analysis~\cite{Seidl:2015lla}, based on an increased luminosity
$\mathcal{L}=159$ fb$^{-1}$, was focused instead on the determination
of the proton/antiproton cross-sections. 
In this study charged pion and kaon measurements were updated,
although they were not intended to be publicly released~\cite{Seidl:2016pcm}.
The second analysis differs from the first in a less dense $z$
binning (particularly in the large-$z$ region), a moderately improved
coverage at small $z$ ($z\sim 0.1$ instead of $z\sim 0.2$), smaller
systematic uncertainties, and a slightly larger center-of-mass energy 
($\sqrt{s}=10.58$ GeV instead of $\sqrt{s}=10.52$ GeV).
Here we include the data from Ref.~\cite{Leitgab:2013qh} for charged
pions and kaons, and the data from Ref.~\cite{Seidl:2015lla} for
protons/antiprotons.

In the case of the OPAL experiment, we have excluded the
proton/antiproton measurements because we experienced difficulties in
providing a satisfactory description of the data.
This approach was also adopted in a previous FF 
analysis~\cite{deFlorian:2007hc}, where the proton/antiproton OPAL data were 
shown to be in tension with other SIA data at the same center-of-mass energy, 
$\sqrt{s}=M_Z$ (see also Ref.~\cite{Hirai:2007cx}).
 
\begin{figure}[t]
\centering
\includegraphics[scale=0.21,angle=270,clip=true,trim=0 1.7cm 0 0]{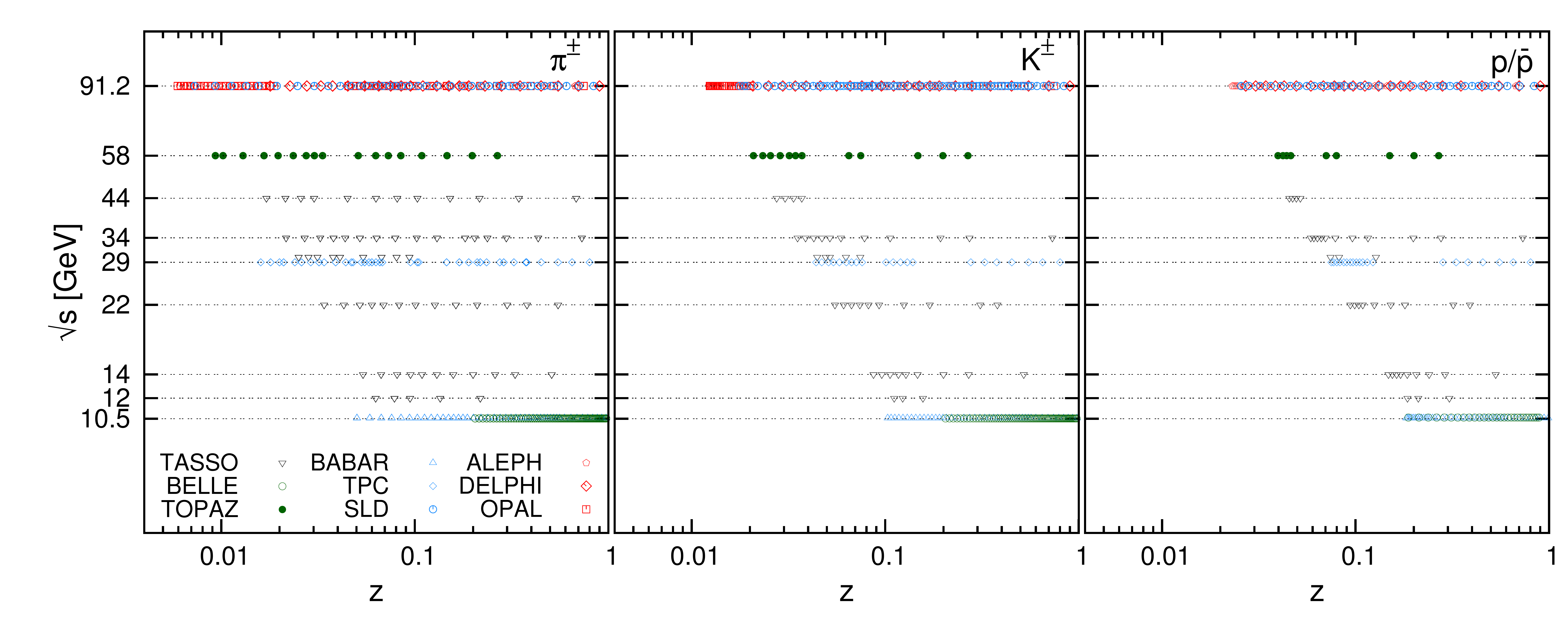}\\
\caption{\small The kinematic coverage in the $(z,\sqrt{s})$ plane of the 
 NNFF1.0 dataset (see Tab.~\ref{tab:datasets}).
 Data are from DESY (black), KEK (green) and CERN (red).}
\label{fig:datakin}
\end{figure}

The dataset included in the NNFF1.0 analysis is summarised in
Tab.~\ref{tab:datasets}, where experiments are ordered by increasing
center-of-mass energy.
In Tab.~\ref{tab:datasets}, we specify the name of the experiment, 
the corresponding publication reference, the measured observable, 
the relative normalisation uncertainty (r.n.u.), 
and the number of data points included in the fit for each hadronic species.
Available datasets that are not included (n.i.) in the NNFF1.0 analysis, 
for the reasons explained above, are also indicated.
The kinematic coverage of the dataset is illustrated in Fig.~\ref{fig:datakin}.

\begin{table}[!t]
\renewcommand{\arraystretch}{2}
\centering
\scriptsize
\begin{tabular}{lclccccc}
\toprule
Exp. & Ref. & Observable & $\sqrt{s}$ [GeV] & 
r.n.u. [\%] & 
$N_{\rm dat}$ ($h=\pi^\pm$) & 
$N_{\rm dat}$ ($h=K^\pm$) & 
$N_{\rm dat}$ ($h=p/\bar{p}$)\\
\midrule
BELLE & \cite{Leitgab:2013qh} 
& $\frac{d\sigma^h}{dz}$                               
& 10.52
& 1.4
& 70 (78)
& 70 (78)
& ---\\
BABAR & \cite{Lees:2013rqd}        
& $\frac{1}{\sigma_{\mathrm{tot}}}\frac{d\sigma^h}{dp_h}$ 
& 10.54
& 0.98
& 40 (45)
& 43 (45)
& 43 (45)\\
BELLE & \cite{Seidl:2015lla} 
& $\frac{d\sigma^h}{dz}$                               
& 10.58
& 1.4
& n.i.
& n.i.
& 29 (29)\\
TASSO12 & \cite{Brandelik:1980iy}    
& $\frac{s}{\beta}\frac{d\sigma^h}{dz}$              
& 12.00
& 20
& 4 (5)
& 3 (3)
& 3 (3)\\
TASSO14 & \cite{Althoff:1982dh}      
& $\frac{s}{\beta}\frac{d\sigma^h}{dz}$              
& 14.00
& 8.5
& 9 (11)
& 9 (9)
& 9 (9)\\
TASSO22 & \cite{Althoff:1982dh}      
& $\frac{s}{\beta}\frac{d\sigma^h}{dz}$              
& 22.00
& 6.3
& 8 (13)
& 6 (10)
& 9 (9)\\
TPC & \cite{Aihara:1988su}       
& $\frac{1}{\beta\sigma_{\mathrm{tot}}}\frac{d\sigma^h}{dz}$
& 29.00
& ---
& 13 (25)
& 13 (21)
& 20 (20)\\
& \cite{Lu:1986mc}
& $\left.\frac{1}{\beta\sigma_{\mathrm{tot}}}\frac{d\sigma^h}{dz}\right|_{uds}$
& 29.00
& ---
& 6 (15)
& ---
& ---\\
& \cite{Lu:1986mc}
& $\left.\frac{1}{\beta\sigma_{\mathrm{tot}}}\frac{d\sigma^h}{dz}\right|_{c}$
& 29.00
& ---
& 6 (15)
& ---
& ---\\
& \cite{Lu:1986mc}
& $\left.\frac{1}{\beta\sigma_{\mathrm{tot}}}\frac{d\sigma^h}{dz}\right|_{b}$
& 29.00
& ---
& 6 (15)
& ---
& ---\\
TASSO30 & \cite{Brandelik:1980iy}    
& $\frac{s}{\beta}\frac{d\sigma^h}{dz}$
& 30.00
& 20 
& -- (9)
& -- (5)
&  2 (5)\\
TASSO34 & \cite{Braunschweig:1988hv} 
& $\frac{1}{\sigma_{\mathrm{tot}}}\frac{d\sigma^h}{dx_p}$
& 34.00
& 6.0 
& 9 (16)
& 5 (11)
& 6 (11)\\
TASSO44 & \cite{Braunschweig:1988hv} 
& $\frac{1}{\sigma_{\mathrm{tot}}}\frac{d\sigma^h}{dx_p}$
& 44.00
& 6.0 
& 6  (12)
& -- (4)
& -- (4)\\
TOPAZ & \cite{Itoh:1994kb}         
& $\frac{1}{\sigma_{\mathrm{tot}}}\frac{d\sigma^h}{d\xi}$
& 58.00
& ---
& 5 (17)
& 3 (12)
& 4 (9)\\
ALEPH & \cite{Buskulic:1994ft}     
& $\frac{1}{\sigma_{\mathrm{tot}}}\frac{d\sigma^h}{dx_p}$
& 91.20
& 3.0 - 5.0
& 23 (39)
& 18 (29)
& 26 (26)\\
DELPHI & \cite{Abreu:1998vq}        
& $\frac{1}{\sigma_{\mathrm{tot}}}\frac{d\sigma^h}{dp_h}$
& 91.20
& ---
& 21 (23)
& 22 (23)
& 22 (23)\\
& \cite{Abreu:1998vq} 
& $\left.\frac{1}{\sigma_{\mathrm{tot}}}\frac{d\sigma^h}{dp_h}\right|_{uds}$
& 91.20
& ---
& 21 (23)
& 22 (23)
& 22 (23)\\
& \cite{Abreu:1998vq} 
& $\left.\frac{1}{\sigma_{\mathrm{tot}}}\frac{d\sigma^h}{dp_h}\right|_{b}$
& 91.20
& ---
& 21 (23)
& 22 (23)
& 22 (23)\\
OPAL & \cite{Akers:1994ez}        
& $\frac{1}{\sigma_{\mathrm{tot}}}\frac{d\sigma^h}{dp_h}$
& 91.20
& ---
& 24 (51)
& 10 (33)
& n.i.\\
SLD & \cite{Abe:2003iy}          
& $\frac{1}{\sigma_{\mathrm{tot}}}\frac{d\sigma^h}{dx_p}$
& 91.20
& 1.0
& 34 (40)
& 35 (36)
& 36 (36)\\
& \cite{Abe:2003iy} 
& $\left.\frac{1}{\sigma_{\mathrm{tot}}}\frac{d\sigma^h}{dx_p}\right|_{uds}$
& 91.20
& 1.0
& 34 (40)
& 35 (36)
& 36 (36)\\
& \cite{Abe:2003iy} 
& $\left.\frac{1}{\sigma_{\mathrm{tot}}}\frac{d\sigma^h}{dx_p}\right|_{c}$
& 91.20
& 1.0
& 34 (40)
& 34 (36)
& 36 (36)\\
& \cite{Abe:2003iy} 
& $\left.\frac{1}{\sigma_{\mathrm{tot}}}\frac{d\sigma^h}{dx_p}\right|_{b}$
& 91.20
& 1.0
& 34 (40)
& 35 (36)
& 35 (35)\\
\midrule
& & & & & 428 (595) & 385 (473) & 360 (382)\\
\bottomrule
\end{tabular}
\caption{\small The dataset included in the NNFF1.0 analysis for charged
 pions, $\pi^\pm=\pi^++\pi^-$, charged kaons, $K^\pm=K^++K^-$, and 
 protons/antiprotons, $p/\bar{p}=p+\bar{p}$.
 For each experiment, we indicate the publication reference, the measured 
 observable, the center-of-mass energy $\sqrt{s}$, the relative normalisation 
 uncertainty (r.n.u.) and the number of data points included, for each hadronic 
 species, after (before) kinematic cuts. 
 Available datasets not included in NNFF1.0 are denoted as n.i., see the text 
 for details.}
\label{tab:datasets}
\end{table}

As one can see from Tab.~\ref{tab:datasets} and
Fig.~\ref{fig:datakin}, the bulk of the NNFF1.0 dataset is composed of
the LEP and SLD measurements, taken at $\sqrt{s}=M_Z$, and of the
$B$-factory measurements, taken at the lower scale $\sqrt{s}\simeq 10$
GeV.
They collectively account for about two thirds of the total dataset
and feature relative uncertainties at the level of a few percent.
The rest of the dataset corresponds to measurements taken at
intermediate energy scales that are typically affected by larger
uncertainties.
From Fig.~\ref{fig:datakin} one also observes that the coverage in $z$ is
limited roughly to the region $0.006\lesssim z \lesssim 0.95$.
As expected from kinematic considerations, experiments at higher
center-of-mass energies provide data at smaller values of $z$, while
experiments at lower center-of-mass energies provide data at larger
values of $z$.

The quantity and the quality of the available data varies depending on the 
hadronic species (see also Figs.~\ref{fig:datatheory1}-\ref{fig:datatheory5}).
Measurements for charged pions, for which the
yield is the largest, are the most abundant and precise.
In comparison, measurements for charged kaons are slightly less abundant and 
precise, while protons/antiprotons measurements are the most sparse and 
the most uncertain among the three hadronic species.
As a consequence, charged pion FFs are better constrained than charged
kaon and proton/antiproton FFs (see Sect.~\ref{sec:ffs}).

We now briefly discuss the differences between the NNFF1.0 dataset
and the dataset included in some of the most recent determinations of FFs.

In comparison to JAM~\cite{Sato:2016wqj}, we do not consider the ARGUS
inclusive~\cite{Albrecht:1989wd}, the HRS
inclusive~\cite{Derrick:1985wd}, and the OPAL fully flavour-tagged
data.
Note that these differences are restricted to the charged pion and charged 
kaon FFs, as proton/antiproton FFs were not determined in the JAM analysis.

In comparison to HKKS~\cite{Hirai:2016loo}, we include the TPC tagged
data for charged pions and remove the HRS data for charged pions and
kaons.
A determination of the proton/antiproton FFs was not performed in
Ref.~\cite{Hirai:2016loo}, but in an earlier analysis based on a
similar framework~\cite{Hirai:2007cx}.
In comparison to this, here we exclude the OPAL inclusive data and
include the BELLE and BABAR data, which were not available when the
analysis in Ref.~\cite{Hirai:2007cx} was performed.

In comparison to DEHSS~\cite{deFlorian:2014xna,deFlorian:2017lwf}, we
include the older TASSO (at $\sqrt{s}=12,14,22,30$ GeV) and TOPAZ
measurements.
These datasets are affected by rather large experimental uncertainties. 
Their effect in the DEHSS fits was deemed negligible and hence they were 
removed.
Note that the DEHSS determinations also include the OPAL fully flavour-tagged
data, not considered here, as well as additional measurements of
hadroproduction in SIDIS and $pp$ collisions.
Similar considerations also apply to their earlier analysis for
proton/antiproton FFs~\cite{deFlorian:2007hc}, in comparison to which
we also include the BELLE and BABAR data.
In the case of proton/antiproton FFs, the NNFF1.0
analysis is the first to include the $B$-factory measurements.

We take into account all the available information on statistical and 
systematic uncertainties, including their correlations.
The full breakdown of bin-by-bin correlated systematics
is provided only by the BABAR experiment. 
No information on correlations among various sources of systematics 
is provided for all the other experiments.
In these cases we sum in quadrature statistical and systematic uncertainties.
Normalisation uncertainties are assumed to be fully correlated across all data 
bins in each experiment.
The asymmetric uncertainties quoted by BELLE  are
symmetrised as described in Ref.~\cite{DelDebbio:2004xtd}.

Systematic uncertainties, with the exception of normalisation uncertainties,
are treated as additive.
Because the naive inclusion of multiplicative normalisation uncertainties in 
the covariance matrix would lead to a biased result~\cite{DAgostini:1993arp}, 
we treat them according to the $t_0$ method~\cite{Ball:2009qv,Ball:2012wy}. 
This method is based on constructing a modified version of the
covariance matrix where the contribution from multiplicative
uncertainties is determined from theory predictions rather than from
the experimental central values for each measurement.
This procedure is iterative, with the results of a fit being used for
the subsequent one until convergence is reached.

The available information on statistical, systematic, and normalisation 
uncertainties is used to construct the covariance matrix associated to each 
experiment.
Following the NNPDF methodology, this covariance matrix is used to generate 
a Monte Carlo sampling of the probability distribution determined by the data.
The statistical sample used in the NNFF1.0 fits is obtained by
generating $N_{\rm rep}=100$ pseudo-data replicas according to a
multi-Gaussian distribution around the data central values and with 
the covariance of the original data~\cite{Ball:2008by}.

\subsection{Physical observables}
\label{sec:observables}

The SIA differential cross-section involving a hadron $h$ in the final
state can be expressed as
\begin{equation}
  \frac{d\sigma^h}{dz}(z,Q)
  =
  \frac{4\pi\alpha^2(Q)}{Q^2}
  F_2^h(z,Q)
  \,\mbox{,}
\label{eq:SIAxsec}
\end{equation}
where $\alpha$ is the Quantum Electrodynamics (QED) running coupling
and $F_2^h$ is the {\it fragmentation} (structure) function, defined in
analogy with the structure function $F_2$ in DIS.
While in the literature $F_2^h$ is often called fragmentation
function, we will denote it as fragmentation structure function in
order to avoid any confusion with the partonic FFs.

The SIA cross-sections used in this analysis are summarised in the third
column of Tab.~\ref{tab:datasets}.
For some experiments, they are presented as multiplicities, {\it i.e.} 
they are normalised to $\sigma_{\rm tot}$, the total 
cross-section for the inclusive electron-positron annihilation into hadrons.
In addition to the normalisation to $\sigma_{\rm tot}$, the format of the
experimental data can differ among the various experiments due to the choice of
scaling variable and/or an additional overall rescaling factor.
These differences are indicated in Tab.~\ref{tab:datasets},
where the following notation is used: $z=E_h/E_b=2E_h/\sqrt{s}$ is the
energy $E_h$ of the observed hadron $h$ scaled to the beam energy
$E_b$; $x_p=|\mathbf{p}_h|/|\mathbf{p}_b|=2|\mathbf{p}_h|/\sqrt{s}$ is
the hadron three-momentum $|\mathbf{p}_h|$ scaled to the beam
three-momentum $|\mathbf{p}_b|$; $\xi=\ln(1/x_p)$; and
$\beta=|\mathbf{p}_h|/E_h$ is the velocity of the observed hadron $h$.

Starting from the measured observables defined in
Tab.~\ref{tab:datasets}, the corresponding data points have been
rescaled by the inverse of $s/\beta$ or $1/\beta$ whenever needed to
match Eq.~(\ref{eq:SIAxsec}), modulo the normalisation to $\sigma_{\rm tot}$.
Corrections depending on the hadron mass $m_h$ are retained according to the
procedure described in Ref.~\cite{Albino:2008fy}.
This implies that the distributions differential in $x_p$, $p_h$ or
$\xi$ are modified by a multiplicative Jacobian factor determined by
the following relations between the scaling variables:
\begin{equation}
  z(p_h)
  =
  2\left(\frac{m_h^2+p_h^2}{s} \right)^{\frac{1}{2}}
  \  
  z(x_p)
  =
  \beta x_p 
  = 
  x_p\left(1+\frac{4}{x_p^2}\frac{m_h^2}{s}\right)^{\frac{1}{2}}
  \ 
  z(\xi)
  = 
  e^{-\xi}\left(1+4\,e^{2\xi}\,\frac{m_h^2}{s}\right)^{\frac{1}{2}}
  \mbox{.}
  \label{eq:scalingvar}
\end{equation}

\begin{figure}[t]
\centering
\includegraphics[scale=0.3,angle=270]{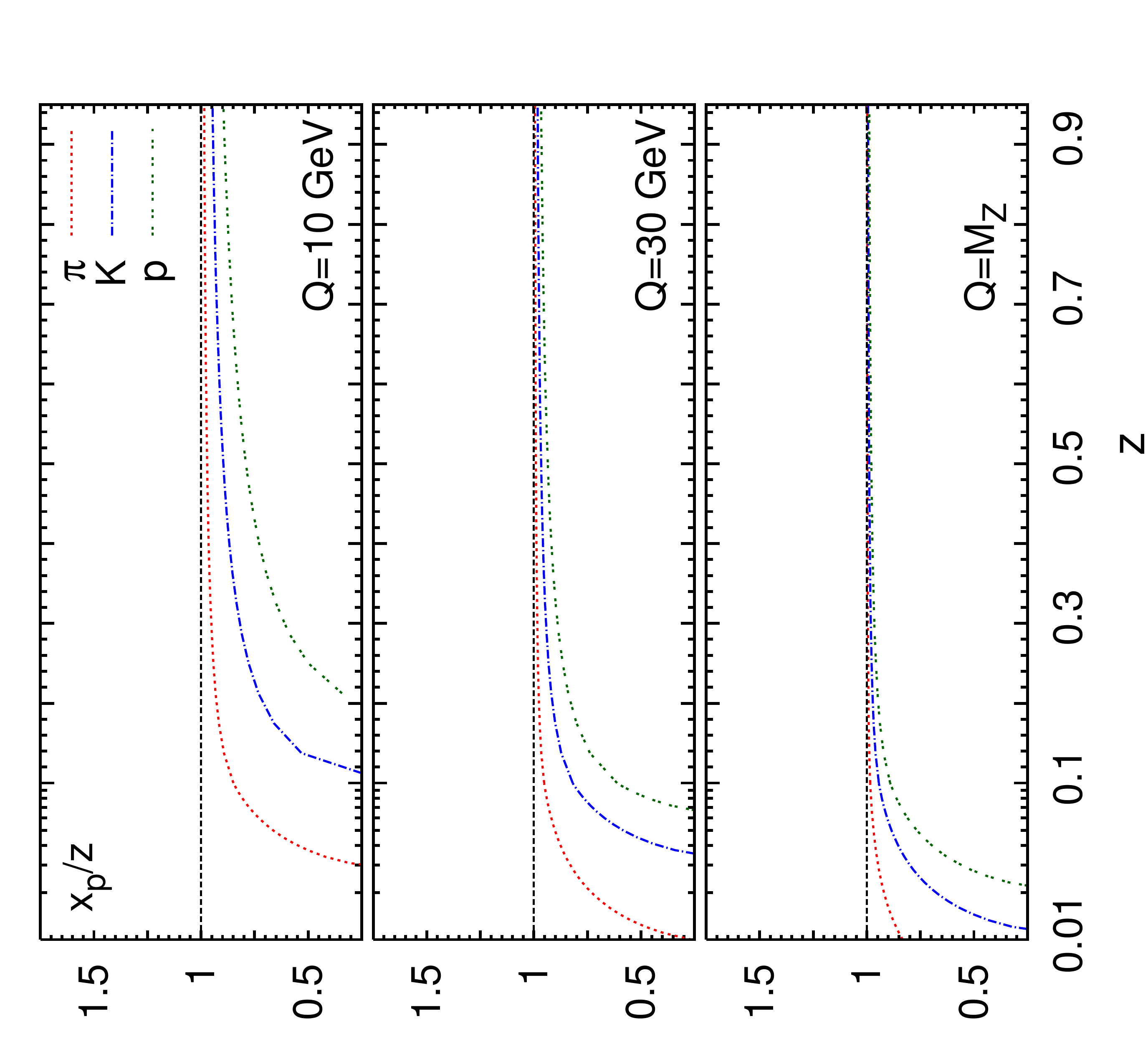}
\includegraphics[scale=0.3,angle=270]{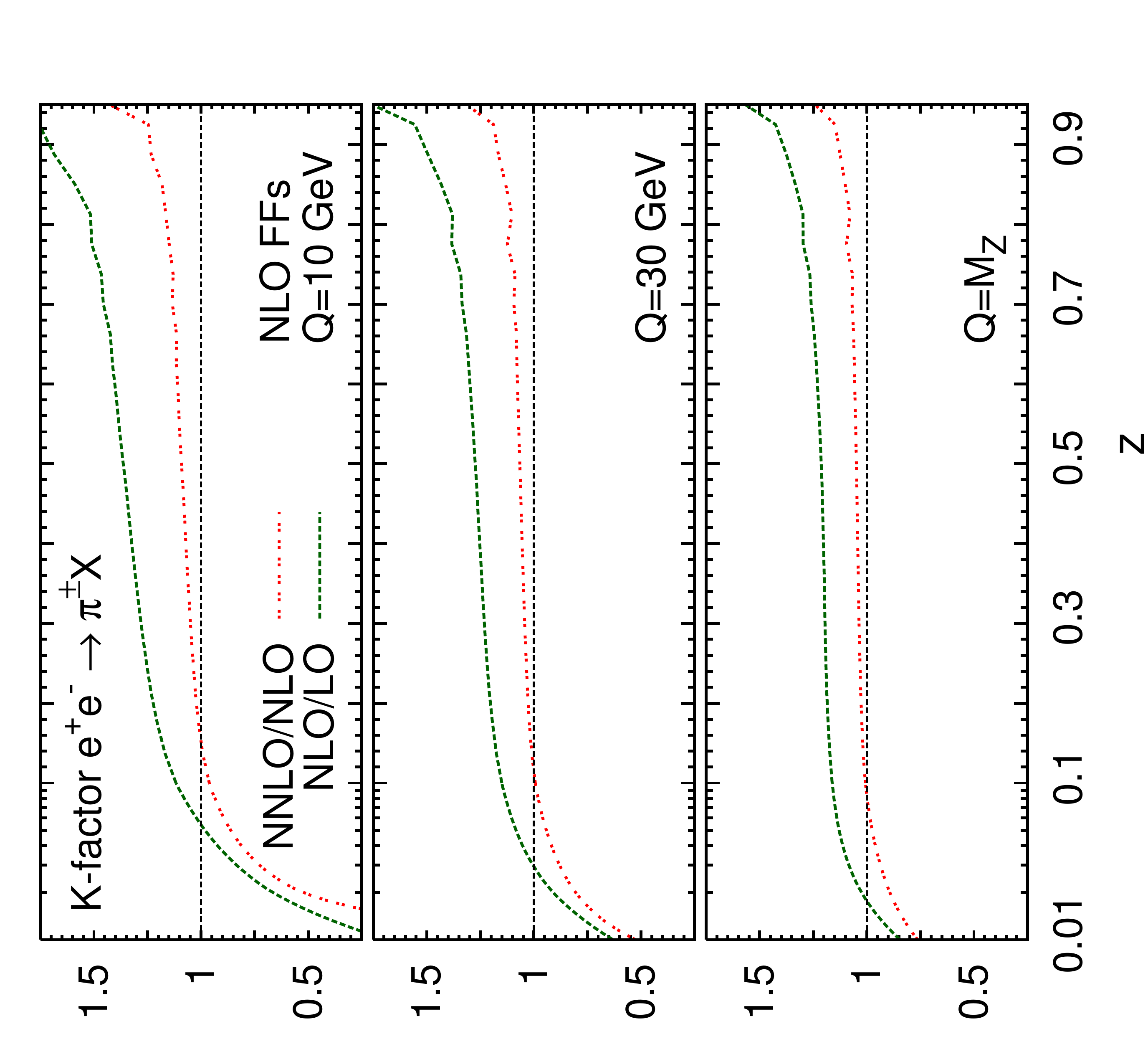}
\caption{\small Left: the scaling variable ratio $x_p/z$
  as a function of $z$, at three representative values of
  $\sqrt{s}$, for pions, kaons and protons.
  Right: the SIA $K$-factor, defined as the ratio 
  $F_2^h({\rm N}^{\rm m}{\rm LO})/F_2^h({\rm N}^{\rm m-1}{\rm LO})$
  for $m=1,2$, at the same three values of $\sqrt{s}$.
  The $K$-factors have been computed with fixed NLO charged pion FFs 
  from the DEHSS determination~\cite{deFlorian:2014xna}.}
\label{fig:corrections}
\end{figure}

The typical size of these hadron-mass corrections is illustrated in the left 
plot of Fig.~\ref{fig:corrections}, where we show the ratio $x_p/z$ as a 
function of $z$, at three representative values of $\sqrt{s}$,
for pions, kaons, and protons.
Hadron-mass corrections become larger when $z$ and/or $\sqrt{s}$ decrease, 
as well as when $m_h$ is increased.
These corrections can become significant in the kinematic region covered by 
the data.
For instance, at $z=0.1$ and $Q=M_Z$ hadron-mass corrections are less
than $10\%$ for all hadronic species, while at $z=0.1$ and $Q=10$ GeV
they rise up to $20\%$ ($70\%$ or more) for pions (kaons and
protons/antiprotons).
For protons/antiprotons, these corrections are already larger than $30\%$ 
around $z=0.4$ at the center-of-mass energy of the $B$-factory data.
Therefore, the inclusion of hadron-mass corrections should improve the
description of the data.

In the case of the BELLE experiment we multiply all data points by a
factor $1/c$, with $c=0.65$ for charged pions and
kaons~\cite{Leitgab:2013dva} and with $c$ a function of $z$ for
protons/antiprotons~\cite{Seidl:2015lla}.
This correction is required in order to treat the BELLE data
consistently with all the other SIA measurements included in NNFF1.0.
The reason is that a kinematic cut on radiative photon events was
applied to the BELLE data sample in the original analysis instead of
unfolding the radiative QED effects.
Specifically, the energy scales in the measured events were kept
within $0.5\%$ of the nominal fragmentation scale $Q/2$; a Monte Carlo
simulation was then performed to estimate the fraction of events with
initial-state (ISR) or final-state radiation (FSR) photon energies
below $0.5\%\times Q/2$.
For each bin, the measured yields are then reduced by these fractions
in order to exclude events with large ISR or FSR contributions.

Finally, note that the $B$-factory measurements correspond to samples
where the effect of bottom-quark production is not included because
they were taken at a center-of-mass energy below the threshold to
produce a $B$-meson pair.
The corresponding theoretical predictions should therefore be computed without 
the bottom-quark contribution, as explained in Sect.~\ref{sec:factorization}.

\subsection{Kinematic cuts}

Our baseline determination of FFs is based on a subset of all the available
data points described above.
Specifically, we impose two kinematic cuts at small and large values
of $z$, $z_{\rm min}$ and $z_{\rm max}$, and retain only the data
points with $z$ in the interval $[z_{\rm min}, z_{\rm max}]$.
These cuts are needed to exclude the kinematic regions where effects
beyond fixed-order perturbation theory should be taken into account
for an acceptable description of the data.
For instance, soft-gluon logarithmic terms proportional to $\ln z$ and
threshold logarithmic terms proportional to $\ln (1-z)$ can
significantly affect the time-like splitting functions and the SIA
coefficient functions below certain values of $z_{\rm min}$ and above
certain values of $z_{\rm max}$.
As a consequence, the convergence of the fixed-order expansion can be spoiled.

While all-order resummation techniques have been developed both at
small~\cite{Vogt:2011jv,Albino:2011si,Albino:2011cm,Kom:2012hd} and
large $z$~\cite{Cacciari:2001cw,Blumlein:2006pj,Moch:2009my,Anderle:2012rq,
Accardi:2014qda}, their inclusion is beyond the scope of the present work.
However, we note that the impact of small- and large-$z$ unresummed
logarithms is alleviated when higher-order corrections are included in
the perturbative expansion of splitting and coefficient functions.
To illustrate the perturbative convergence of the SIA structure function in 
Eq.~(\ref{eq:SIAxsec}), we show in Fig.~\ref{fig:corrections} the SIA 
$K$-factors at three representative values of $\sqrt{s}$.
They are defined as the ratios
$F_2^h({\rm N}^{\rm m}{\rm LO})/F_2^h({\rm N}^{\rm m-1}{\rm LO})$,
for $m=1,2$, and have been computed with fixed NLO charged pion
FFs taken from the DEHSS determination~\cite{deFlorian:2014xna}.
The NNLO/NLO $K$-factors are significantly smaller than the NLO/LO
ones for most of the kinematic range, except at small $z$ and
$\sqrt{s}$, where they become comparable.
For most of the values of $z$ in the kinematic range of the data, the
NNLO corrections are at the level of about $10\%$ or less, except
below $z\sim 0.02$ ($z\sim 0.07$) and above $z\sim 0.9$ at $Q=M_Z$
($Q=10$ GeV), where they become larger.
This suggests that in these regions large logarithms can spoil the
convergence of the truncated perturbative series even at NNLO,
indicating the need of resumming them.

In general the values of $z_{\rm min}$ and $z_{\rm max}$ can vary
with the center-of-mass energy $\sqrt{s}$.
Based on the considerations above,
here we choose the following values of $z_{\rm min}$ and
$z_{\rm max}$: $z_{\rm min}=0.02$ for experiments at $\sqrt{s}=M_Z$;
$z_{\rm min}=0.075$ for all other experiments; and $z_{\rm max}=0.9$
for all experiments.
The same kinematic cuts are applied to the three hadronic species.
The number of data points before applying these kinematic cuts is
reported in parentheses in Tab.~\ref{tab:datasets}.
Further motivation for our choice of $z_{\rm min}$ is provided by studying the
deterioration of the fit quality upon its variation, as we will
discuss in detail in Sect.~\ref{sec:kincutdep}.
Specifically, we find that our choice of $z_{\rm min}$ leads to the smallest 
total $\chi^2$.
The value of $z_{\rm max}$ used here also minimises possible tensions between 
different datasets in the large-$z$ region.

\section{From fragmentation functions to physical observables}
\label{sec:theory}

In this section we review the collinear factorisation of the fragmentation 
structure function and the time-like DGLAP evolution of FFs.
We also provide the details of the numerical computation of the SIA
cross-sections, including our choice of the theoretical settings and
the physical parameters.

\subsection{Factorisation and evolution}
\label{sec:factorization}

The factorised expression of the inclusive fragmentation structure
function $F_2^h(z,Q)$ in Eq.~(\ref{eq:SIAxsec}) is given as a
convolution between coefficient functions and FFs,
\begin{align}
  F_2^{h}(z,Q)
  =
  \left\langle \hat{e}^2(Q) \right\rangle 
  &\left[
    C_{2,q}^{\rm S}\left(z,\alpha_s(Q)\right)
    \otimes 
    D_\Sigma^{h}(z,Q) 
    +
    C_{2,q}^{\rm NS}\left(z,\alpha_s(Q)\right)
    \otimes 
    D_{\rm NS}^{h}(z,Q)
    \right. 
    \nonumber\\
  &+
    \left.
    C_{2,g}^{\rm S}\left(z,\alpha_s(Q)\right)
    \otimes 
    D_g^{h}(z,Q)
    \right]
    \,\mbox{,}
\label{eq:F2convolution}
\end{align}
where both factorisation and renormalisation scales are set equal to
the center-of-mass energy of the collision, $\mu_F=\mu_R=\sqrt{s}=Q$.
In Eq.~(\ref{eq:F2convolution}) $\otimes$ denotes the usual
convolution integral with respect to $z$,
\begin{equation}
  f(z)\otimes g(z) 
  \equiv 
  \int_z^1 \frac{dy}{y}f(y)g\left(\frac{z}{y}\right)
  \,\mbox{,}
\label{eq:convolution}
\end{equation}
and
\begin{equation}
  \left\langle \hat{e}^2(Q)\right\rangle
  \equiv 
  \frac{1}{n_f}\sum_q^{n_f}\hat{e}_q^2(Q)
\label{eq:ewcharges}
\end{equation}
is the average of the effective quark electroweak charges $\hat{e}_q$
(see {\it e.g.}  Appendix~A of Ref.~\cite{deFlorian:1997zj} for their
definition) over the $n_f$ active flavours at the scale $Q$;
$\alpha_s$ is the QCD running coupling; and $C_{2,q}^{\rm S}$,
$C_{2,q}^{\rm NS}$, $C_{2,g}^{\rm S}$ are the coefficient functions
corresponding respectively to the singlet and nonsinglet combinations
of FFs,
\begin{equation}
  D^{h}_\Sigma(z,Q) 
  \equiv 
  \sum_q^{n_f} D^{h}_{q^+}(z,Q)
  \,\mbox{,}
  \ \ \ \ \ \ \ \ \ \
  D_{\rm NS}^{h} (z,Q) 
  \equiv  
  \sum_q^{n_f}\left(\frac{\hat{e}_q^2}{\langle \hat{e}^2\rangle}-1\right) 
  D^{h}_{q^+}(z,Q)
  \,\mbox{,}
\label{eq:Dsingnonsing}
\end{equation}
and to the gluon FF, $D_g^{h}$.  
The notation $D_{q^+}^h \equiv D_q^h+D_{\bar{q}}^h$ has been used.

The total cross-section for $e^+e^-$ annihilation into hadrons
$\sigma_{\rm tot}$, required to normalise the differential
cross-section in Eq.~(\ref{eq:SIAxsec}) in the case of multiplicities,
is
\begin{equation}
  \sigma_{\rm tot}(Q) 
  = 
  \frac{4\pi\alpha^2(Q)}{Q^2}\left(\sum_q^{n_f}\hat{e}^2_q(Q)\right)
  \left(1+\alpha_s K_{\rm QCD}^{(1)}+\alpha_s^2 K_{\rm QCD}^{(2)}+\dots\right)
  \,\mbox{.}
\label{eq:sigmatot}
\end{equation}
The coefficients $K_{\rm QCD}^{(i)}$ indicate the QCD perturbative
corrections to the LO result and are currently known up to
$\mathcal{O}(\alpha_s^3)$~\cite{Gorishnii:1990vf}.

The evolution of FFs with the energy scale $Q$ is governed by the
DGLAP equations~\cite{Gribov:1972ri,Lipatov:1974qm,Altarelli:1977zs,
Dokshitzer:1977sg}
\begin{equation}
  \frac{\partial}{\partial\ln Q^2} D_i^h(z,Q)
  =
  \sum_j 
  P_{ji}\left(z,\alpha_s(Q)\right)
  \otimes 
  D_j\left(z,Q\right)
  \,\mbox{,}
  \ \ \ \ \ \ \ \ \ 
  i,j=q,\bar{q},g
  \,\mbox{,}
\label{eq:DGLAP}
\end{equation}
where $P_{ji}$ are the time-like splitting functions.
The choice of FF combinations defined in
Eq.~(\ref{eq:Dsingnonsing}) allows one to rewrite Eq.~(\ref{eq:DGLAP})
as a decoupled evolution equation
\begin{equation}
  \frac{\partial}{\partial\ln Q^2} D_{\rm NS}^h(z,Q)
  =
  P^+\left(z,\alpha_s(Q)\right)\otimes D_{\rm NS}^h(z,Q)
\label{eq:evNS}
\,\mbox{,}
\end{equation}
for the nonsinglet combination of FFs, and a system of two coupled
equations
\begin{equation}
  \frac{\partial}{\partial\ln Q^2}
  \left(
    \begin{array}{c}
      D_{\Sigma}^h\\
      D_g^h
\end{array}
\right)(z,Q)
=
\left(
  \begin{array}{cc}
    P^{qq} & 2n_fP^{gq}\\
    \frac{1}{2n_f}P^{qg} & P^{gg}
\end{array}
\right)\left(z,\alpha_s(Q)\right)
\otimes
\left(
\begin{array}{c}
D_\Sigma^h\\
D_g^h
\end{array}
\right)(z,Q)
\,\mbox{,}
\label{eq:evSI}
\end{equation}
for the singlet combination of quark FFs and the gluon FF.
In comparison to the space-like case, the off-diagonal splitting
functions are interchanged and multiplied by an extra colour factor.

Both the coefficient functions in Eq.~(\ref{eq:F2convolution}) and
the splitting functions in Eqs.~(\ref{eq:evNS})-(\ref{eq:evSI}) allow for
a perturbative expansion in powers of the QCD coupling
\begin{equation}
  C_{2,i}^{\rm S,NS}\left(z,\alpha_s\right)
  =
  \sum_{k=0}a_s^k\, C_{2,i}^{{\rm S,NS}\,(k)}(z)
  \,\mbox{,}
  \ \ \ \ \ 
  P^{ji,+}\left(z,\alpha_s\right)
  =
  \sum_{k=0}a_s^{k+1}\, P^{ji,+\,(k)}(z)
  \,\mbox{,}
\label{eq:pertexp}
\end{equation}
where $i,j=q,g$ and $a_s \equiv \alpha_s/(4\pi)$.
The SIA coefficient functions have been computed up to
$\mathcal{O}\left(a_s^2\right)$
($k=2$)~\cite{Rijken:1996vr,Rijken:1996ns,
  Rijken:1996npa,Mitov:2006wy,Blumlein:2006rr}, and the time-like
splitting functions up to $\mathcal{O}\left(a_s^3\right)$
($k=2$)~\cite{Mitov:2006ic,Moch:2007tx, Almasy:2011eq}, both in the
$\overline{\rm MS}$ scheme.
A residual theoretical uncertainty on the exact form of $P^{qg,(2)}$
still remains, though this is unlikely to have any phenomenological
relevance~\cite{Moch:2007tx}.
Note that space- and time-like splitting functions are identical at LO,
while they differ at NLO and beyond.

Expressing the SIA fragmentation structure function $F_2^h$ in
Eq.~(\ref{eq:F2convolution}) in terms of the quark
flavour singlet and non-singlet combinations of FFs
defined in Eq.~(\ref{eq:Dsingnonsing}) allows one to identify some of
the limitations that affect a determination of FFs based exclusively on 
SIA data.
These include the following.
\begin{itemize}
\item Quark and antiquark FFs always appear through the combinations 
  $D_{q^+}^h$ in Eqs.~(\ref{eq:F2convolution}) and (\ref{eq:Dsingnonsing}). 
  Therefore, SIA measurements are not sensitive to the separation
  between quark and antiquark FFs.

\item The leading contribution to the gluon coefficient function
  $C_{2,g}^{\rm S}$ in Eq.~(\ref{eq:F2convolution}) is
  $\mathcal{O}(a_s)$, hence the gluon FF directly enters the
  fragmentation structure function starting at NLO.

\item The separation between different light quark flavour FFs is
  probed indirectly via the dependence of the electroweak charges
  $\hat{e}_q$ on the energy scale $Q$.
  For instance, at the scale of the LEP/SLC data, $Q=M_Z$, the
  fragmentation structure function in Eq.~(\ref{eq:F2convolution})
  receives its leading contribution from a $Z$-boson exchange, which
  couples almost equally to up- and down-type quarks.
  At this scale, the term
  $\hat{e}_q^2/\left\langle \hat{e}^2 \right\rangle$ that appears in
  Eq.~(\ref{eq:Dsingnonsing}) is close to one, therefore the 
  quark nonsinglet contribution to the structure function is suppressed.

  Conversely, at the typical scale of the $B$-factory measurements,
  $Q\sim 10$ GeV, SIA largely proceeds via a photon exchange, which couples 
  differently to up- and down-type quarks.
  Therefore the term
  $\hat{e}_q^2/\left\langle \hat{e}^2 \right\rangle$ is significantly
  different from one, and the relative contribution of the quark
  nonsinglet combination to $F_2^h$ is sizeable.

\item A direct handle on the separation between light- and heavy-quark
  flavour FFs is provided by the heavy-flavour tagged data from the
  LEP, SLC, and TPC experiments.
\end{itemize}

\subsection{Computation of SIA cross-sections}
\label{sec:numericalimplementation}

The computation of the SIA cross-sections and the DGLAP evolution of
the FFs are performed in the $\overline{\rm MS}$ factorisation scheme
using the $z$-space public code {\tt APFEL}~\cite{Bertone:2013vaa}.
The numerical solution of the time-like evolution equations,
Eqs.~(\ref{eq:evNS})-(\ref{eq:evSI}), has been extensively benchmarked up to 
NNLO QCD in {\tt APFEL}, see Ref.~\cite{Bertone:2015cwa}.
The {\tt FastKernel} method, introduced in Refs.~\cite{Ball:2010de,
Ball:2012cx} and revisited in Ref.~\cite{Ball:2014uwa}, is used to
ensure a fast evaluation of the theoretical predictions.
We include QED running coupling effects in these calculations.

Concerning the treatment of heavy-quark mass effects, here we adopt the 
zero-mass variable-flavour-number (ZM-VFN) scheme in which they are neglected.
Taking into account such effects would require the use of a general-mass
variable-flavour-number (GM-VFN) scheme, as customarily done in the
case of unpolarised PDF fits~\cite{Forte:2010ta,Guzzi:2011ew,Thorne:2000zd}.
Their inclusion into a fit of FFs could improve the description of
some of the SIA datasets, particularly the BELLE and BABAR measurements
whose center-of-mass energy is not far above the bottom-quark
mass~\cite{Epele:2016gup}.
We leave a possible extension of our analysis along the lines of
Ref.~\cite{Epele:2016gup} for a future work.

Whenever multiplicities should be computed, the differential cross-section
in Eq.~(\ref{eq:SIAxsec}) is normalised to the total cross-section
for electron-positron annihilation
$\sigma_{\rm tot}$ defined in Eq.~(\ref{eq:sigmatot}).
In the calculation of $\sigma_{\rm tot}$, perturbative corrections are 
consistently included up to $\mathcal{O}(1)$, $\mathcal{O}(a_s)$, 
and $\mathcal{O}(a_s^2)$ in the LO, NLO and NNLO fits, respectively.

As mentioned in Sect.~\ref{sec:dataset}, the BELLE and BABAR
measurements correspond to an observable for which the contribution
from bottom quarks is explicitly excluded.
This is taken into account in the corresponding theoretical calculation
by setting to zero the bottom-quark electroweak charge appearing in 
Eq.~(\ref{eq:F2convolution}).
Analogously, only the contributions proportional to the electroweak charges 
of the relevant flavours are retained in Eq.~(\ref{eq:F2convolution}) for the 
computation of the light- and heavy-quark tagged cross-sections.

The values of the physical parameters used in the computation of the
SIA cross-sections and in the evolution of FFs are the same as those used in
the NNPDF3.1 global analysis of unpolarised PDFs~\cite{Ball:2017nwa},
supplemented with the PDG averages~\cite{Olive:2016xmw}
for those parameter values not specified there.
Specifically, we use: $M_Z=91.1876$ GeV for the $Z$-boson mass,
$\alpha_s(M_Z)=0.118$ and $\alpha(M_Z)=1/128$ as reference values for
the QCD and QED couplings, and $m_c=1.51$ GeV and $m_b=4.92$ GeV for
the charm- and bottom-quark pole masses.
For the weak mixing angle and $Z$-boson decay width, entering the
definition of the electroweak charges $\hat{e}_q$, we use
$\sin^2\theta_W=0.231$ and $\Gamma_Z=2.495$ GeV.
Finally, the values of the hadron masses used to correct the scaling
variables in Eq.~(\ref{eq:scalingvar}) are $m_\pi=0.140$ GeV,
$m_K=0.494$ GeV, and $m_p=0.938$ GeV.

\section{Fitting methodology}
\label{sec:methodology}

The NNPDF fitting methodology has been described at length in previous 
publications~\cite{Forte:2002fg,DelDebbio:2004xtd,DelDebbio:2007ee,Ball:2008by,
Ball:2009mk,Ball:2010de,Ball:2011mu,Ball:2011uy}.
In this section, we present its aspects specific to the 
NNFF1.0 analysis, some of which are introduced here for the first time.
We first discuss the parametrisation of FFs in terms of neural
networks, then the minimisation strategy to optimise their parameters,
and finally a comprehensive validation of the whole methodology by
means of closure tests.

\subsection{Neural network parametrisation}
\label{sec:parametrization}

As discussed in Sect.~\ref{sec:factorization}, inclusive SIA data allow for 
the determination of only three independent combinations of FFs, namely the 
singlet and nonsinglet combinations in Eq.~(\ref{eq:Dsingnonsing}) and the 
gluon FF.
In addition, charm- and bottom-quark tagged data make possible to constrain 
two additional nonsinglet combinations involving heavy-quark FFs, adding up 
to a total of five independent combinations of FFs.

We adopt the parametrisation basis
\begin{equation}
  \left\{
    D^h_{u^+},\ D^h_{d^++s^+},\ D^h_{c^+},\ D^h_{b^+},\ D^h_{g}
  \right\}
  \,\mbox{,}
\label{eq:parambasis}
\end{equation} 
where the light-flavour combinations of quark FFs in
Eq.~(\ref{eq:Dsingnonsing}) have been separated according to the
values of the corresponding electroweak charges.
In Eq.~(\ref{eq:parambasis}) we have used the shorthand notation
$D^h_{d^++s^+}=D^h_{d^+}+D^h_{s^+}$.
This parametrisation basis is used for all hadronic species. 
The superscript $h$ in Eq.~(\ref{eq:parambasis}) denotes in turn the
sum of positive and negative pions, $h=\pi^\pm=\pi^++\pi^-$, of
positive and negative kaons, $h=K^\pm=K^++K^-$, or of protons and
antiprotons, $h=p/\bar{p}=p+\bar{p}$.
The choice of the parametrisation basis is of course not unique.
Any linear combination of the FFs in Eq.~(\ref{eq:parambasis}) could
be used.

Each FF in the basis defined in Eq.~(\ref{eq:parambasis}) is
parametrised at the initial scale $Q_0$ as
\begin{equation}
  D_i^h(z,Q_0) = \left[ {\rm NN}_i(z) - {\rm NN}_i(1) \right]
  \,\mbox{,}
  \ \ \ \ \ \
  i=u^+,d^++s^+,c^+,b^+,g
  \,\mbox{,}
\label{eq:paramform}
\end{equation}
where ${\rm NN}_i$ is a neural network, specifically a multi-layer
feed-forward perceptron~\cite{Forte:2002fg,DelDebbio:2007ee}.
It consists of a set of nodes organised into sequential layers, in
which the output $\xi_i^{(l)}$ of the $i$-th node of the $l$-th layer
is
\begin{equation}
  \xi_i^{(l)}=g\left(\sum_j\omega_{ij}^{(l)}\xi_j^{(l-1)}+\theta_i^{(l)} \right)
  \,\mbox{,}
\label{eq:actfunc}
\end{equation}
where the function $g$ is a given \textit{activation function}. 
The parameters
$\left\{\omega_{ij}^{(l)},\theta_i^{(l)}\right\}$, known
as {\it weights} and {\it thresholds} respectively, are determined during
the minimisation procedure.

As in previous NNPDF analyses, here we use the same neural network
architecture for all the fitted FFs, namely 2-5-3-1 (that is, four layers
with 2, 5, 3 and 1 nodes each).
This corresponds to 37 free parameters for each FF, and to a total of 185
free parameters for each hadronic species.
This is to be compared to about 15-30 free parameters per hadronic
species typically used in other determinations of FFs.
The 2-5-3-1 architecture is sufficiently flexible to avoid any
parametrisation bias, as we will demonstrate by means of closure tests
in Sect.~\ref{sec:closuretests}.

In contrast with previous NNPDF analyses, here we do not supplement the
neural network parametrisation with a preprocessing function of the form
$z^\alpha(1-z)^\beta$.
Such a function is used in previous NNPDF fits in order to speed up the
minimisation process.
By absorbing in this prefactor the bulk of the behaviour in the
extrapolation regions, the neural network only has to fit deviations
from it.
In order to minimise potential biases in the final result, the values of the 
exponents $\alpha$ and $\beta$ are chosen for each replica at random within a
suitable range determined iteratively~\cite{Ball:2013lla,Ball:2014uwa}.

A typical fit of FFs in this analysis is by far computationally less expensive
than a typical NNPDF fit of PDFs.
This is because the quantity, the quality, and the variety of the data
are much more limited in the former case than in the latter.
Therefore, removing the preprocessing function from the FF parametrisation 
does not significantly affect the efficiency of the fitting procedure. 
Moreover, it avoids the need for the iterative determination of the
preprocessing exponents.
However, the absence of the preprocessing function in
Eq.~(\ref{eq:paramform}) affects the behaviour of the FFs in the
extrapolation regions at small and large values of $z$, and requires
further modifications of the parametrisation.

At large $z$, it is sufficient to explicitly enforce the constraint
that $D_i^h(z,Q_0)$ vanishes in the limit $z\to 1$ by subtracting the
term ${\rm NN}_i(1)$ in Eq.~(\ref{eq:paramform}).
Indeed, in the case of FFs, the large-$z$ extrapolation region is significantly
reduced as compared to the PDF case.
Data from the $B$-factory and LEP/SLD experiments ensure a kinematic
coverage up to the large-$z$ kinematic cut $z_{\rm max}=0.9$ for all
hadronic species (see Fig.~\ref{fig:datakin}).

At small $z$ instead, the available experimental information
becomes more sparse as the lower kinematic cut $z_{\rm min}=0.02$
($z_{\rm min}=0.075$) at $\sqrt{s}=M_Z$ ($\sqrt{s}<M_Z$) is approached.
Therefore, without the preprocessing function, the behaviour of the FFs in the 
small-$z$ extrapolation region will exhibit a significant dependence on the 
choice of the activation function $g$ in Eq.~(\ref{eq:actfunc}).
For instance, if $g$ is chosen to be the sigmoid function,
$g(x)=\left[1+\exp(-x)\right]^{-1}$, as usual in NNPDF fits, all FF replicas 
freeze out to a constant in a region close and below $z_{\rm min}$.
Such a behaviour is clearly unphysical, thus it should be avoided.

In order to overcome this issue, an activation function that preserves 
the features of the sigmoid in the data region and avoids its 
limitations in the extrapolation region should be adopted.
Specifically, we choose $g$ as 
\begin{equation}
  g(a) = {\rm sign}(a)\ln\left(|a|+1\right)
  \,\mbox{.}
\label{eq:actFF}
\end{equation}
Like the sigmoid, the function in Eq.~(\ref{eq:actFF}) exhibits two
different regimes: it is linear for values of $a$ close to zero and
becomes nonlinear for values of $a$ far from zero.
In contrast with the sigmoid, the function in Eq.~(\ref{eq:actFF})
does not saturate asymptotically to zero (one) for large negative
(positive) values of $a$.
This feature prevents FFs from freezing out to a constant for values
of $z$ close and below the low-$z$ cut $z_{\rm min}$.
We have explicitly verified that the choice of activation function
does not affect the behaviour of the fitted FFs in the kinematic
region covered by data.

An important theoretical requirement on FFs is that they must satisfy
positivity, {\it i.e.} any physical cross-section computed from them
must be positive.
At LO, this simply implies that FFs are positive definite.
Beyond LO FFs do not need to be positive definite, and positivity could be 
imposed for instance by requiring a set of at least as many independent 
observables as the parametrised FFs to be positive.
However, for simplicity, here we impose positivity by requiring that
FFs are positive definite at all orders, as it is customarily assumed
in all other analyses of FFs.
This is achieved by squaring the output of Eq.~(\ref{eq:paramform}).
We have explicitly checked that this choice does not bias our
determination, in that the differences with a fit in which positivity
is not imposed at all are negligible.
This suggests that the quality of the dataset included in our analysis is 
good enough to ensure the positivity of FFs for most of the relevant kinematics.

Finally, we should mention that the initial parametrisation scale in
Eq.~(\ref{eq:paramform}) is $Q_0=5$ GeV.
This value is both above the bottom-quark threshold ($m_b=4.92$ GeV)
and below the lowest center-of-mass energy of the data included in the
fits ($\sqrt{s}=10.52$ GeV).
This choice is advantageous for two reasons.
First, it implies that no heavy-quark thresholds have to be
crossed in the evolution between the initial scale and the
scale of the data. 
Therefore, the number of active flavours is always $n_f=5$ and no
matching is required.
This is advantageous because time-like matching conditions are
currently known only up to NLO~\cite{Cacciari:2005ry}.
Second, in a VFN scheme, our choice implies that charm- and
bottom-quark FFs are parametrised on the same footing as the
light-quark and gluon FFs.
This is beneficial because heavy-quark FFs receive large
non-perturbative contributions. 
Indeed, perturbatively generated heavy-quark FFs lead to a poor
description of the data, in particular of heavy-quark tagged data.

\subsection{Optimisation of neural network parameters}
\label{sec:cmaes}

The determination of neural network parameters in a fit of FFs to
experimental data is a fairly involved optimisation problem.
In our analysis, it requires the minimisation of the $\chi^2$ estimator
\begin{equation}
  \chi^2
  =
  \sum_{i,j}^{N_{\rm dat}}
  (T_i[f]-D_i) (C_{ij}^{t_0})^{-1} (T_j[f]-D_j)
  \,\mbox{,}
\label{eq:chi2def}
\end{equation}
where $i$ and $j$ run over the number of experimental data points
$N_{\rm dat}$, $D_i$ are their measured central values, $T_i$ are the
corresponding theoretical predictions computed with a given set of FFs $f$,
and $C_{ij}^{t_0}$ is the $t_0$ covariance matrix discussed in
Sect.~\ref{sec:dataset}.

In most situations where neural networks are applied, optimisation is 
performed by means of some variation of simple gradient descent.
In order to optimise the model parameters in this way, it is necessary
to be able to straightforwardly compute the gradient of $\chi^2$ 
with respect to model parameters,
\begin{equation}
  \frac{\partial \chi^2}{\partial w_{ij}^{(l)}}
  \,\mbox{,} 
  \quad 
  \frac{\partial \chi^2}{\partial \theta_{i}^{(l)}}
  \,\mbox{.}
\label{eq:param_gradients}
\end{equation}
Computing these gradients in the case of PDF or FF fits is complicated by 
the non-trivial relationship between PDFs/FFs and physical observables.

In previous NNPDF analyses of PDFs, minimisation was performed by means
of a simple example of a {\it genetic algorithm}.
At each iteration of the fit, variations (or mutants) of PDFs are generated 
by random adjustment of the previous best-fit neural-network parameters.
The mutant PDF parameters with the lowest $\chi^2$ to data are then
selected as the best-fit for the next iteration.
Such a procedure is blind to the higher-order structure of the problem
in parameter space and does not require the computation of the
gradients in Eq.~(\ref{eq:param_gradients}).
The most prominent drawback of such a procedure is that it is
considerably less efficient than standard gradient descent.
Furthermore, while this basic procedure is adequate in the case of PDF
fits to a global dataset, it can be sensitive to the noise in the
$\chi^2$ driven by noisy experimental data.

In the present fit of FFs, the dataset is much more limited than in
typical global PDF fits.
It is therefore worth considering alternative minimisation strategies
that may be less sensitive to such effects.
There are a great deal of strategies available in the
literature for the optimisation of problems where standard gradient
descent methods are difficult or impossible to apply.
One such strategy, the Covariance Matrix Adaption - Evolutionary Strategy 
(CMA-ES) family of algorithms~\cite{Hansen2006,DBLP:journals/corr/Hansen16a},
finds regular application in this context.

In this analysis  we apply a standard variant of the CMA-ES procedure for
the minimisation of the neural network parameters.
While we leave the details and the specification of algorithm
parameters to Ref.~\cite{DBLP:journals/corr/Hansen16a}, we will 
outline the procedure schematically here.
We denote the set of fit parameters
$\left\{\omega_{ij}^{(l)},\theta_i^{(l)}\right\}$ as a single vector
$\mathbf{a}^{(i)}$.
In all relevant quantities described here, the superscript $i$ indicates 
the values at the $i^{th}$ iteration of the algorithm.
The fit parameters are initialised at the start of the fit according
to a multi-Gaussian distribution $\mathcal{N}$ with zero mean and unit
covariance
\begin{equation}
  \mathbf{a}^{(0)} \sim \mathcal{N}(0,\mathbf{C}^{(0)})
  \,\mbox{,} 
  \quad 
  \mathbf{C}^{(0)} = \mathbf{I}
  \,\mbox{.}
\label{eq:avect}
\end{equation}
where we use $\sim$ to denote the distribution of the random vector. 
This vector is then used as the centre of a search distribution in parameter space.
At every iteration of the algorithm, $\lambda=80$ mutants
$\mathbf{x}_1, \ldots, \mathbf{x}_\lambda$ of the parameters are generated as
\begin{equation}
  \mathbf{x}_k^{(i)} 
  \sim \mathbf{a}^{(i-1)} + \sigma^{(i-1)}\mathcal{N}(0,\mathbf{C}^{(i-1)})
  \,\mbox{,} 
  \quad 
  \text{for } k = 1\,\mbox{,} \ldots \mbox{,} \lambda  \, ,
\label{eq:mutgen}
\end{equation}
that is, mutants are generated around the search centre according to a
multi-Gaussian $\mathcal{N}$ with covariance $\mathbf{C}^{(i)}$ and
according to a step-size $\sigma^{(i)}$ initialised as
$\sigma^{(0)}=0.1$.
The mutants are then sorted according to their fitness such that
$\chi^2(\mathbf{x}_k) < \chi^2(\mathbf{x}_{k+1})$ and the new search
centre is computed as a weighted average over the $\mu=\lambda/2$ best
mutants
\begin{equation}
  \mathbf{a}^{(i)} 
  = 
  \mathbf{a}^{(i-1)} + \sum_{k=1}^{\mu} w_k 
  \left( \mathbf{x}_k^{(i)} - \mathbf{a}^{(i-1)} \right)
  \,\mbox{,}
\label{eq:bestmut}
\end{equation}
where the weights $\{w_k\}$ are internal parameters of the CMA-ES algorithm.

A key feature of the CMA-ES algorithms is that both the step
size $\sigma^{(i)}$ and the search distribution covariance matrix
$\mathbf{C}^{(i)}$ are optimised by the fit procedure.
To this purpose, the information present in the ensemble of mutants 
is used to learn preferred directions in parameter space without the need for 
the explicit computation of gradients.
This adaptive behaviour improves the efficiency of the minimisation procedure 
in comparison to the genetic algorithm adopted in all previous NNPDF fits.
Also, we implement a \emph{non-elitist} version of the
CMA-ES, whereby each iteration's best fit is computed by means of
weighted average over some number of mutants.
In contrast an \emph{elitist} selection would take only the best mutant from 
each iteration.
In this way the effect of the noise induced in the $\chi^2$ by a 
relatively small dataset should be reduced.

The procedure outlined in Eqs.~(\ref{eq:mutgen})-(\ref{eq:bestmut}) is
iterated until the optimal fit is achieved.
As in previous NNPDF analyses, the stopping point is determined by means of a 
cross-validation method~\cite{Ball:2008by}, based on the separation of the 
whole dataset into two subsets: a training set and a validation set.
Equal training and validation data fractions are chosen for each
experimental dataset, except for those datasets with less than $10$
data points.
In this case, 80\% of the data are included in the training set and
the remaining 20\% in the validation set.
The $\chi^2$ of the training set is then minimised while the $\chi^2$
of the validation set is monitored.
The best-fit configuration is determined according to the {\it look-back} 
criterion~\cite{Ball:2014uwa}, according to which the stopping point
is identified as the absolute minimum of the validation $\chi^2$
within a maximum number of generations, $N_{\rm gen}^{\rm max}$.
Here we take $N_{\rm gen}^{\rm max}=4\times10^4$.
This value is large enough to
guarantee that the best-fit FFs do not depend on it.

Finally, as in the NNPDF3.0~\cite{Ball:2014uwa} and
NNPDF3.1~\cite{Ball:2017nwa} PDF fits, an {\it a posteriori} selection on
the resulting sample of Monte Carlo replicas is performed for each fit.
Specifically, replicas whose $\chi^2$ is more than four-sigma
away from its average value are discarded and replaced by other replicas
which instead satisfy this condition.
This ensures that outliers, which might be present in the Monte Carlo ensemble 
due to residual inefficiencies of the minimisation procedure, are removed.

\subsection{Closure testing fragmentation functions}
\label{sec:closuretests}

The determination of FFs through a fit to experimental data is a
procedure that necessarily implies a number of assumptions, mostly
concerning their parametrisation and the propagation of the
experimental uncertainties into them.
Therefore it is crucial to systematically validate the fitting
methodology in order to avoid any procedural bias that could limit the
reliability of the fitted quantities.

As discussed in Ref.~\cite{Ball:2014uwa}, the robustness of the
fitting procedure used in a global QCD analysis can be assessed by means
of \textit{closure tests}.
The basic idea of a closure test is to perform a fit of FFs to a set
of pseudo-data generated using theoretical predictions obtained with a
pre-existing set of FFs as an input.
In such a scenario, the underlying physical behaviour of the FFs is known by
construction. 
Therefore, it is possible to assess the reliability of the fitting
methodology by comparing the distributions obtained from the fit to
those used as an input.
We refer the reader to Ref.~\cite{Ball:2014uwa} for a thorough
description of the various levels of closure tests 
and of the statistical estimators designed to validate
different aspects of the fitting methodology.
Here we focus on two types of closure tests.
\begin{itemize}

\item \textbf{Level 0} (L0). Pseudo-data in one-to-one correspondence with
  the data discussed in Sect.~\ref{sec:dataset} are generated using the
  theoretical predictions obtained with a given set of FFs; no random
  noise is added at this level.
  Then $N_{\rm rep}$ fits, each to exactly the same set of
  pseudo-data, are performed.
  In order to take into account correlations, the error function that
  is minimised ($i.e.$ the $\chi^2$ evaluated for each replica) is
  still computed using the covariance matrix of the real data, even
  though the pseudo-data have zero uncertainty.

  In a L0 closure test, provided a sufficiently flexible parametrisation
  and a sufficiently efficient minimisation algorithm,
  the fit quality can become arbitrarily good. 
  The $\chi^2$ should decrease to arbitrarily small values, and the
  resulting FFs should coincide with the input ones in the kinematic
  region covered by the pseudo-data.

\item \textbf{Level 2} (L2).  Exactly as in the case of the fits to real data,
  $N_{\rm rep}$ Monte Carlo replicas of the data are
  generated applying the standard NNPDF procedure. 
  The only difference is that the central values of the single measurements are
  replaced by the respective theoretical predictions obtained using 
  the input FFs.
  Then a set of FFs is fitted to each replica.
  
  In a L2 closure test, the final $\chi^2$ should be close to
  unity, provided that the fitted procedure correctly propagates the
  fluctuations of the pseudo-data, due to experimental statistical,
  systematic, and normalisation uncertainties, into the FFs.
  In the kinematic region covered by the data, the input FFs should fall 
  inside the one-$\sigma$ band of their fitted counterparts with a probability 
  of around 68\%.
\end{itemize}

In summary, the goal of a L0 closure test is to assess whether the
fitting methodology, including the parametrisation form and the
minimisation algorithm, is such to avoid any procedural bias.
The goal of a L2 closure test, instead, is to verify whether the fitting
methodology allows for a faithful propagation of the data uncertainties 
into the FFs.

Here we present results for the L0 and L2 closure tests
applied to the determination of the charged pion FFs at NLO. 
We use as input FFs the central distributions of the HKNS07
set~\cite{Hirai:2007cx}.
We have also verified that closure tests are successful for all the hadronic
species considered in the NNFF1.0 analysis, when either the HKNS07 or the 
DSS07 sets~\cite{deFlorian:2007aj,deFlorian:2007hc} at LO or NLO are used as 
an input.

\begin{figure}[!t]
\centering
\includegraphics[scale=0.6,angle=270,clip=true,trim=0 0 1.5cm 0]
{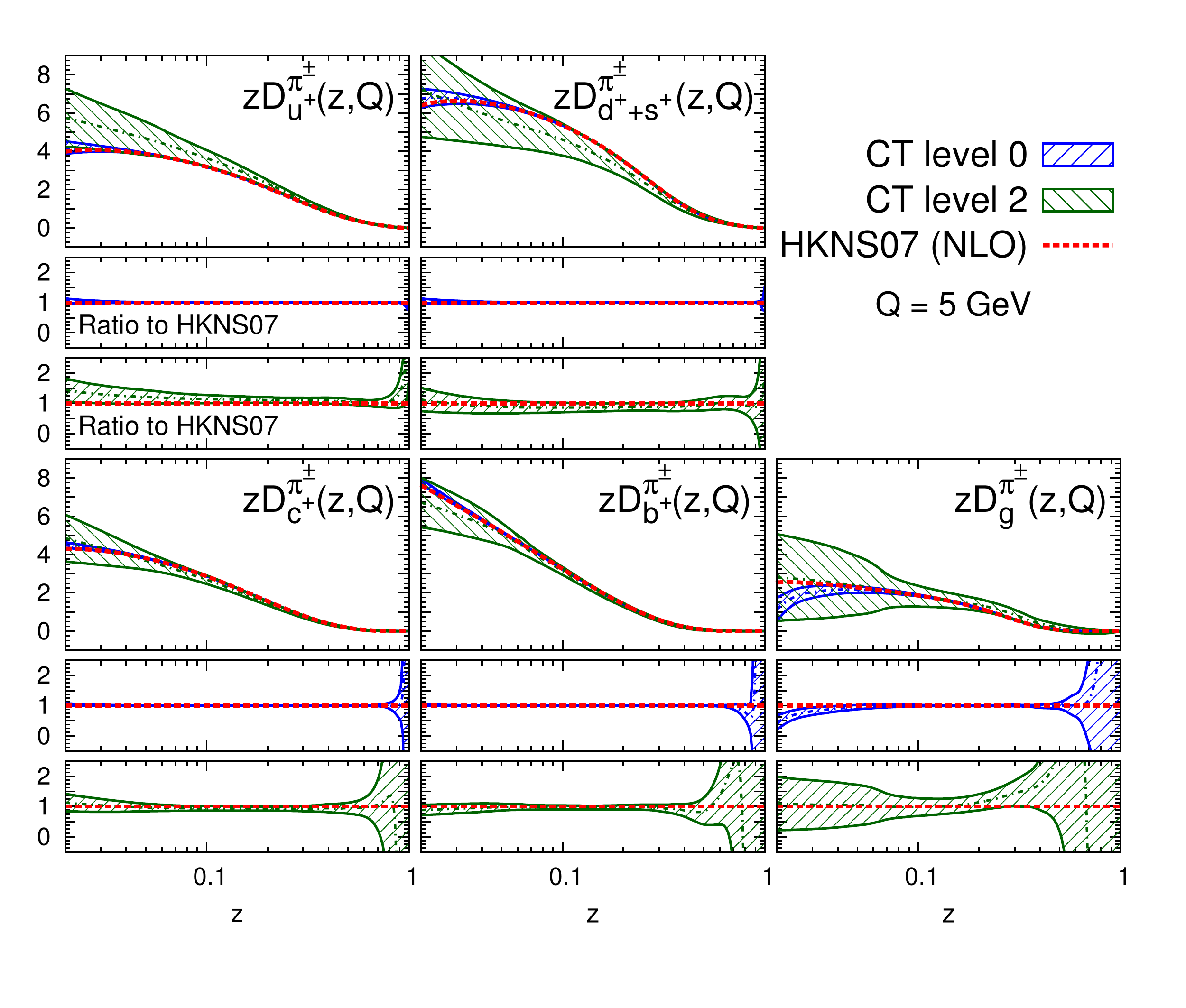}\\
\caption{\small Comparison between the central value of the charged pion HKNS07 
  FFs~\cite{Hirai:2007cx} (red dashed line) and the corresponding FFs 
  obtained from the L0 (blue bands) and L2 (green bands) 
  closure tests. 
  The five panels show the 
  $D_g^{\pi^\pm}$, $D_{u^+}^{\pi^\pm}$, $D_{d^++s^+}^{\pi^\pm}$,
  $D_{c^+}^{\pi^\pm}$, and $D_{b^+}^{\pi^\pm}$ FFs at $Q=5$ GeV,
  in the $z$ range which roughly
  matches the kinematic coverage of the fitted data.
  Shaded bands indicate their one-$\sigma$ uncertainties.
  The  central (lower) inset show the ratio of the L0 (L2) FFs to the 
  HKNS07 FFs.
}
\label{fig:closuretest}
\end{figure}

The value of the total $\chi^2/N_{\rm dat}$ resulting from the L0 and L2 closure 
tests is displayed in Tab.~\ref{tab:closuretestchi2}. 
In Fig.~\ref{fig:closuretest} we compare the input FFs from the HKNS07 set 
and the corresponding fitted FFs from the L0 and L2 closure tests.
The comparison is shown for the five combinations of FFs parametrised in
our analysis, see Eq.~(\ref{eq:parambasis}), at the input scale $Q = 5$ GeV,
and in a range of $z$ which roughly corresponds to the kinematic coverage of 
the data included in the fits.
The upper panel of the plots in Fig.~\ref{fig:closuretest} displays the 
absolute distributions, while the central and the lower panels show the ratio 
of the L0 and L2 FFs to the input HKNS07 FFs, respectively.
The shaded bands for the L0 and L2 distributions indicate the one-$\sigma$ 
FF uncertainty.

\begin{table}[!t]
\scriptsize
\renewcommand{\arraystretch}{1.3}
\centering
\begin{tabular}{lcc}
\toprule
& Level 0 & Level 2  \\
\midrule
$\chi^2/N_{\rm dat}$  & 0.0001 & 1.0262\\
\bottomrule
\end{tabular}
\caption{\small
  The total  $\chi^2/N_{\rm dat}$   
  obtained in the L0 and L2 closure test fits to charged pion
  pseudo-data generated using the HKNS07 NLO FFs as input.}
\label{tab:closuretestchi2}
\end{table}

From Tab.~\ref{tab:closuretestchi2}, as expected, the $\chi^2/N_{\rm dat}$ is 
close to zero for the L0 closure test and close to one for the L2 closure test.
These results indicate that our fitting methodology is adequate to reproduce 
the input FFs without introducing any significant procedural bias.

From Fig.~\ref{fig:closuretest}, it is evident that the FFs of the L0
closure test are almost identical to the input HKNS07 FFs all over the
range in $z$, hence the $\chi^2$ close to zero.
However, we also observe a spread of the uncertainty bands in the very
large-$z$ region.
This is due to the upper kinematic cut ($z_{\rm max} = 0.9$) imposed
on the fitted dataset, such that distributions at large values of $z$
remain unconstrained.
This effect is more enhanced for the gluon, due to the reduced
sensitivity of the data included in the fit to this distribution and
to the smallness of its input FF in that region.
Analogously, in the small-$z$ region, where the data included in the
fit are rather sparse, the fitted FFs display an increase of the
uncertainties.
This confirms that the fitting methodology used here can faithfully reproduce 
the input FFs in the region where the data are sufficiently constraining.

From Fig.~\ref{fig:closuretest}, it is also apparent that the fitted FFs 
in the L2 closure test are in good agreement with the input FFs within
their uncertainties, hence the $\chi^2$ close to one.
This indicates that our fitting methodology correctly propagates the
experimental uncertainty of the data into the uncertainties of the fitted FFs.
As in the case of the L0 closure test, we note that the uncertainty
bands of the FFs in the large- and small-$z$ regions inflate.

In the light of the results of the L0 and L2 closure tests, we
conclude that the fitting methodology adopted here for the
determination of FFs is suitable to ensure a negligible procedural
bias and a faithful representation of their uncertainties.

\section{The NNFF1.0 fragmentation functions}
\label{sec:results}

In this section we present the main results of this work, namely the
NNFF1.0 sets of FFs for charged pions, charged
kaons, and protons/antiprotons at LO, NLO, and NNLO.
First we discuss the quality of the fits and compare the NNFF1.0 predictions 
to the fitted dataset.
Then we show the resulting FFs and their uncertainties, focusing on
their perturbative convergence upon inclusion of higher-order QCD corrections, 
on a comparison of the NLO pion and kaon FFs with their counterparts in the 
DEHSS and JAM analyses, and on the momentum sum rule.

\subsection{Fit quality and data/theory comparison}
\label{sec:fitquality}

\begin{table}[!t]
\renewcommand{\arraystretch}{1.4}
\centering
\scriptsize
\begin{tabular}{lccccccccc}
\toprule
     & \multicolumn{3}{c}{$\chi^2/N_{\rm dat}$ ($h=\pi^\pm$)}
     & \multicolumn{3}{c}{$\chi^2/N_{\rm dat}$ ($h=K^\pm$)}
     & \multicolumn{3}{c}{$\chi^2/N_{\rm dat}$ ($h=p/\bar{p}$)}\\
Exp. & LO & NLO & NNLO & LO & NLO & NNLO &LO & NLO & NNLO \\
\midrule
BELLE               & 0.60 & 0.11 & 0.09 
                    & 0.21 & 0.32 & 0.33 
                    & 0.10 & 0.31 & 0.50 \\
BABAR               & 1.91 & 1.77 & 0.78 
                    & 2.86 & 1.11 & 0.95 
                    & 4.74 & 3.75 & 3.25 \\
TASSO12             & 0.70 & 0.85 & 0.87 
                    & 1.10 & 1.03 & 1.02 
                    & 0.69 & 0.70 & 0.72 \\
TASSO14             & 1.55 & 1.67 & 1.70 
                    & 2.17 & 2.13 & 2.07 
                    & 1.32 & 1.25 & 1.22 \\
TASSO22             & 1.64 & 1.91 & 1.91 
                    & 2.14 & 2.77 & 2.62 
                    & 0.98 & 0.92 & 0.93 \\
TPC (incl.)         & 0.46 & 0.65 & 0.85 
                    & 0.94 & 1.09 & 1.01 
                    & 1.04 & 1.10 & 1.08 \\
TPC ($uds$ tag)     & 0.78 & 0.55 & 0.49 
                    & ---  & ---  & ---  
                    & ---  & ---  & ---  \\
TPC ($c$ tag)       & 0.55 & 0.53 & 0.52 
                    & ---  & ---  & ---  
                    & ---  & ---  & ---  \\
TPC ($b$ tag)       & 1.44 & 1.43 & 1.43 
                    & ---  & ---  & ---  
                    & ---  & ---  & ---  \\
TASSO30             & ---  & ---  & ---  
                    & ---  & ---  & ---  
                    & 0.25 & 0.19 & 0.18 \\
TASSO34             & 1.16 & 0.98 & 1.00 
                    & 0.27 & 0.44 & 0.36 
                    & 0.82 & 0.81 & 0.78 \\
TASSO44             & 2.01 & 2.24 & 2.34 
                    & ---  & ---  & ---  
                    & ---  & ---  & ---  \\
TOPAZ               & 1.04 & 0.82 & 0.80 
                    & 0.61 & 1.19 & 0.99 
                    & 0.79 & 1.21 & 1.19 \\
ALEPH               & 1.68 & 0.90 & 0.78 
                    & 0.47 & 0.55 & 0.56 
                    & 1.36 & 1.43 & 1.28 \\
DELPHI (incl.)      & 1.44 & 1.79 & 1.86 
                    & 0.28 & 0.33 & 0.34 
                    & 0.48 & 0.49 & 0.49 \\
DELPHI ($uds$ tag)  & 1.30 & 1.48 & 1.54 
                    & 1.38 & 1.49 & 1.32 
                    & 0.47 & 0.46 & 0.45 \\
DELPHI ($b$ tag)    & 1.21 & 0.99 & 0.95 
                    & 0.58 & 0.49 & 0.52 
                    & 0.89 & 0.89 & 0.91 \\
OPAL                & 2.29 & 1.88 & 1.84 
                    & 1.67 & 1.57 & 1.66 
                    & ---  & ---  & ---  \\
SLD (incl.)         & 2.33 & 1.14 & 0.83 
                    & 0.86 & 0.62 & 0.57 
                    & 0.66 & 0.65 & 0.64 \\
SLD ($uds$ tag)     & 0.95 & 0.65 & 0.52 
                    & 1.31 & 1.02 & 0.93 
                    & 0.77 & 0.76 & 0.78 \\
SLD ($c$ tag)       & 3.33 & 1.33 & 1.06 
                    & 0.92 & 0.47 & 0.38 
                    & 1.22 & 1.22 & 1.21 \\
SLD ($b$ tag)       & 0.45 & 0.38 & 0.36 
                    & 0.59 & 0.67 & 0.62 
                    & 1.12 & 1.29 & 1.33 \\
\midrule
Total dataset       & \bf 1.44 & \bf 1.02 & \bf 0.87 
                    & \bf 1.02 & \bf 0.78 & \bf 0.73 
                    & \bf 1.31 & \bf 1.23 & \bf 1.17 \\
\bottomrule
\end{tabular}
\caption{\small The  values of $\chi^2/N_{\rm dat}$ for each hadronic species, 
  perturbative order and experiment included in the NNFF1.0 analysis. 
  The number of data points $N_{\rm dat}$ in each case is reported in 
  Tab.~\ref{tab:datasets}.
}
\label{tab:chi2s}
\end{table}

In Tab.~\ref{tab:chi2s} we report the values of the $\chi^2$ per data
point, $\chi^2/N_{\rm dat}$, for both the individual and the total
datasets included in the NNFF1.0 analysis. 
The values are shown at LO, NLO, and NNLO for all the hadronic species.

Concerning the fit quality of the total dataset, the most noticeable
feature is the sizeable improvement upon inclusion of higher-order
corrections.
The improvement of the total $\chi^2/N_{\rm dat}$ is particularly pronounced when 
going from LO to NLO, and more moderate, but still significant, when going from 
NLO to NNLO.
This demonstrates that the inclusion of the NNLO corrections 
improves the description of the data.
This effect is particularly evident for the charged pion fits,
which are based on the most abundant and accurate dataset.

Concerning the fit quality of the individual experiments, the general
trend of the $\chi^2/N_{\rm dat}$ is the same as that of the total
$\chi^2/N_{\rm dat}$, with two main exceptions.
First, the $\chi^2/N_{\rm dat}$ value for charged kaons and protons/antiprotons
data in the BELLE experiment, despite remaining good, increases as higher-order
QCD corrections are included.
Such an increase is accompanied by a decrease of the $\chi^2/N_{\rm dat}$ value 
in the BABAR experiment.
Since the kinematic coverage of these two experiments largely overlaps, 
and given the precision of the corresponding measurements, the opposite 
trend of the $\chi^2/N_{\rm dat}$ suggests a possible tension between the two.
Such a tension was already reported in Ref.~\cite{Hirai:2016loo},
although its origin is still not completely understood.
Second, the $\chi^2/N_{\rm dat}$ value for inclusive and light-tagged charged 
pion data in the DELPHI experiment
deteriorates as higher-order QCD corrections are included.
This behaviour indicates a possible tension between the DELPHI
measurements and the other datasets at the same scale
($\sqrt{s}=M_Z$), whose description instead significantly improves
when going from LO to NNLO.
The origin of such a tension arises mostly from the large-$z$ region, where
the DELPHI inclusive and light-tagged measurements for charged pions
are undershot by the theo\-re\-ti\-cal predictions.

From Tab.~\ref{tab:chi2s} we also observe that in all our fits the
$\chi^2/N_{\rm dat}$ value for the BELLE experiment is anomalously small.
This result was already observed in previous
analyses~\cite{deFlorian:2014xna,deFlorian:2017lwf,Hirai:2016loo,Sato:2016wqj}
and is likely to be due to an overestimate of the uncorrelated 
systematic uncertainty.

We also notice that for some datasets the $\chi^2/N_{\rm dat}$ is poor
even at NNLO: this happens specifically for the TASSO14, TASSO22, TASSO44
and OPAL experiments in the case of charged pions, for the TASSO14, TASSO22, 
and OPAL experiments in the case of charged kaons, and for the BABAR 
experiment in the case of protons/antiprotons.
As we will show below, the experimental data points for the TASSO
datasets fluctuate around the theoretical predictions by an amount
that is typically larger than their uncertainties. 
This explains the poor $\chi^2/N_{\rm dat}$ values reported in 
Tab.~\ref{tab:chi2s} at all perturbative orders.
For charged kaons and pions, the large $\chi^2/N_{\rm dat}$ associated
to the OPAL data comes from the large-$z$ region, where theoretical
predictions overshoot the data.
For charged protons/antiprotons, the large $\chi^2/N_{\rm dat}$ of the BABAR
experiment is driven by a genuine tension between BABAR and TPC/TASSO34 
data below $z=0.2$.
Indeed, if the TPC and TASSO34 data are removed from the fits, the
value of the $\chi^2/N_{\rm dat}$ for the BABAR experiment improves
significantly (see Sect.~\ref{sec:datdep}).

\begin{figure}[!t]
\centering
\includegraphics[scale=0.16,angle=270,clip=true,trim=0 0 4cm 0]
{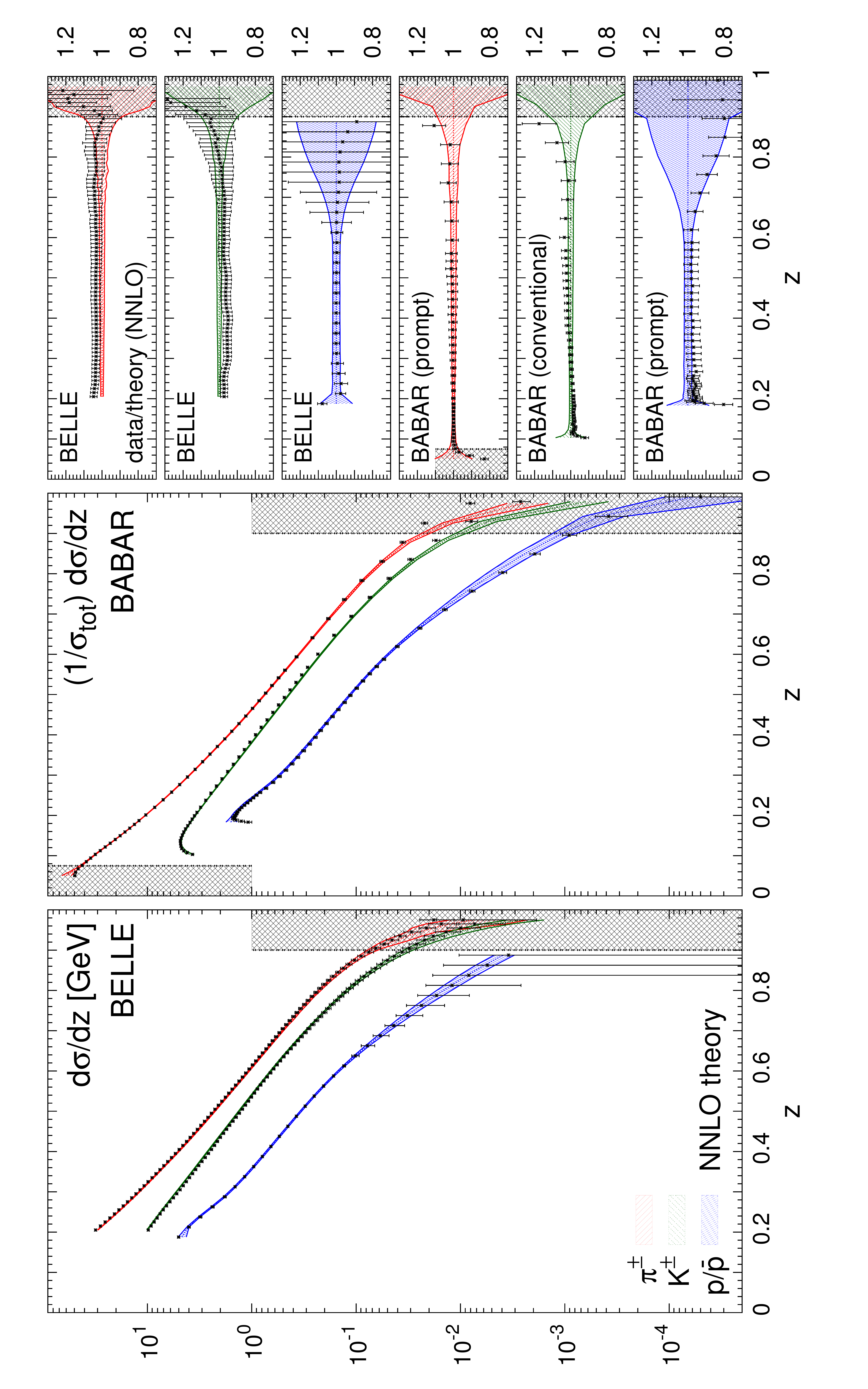}\\
\caption{\small Comparison between the dataset for
  charged pions, charged kaons, and protons/antiprotons
  from the BELLE and BABAR experiments and the corresponding NNLO
  theoretical predictions using our best-fit NNLO FFs.
  We show both the absolute distributions (left and central panel)
  and the data/theory ratios (right panel).  
  Shaded areas indicate the kinematic regions excluded by our cuts,
  and the bands correspond to one-$\sigma$ FF uncertainties.}
\label{fig:datatheory1}
\end{figure}

\begin{figure}[!t]
\centering
\includegraphics[scale=0.16,angle=270,clip=true,trim=0 0 15cm 0]
{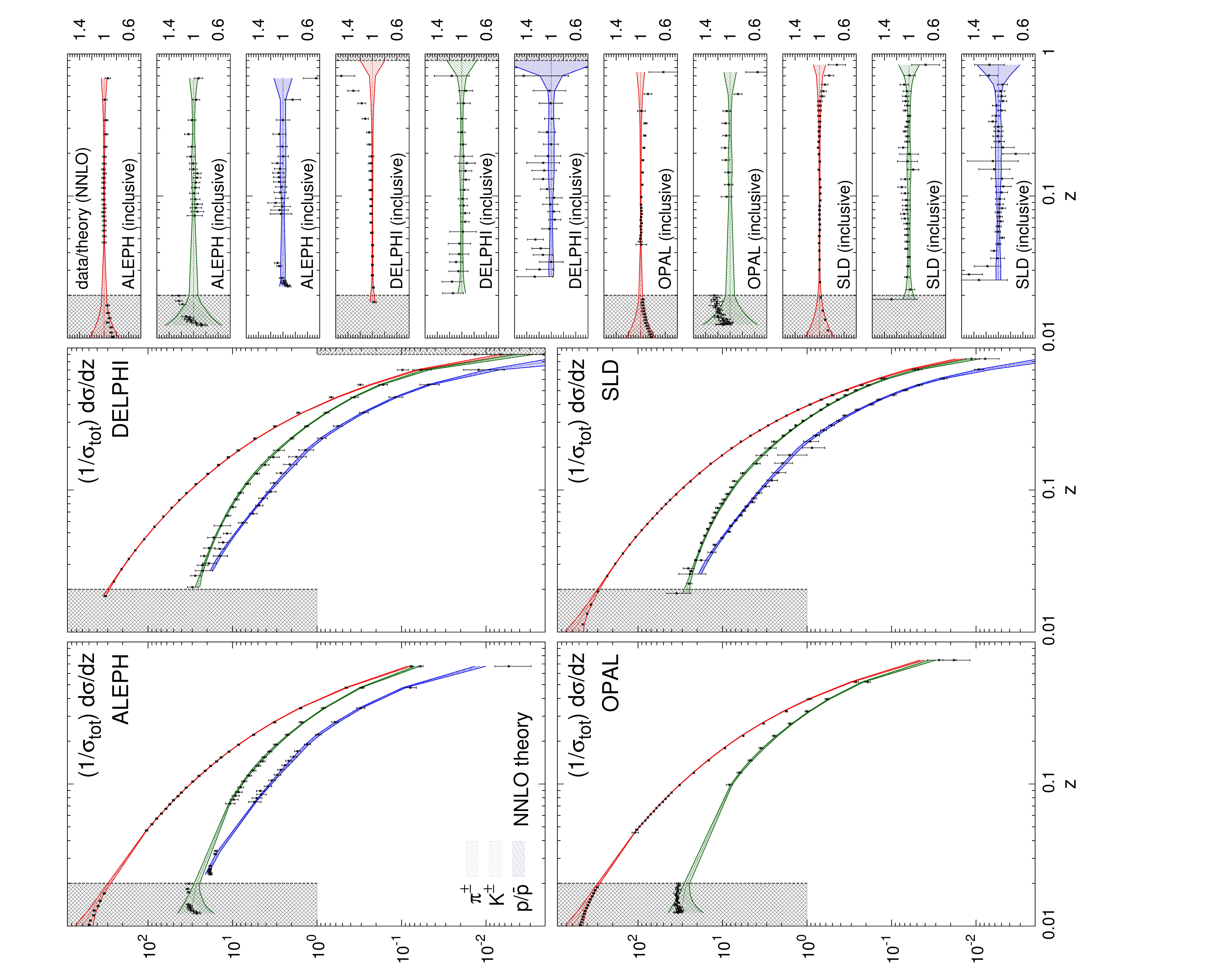}\\
\caption{\small Same as Fig.~\ref{fig:datatheory1} for the ALEPH, DELPHI,
  OPAL and SLD inclusive measurements.}
\label{fig:datatheory2}
\end{figure}

\begin{figure}[!t]
\centering
\includegraphics[scale=0.16,angle=270,clip=true,trim=0 0 4cm 0]
{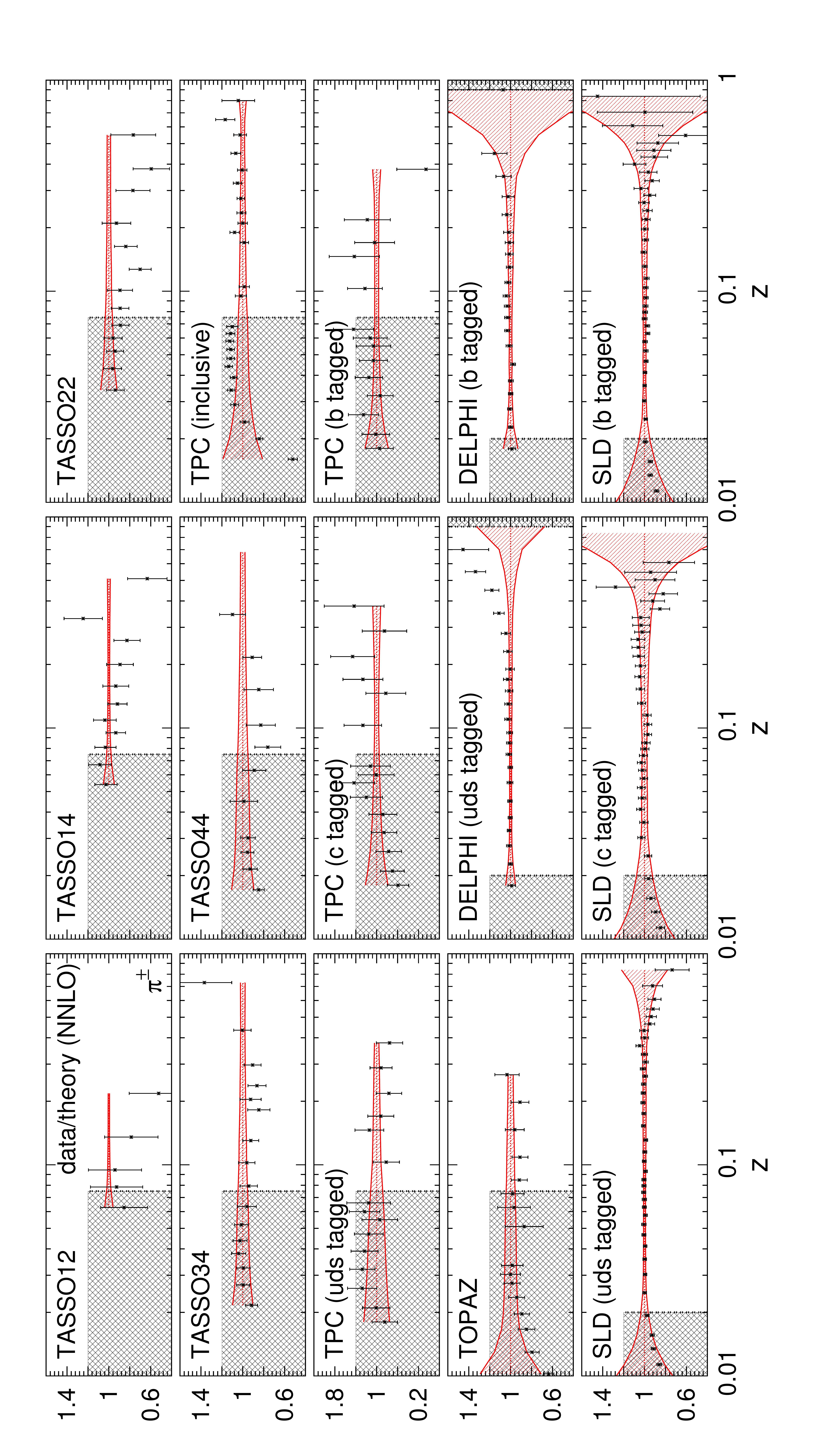}\\
\caption{\small The data/theory ratio for the charged pion 
  data included in the NNFF1.0 fit and not accounted for
  in Figs.~\ref{fig:datatheory1} and~\ref{fig:datatheory2}.
  As before, theoretical predictions are computed at NNLO with our best-fit 
  NNLO FFs, shaded areas indicate the regions excluded by kinematic cuts,
  and the bands correspond to one-$\sigma$ FF uncertainties.}
\label{fig:datatheory3}
\end{figure}

\begin{figure}[!t]
\centering
\includegraphics[scale=0.16,angle=270,clip=true,trim=0 0 14cm 0]
{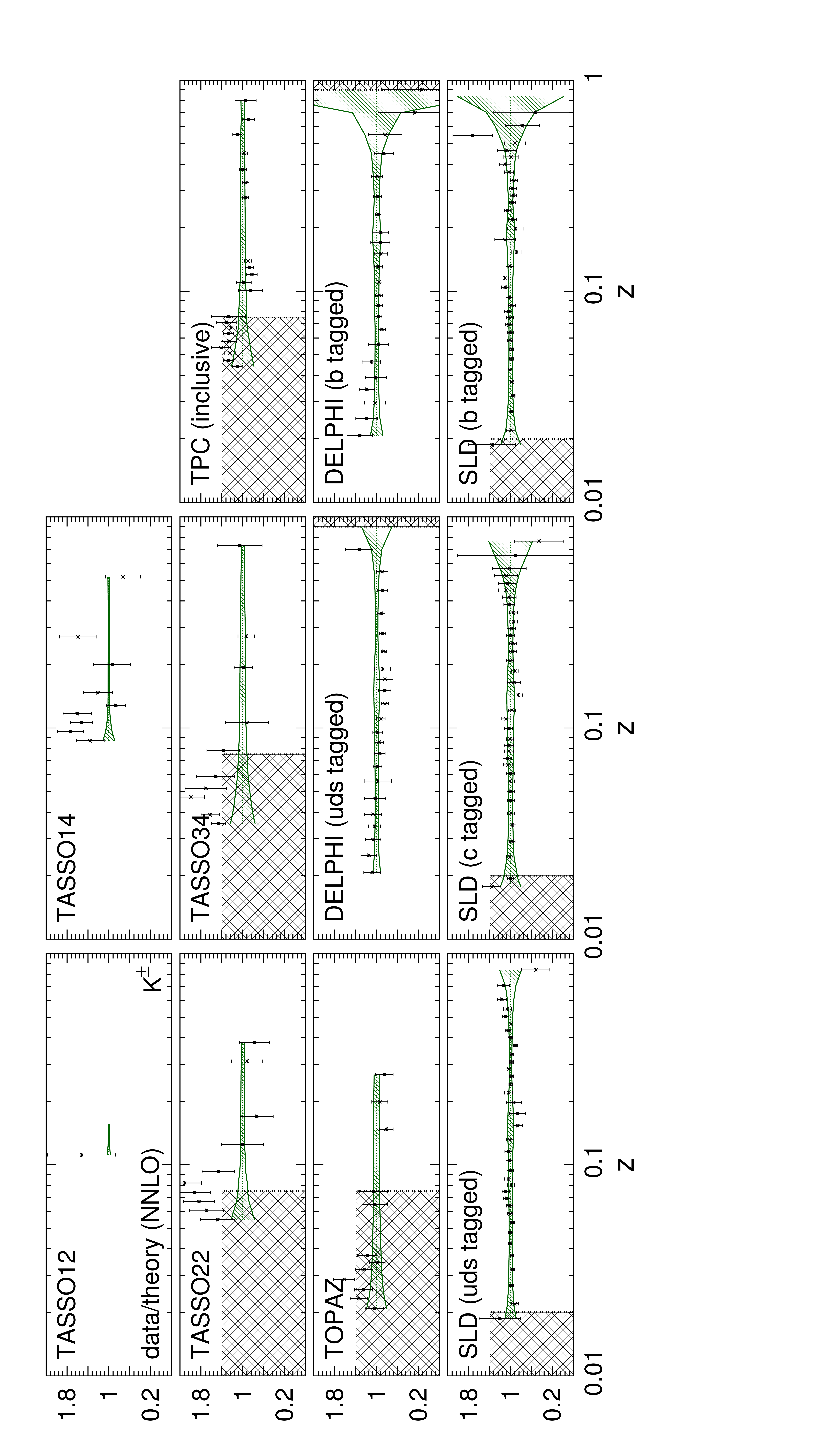}\\
\caption{\small Same as Fig.~\ref{fig:datatheory3} but for charged
kaons.}
\label{fig:datatheory5}
\end{figure}

\begin{figure}[!t]
\centering
\includegraphics[scale=0.16,angle=270,clip=true,trim=0 0 14cm 0]
{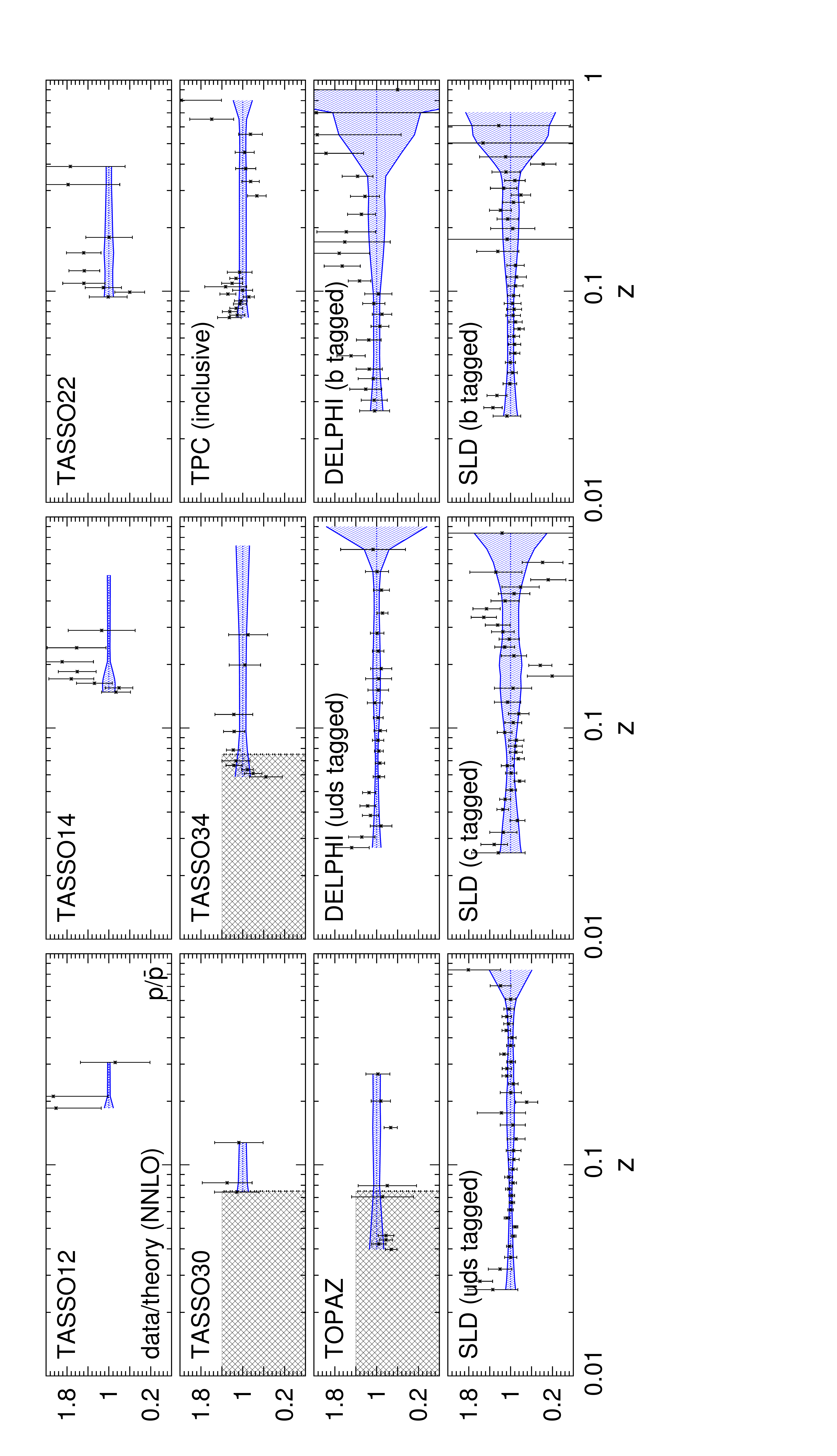}\\
\caption{\small Same as Fig.~\ref{fig:datatheory3} but for 
protons/antiprotons, $p/\bar{p}$.}
\label{fig:datatheory4}
\end{figure}

In order to give further weight to these considerations, we present a
comparison of the dataset used in this analysis to the corresponding
NNLO theoretical predictions obtained using the NNLO FFs from our
fits.
In Fig.~\ref{fig:datatheory1} such a comparison is displayed for the BELLE and 
BABAR data for charged pions, charged kaons, and protons/antiprotons. 
The plots on the r.h.s. of Fig.~\ref{fig:datatheory1} display the corresponding 
data/theory ratios.
The uncertainty bands indicate the one-$\sigma$ FF uncertainty, while
the shaded areas represent the regions excluded by kinematic cuts (see
Sect.~\ref{sec:observables}).
In Fig.~\ref{fig:datatheory2} we show the same comparison as in
Fig.~\ref{fig:datatheory1} for all the inclusive measurements at
$\sqrt{s}=M_Z$.
To complete the picture, we display the data/theory ratios for all the
remaining datasets: in Fig.~\ref{fig:datatheory3} for charged pions,
in Fig.~\ref{fig:datatheory5} for charged kaons, and in
Fig.~\ref{fig:datatheory4} for protons/antiprotons.

In general, an overall good agreement between data and theoretical
predictions is achieved for all experiments, consistently with the
$\chi^2/N_{\rm dat}$ values reported in Tab.~\ref{tab:chi2s}.
Remarkably, theoretical predictions and data are in reasonable agreement also 
in the small- and large-$z$ extrapolation regions excluded by kinematic cuts, 
although the uncertainties of the predictions inflate in these regions.

A few remarks concerning the individual datasets are in order.
A significant deviation from the theoretical predictions is observed
for the low-$z$ proton/antiproton measurements from the BABAR experiment.
This is the origin of the large $\chi^2$ reported in Tab.~\ref{tab:chi2s}.
As already mentioned and as we will further demonstrate in
Sect.~\ref{sec:datdep}, this is a consequence of the tension between
the BABAR and TPC/TASSO34 measurements.
We have explicitly verified that the low-$z$ BABAR data can be
satisfactorily described if the TPC and TASSO34 datasets are removed
from the fit.
However, we have chosen to keep these two experiments in our baseline
dataset because FFs turn out to be very stable irrespective of their
inclusion (see Sect.~\ref{sec:datdep}).

The BABAR measurements for pions and kaons tend to be overshot by the
NNLO theoretical predictions at large $z$.
This is not the case for the BELLE data that cover a similar large-$z$ region 
and are fairly described by the predictions.
This points to a tension between the BELLE and BABAR measurements in
that region.
We also note that, as compared to pions and kaons, the BELLE and BABAR
proton/antiproton measurements are affected by larger experimental
uncertainties, especially at $z\gtrsim 0.6$.
This consistently propagates into 
larger uncertainty bands for the corresponding predictions.

In the case of the DELPHI experiment, theoretical predictions undershoot the 
data for charged pions at $z\gtrsim 0.3$.
This is the reason of the large $\chi^2/N_{\rm dat}$ value reported in
Tab.~\ref{tab:chi2s} for this experiment.
The tension between DELPHI and the other experiments at the same value
of $\sqrt{s}$ (ALEPH, OPAL, and SLD), which are instead well described
by our FFs, is apparent from Fig.~\ref{fig:datatheory2}.

Some of the observations made in this section on possible tensions
between different experiments in certain kinematic regions will be
quantified in Sect.~\ref{sec:datdep}, where a thorough study of the
stability of our fits upon variations of the dataset will be
presented.

\subsection{Fragmentation functions}
\label{sec:ffs}

We now turn to discuss the NNFF1.0 sets.
In order to study the perturbative convergence of the FFs upon
inclusion of higher-order QCD corrections, we first compare our LO,
NLO, and NNLO determinations among each other.
Then we compare our best-fit NLO pion and kaon FFs to their
counterparts in the DEHSS and JAM analyses.
Finally, we conclude with a comment on the momentum sum rule.

\subsubsection{Perturbative stability}

\begin{figure}[!t]
\centering
\includegraphics[scale=0.6,angle=270,clip=true,trim=0 0 1.5cm 0]
{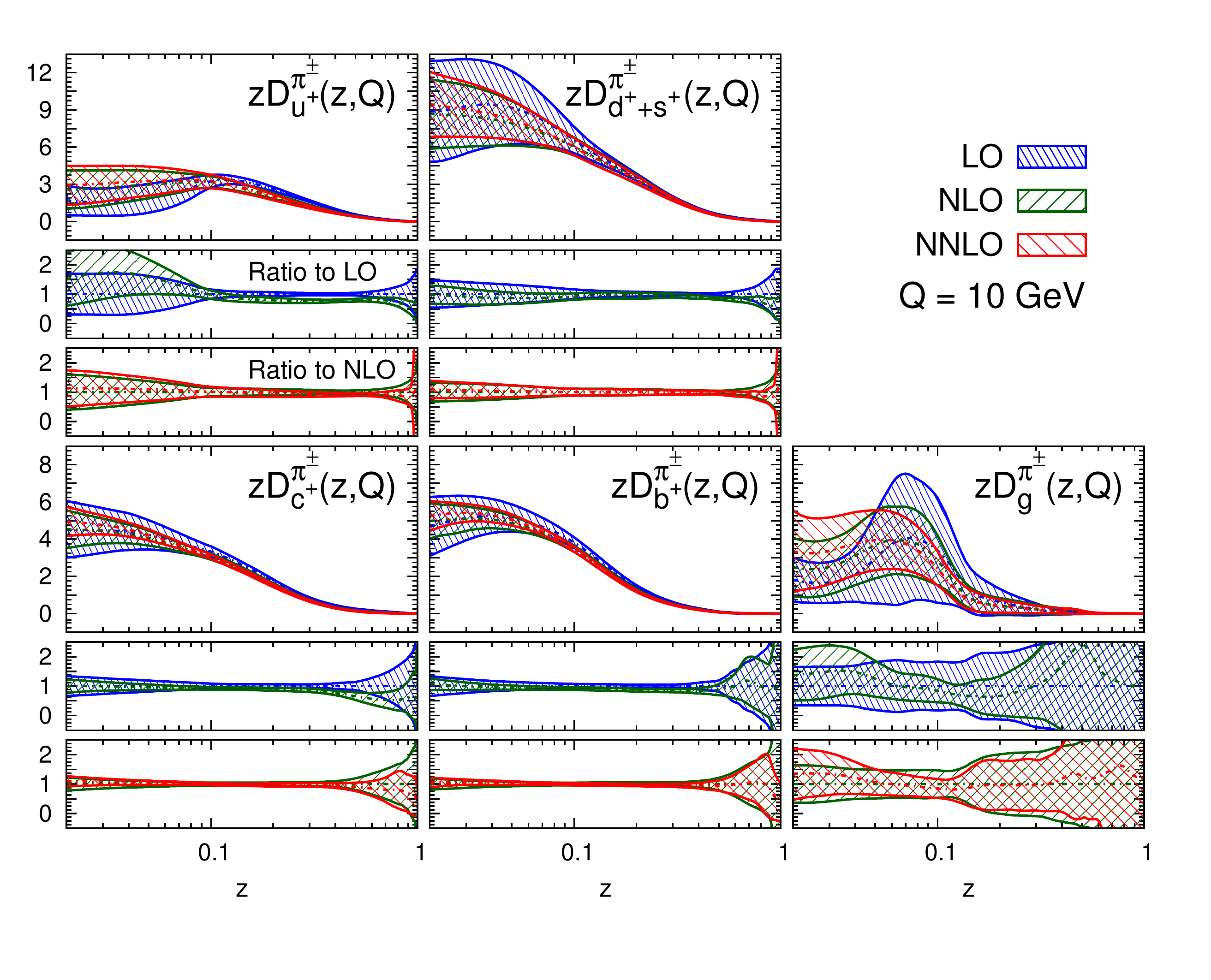}\\
\caption{\small Comparison among the LO, NLO, and NNLO NNFF1.0 charged pion
  FFs, together with their one-$\sigma$ uncertainties,
  in the parametrisation basis of Eq.~(\ref{eq:parambasis}) at $Q=10$ GeV.
  The corresponding NLO/LO and NNLO/NLO ratios are displayed in the
  two insets below each FF.}
\label{fig:pertPI}
\end{figure}

\begin{figure}[!t]
\centering
\includegraphics[scale=0.6,angle=270,clip=true,trim=0 0 1.5cm 0]
{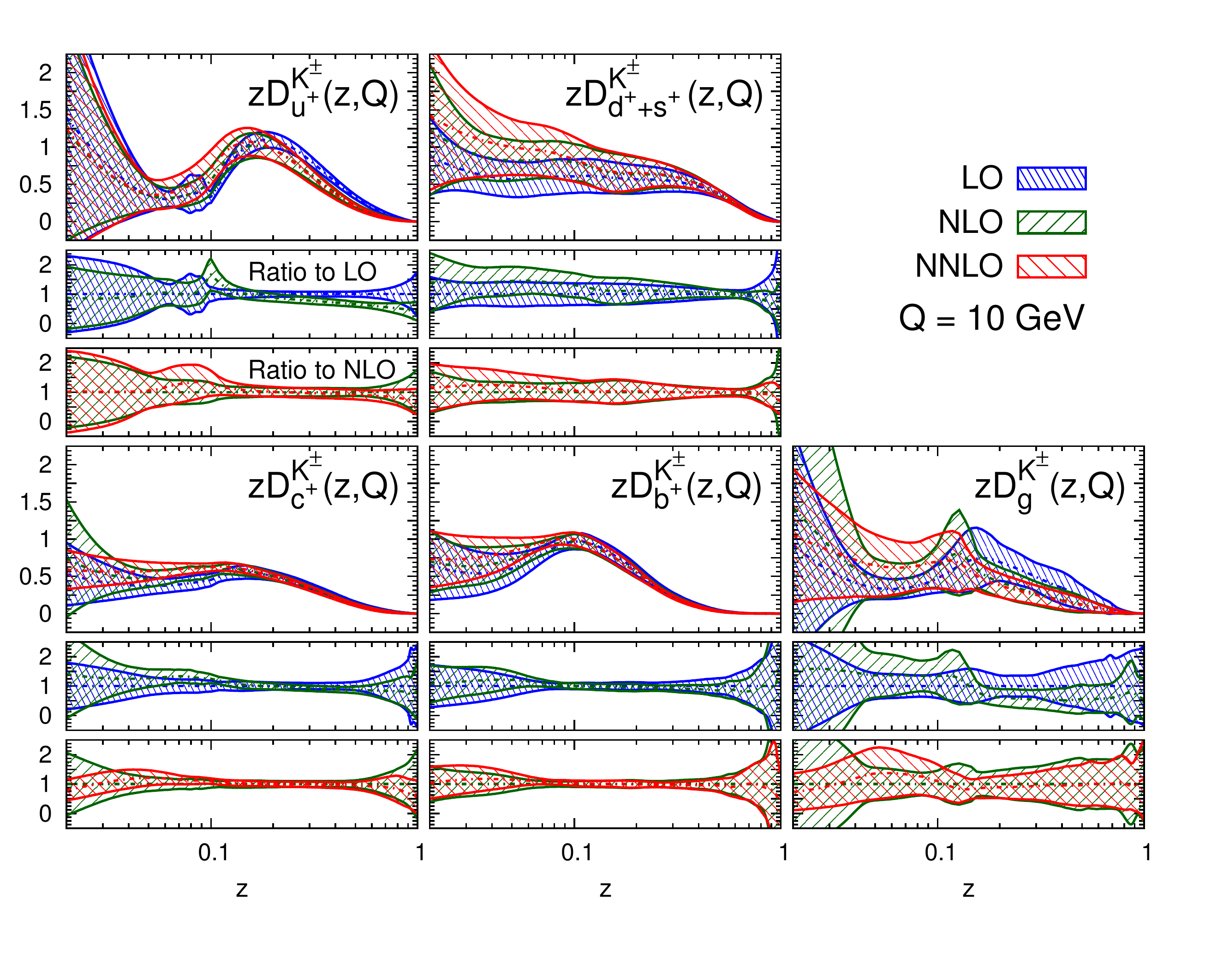}\\
\caption{\small Same as Fig.~\ref{fig:pertPI} but for the sum of charged 
   kaons, $K^\pm$.}
\label{fig:pertKA}
\end{figure}

\begin{figure}[!t]
\centering
\includegraphics[scale=0.6,angle=270,clip=true,trim=0 0 1.5cm 0]
{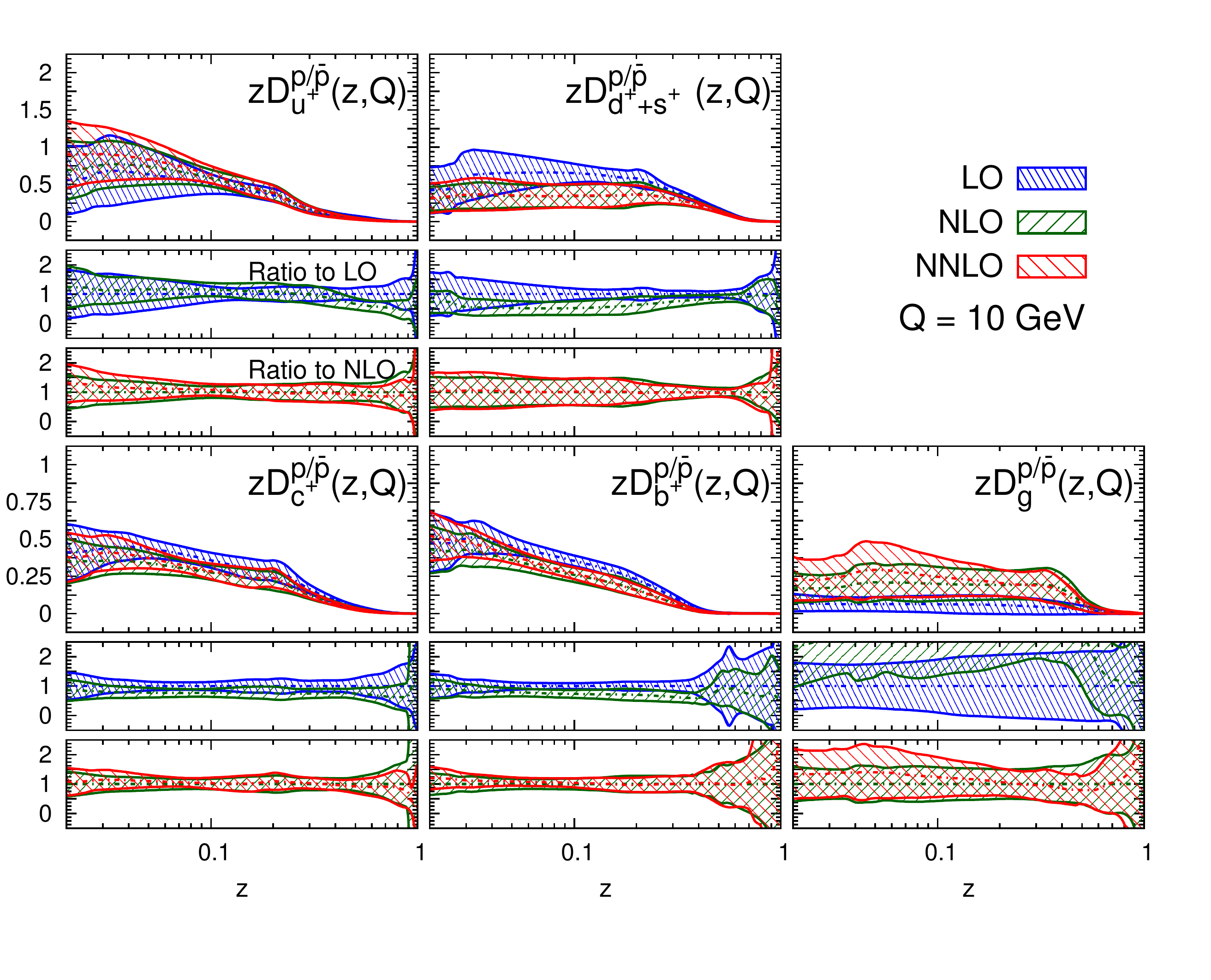}\\
\caption{\small Same as Fig.~\ref{fig:pertPI} but for the sum of protons and 
   antiprotons, $p/\bar{p}$.}
\label{fig:pertPR}
\end{figure}

We display the five FF combinations parametrised in our fits,
Eq.~(\ref{eq:parambasis}), and their one-$\sigma$ uncertainties in
Figs.~\ref{fig:pertPI},~\ref{fig:pertKA}, and~\ref{fig:pertPR} for
charged pions, charged kaons and protons/antiprotons respectively.
For each hadronic species, the FFs are shown at LO, NLO, and NNLO as
functions of $z$ at $Q=10$ GeV.
The upper panel of each plot displays the absolute distributions,
while the central and the lower panels display the NLO/LO and NNLO/NLO
ratios.

A remarkable feature of the distributions shown in 
Figs.~\ref{fig:pertPI}-\ref{fig:pertPR} is their perturbative convergence.
While LO and NLO distributions can in some cases differ by more than one 
standard deviation (see for example $D_{u^+}^h$ for the three hadronic species 
in the medium-/large-$z$ region), the differences between NLO and NNLO FFs are 
small.
This is consistent with the perturbative convergence of the global
$\chi^2$ discussed in Sect.~\ref{sec:fitquality}.

A further noticeable aspect of the comparison in 
Figs.~\ref{fig:pertPI}-\ref{fig:pertPR} 
is related to the size of the FF uncertainties.
While the NLO and NNLO uncertainties are similar in size, the LO
uncertainty bands are in general visibly larger, particularly those of
the gluon FFs.
This was expected because LO predictions for SIA data are only
indirectly sensitive to the gluon FF through DGLAP evolution.
This entails a broadening of the uncertainties of all FFs due to the
cross-talk induced by DGLAP evolution.

\subsubsection{Comparison with other FF sets}

\begin{figure}[!t]
\centering
\includegraphics[scale=0.6,angle=270,clip=true,trim=0 0 1.5cm 0]
{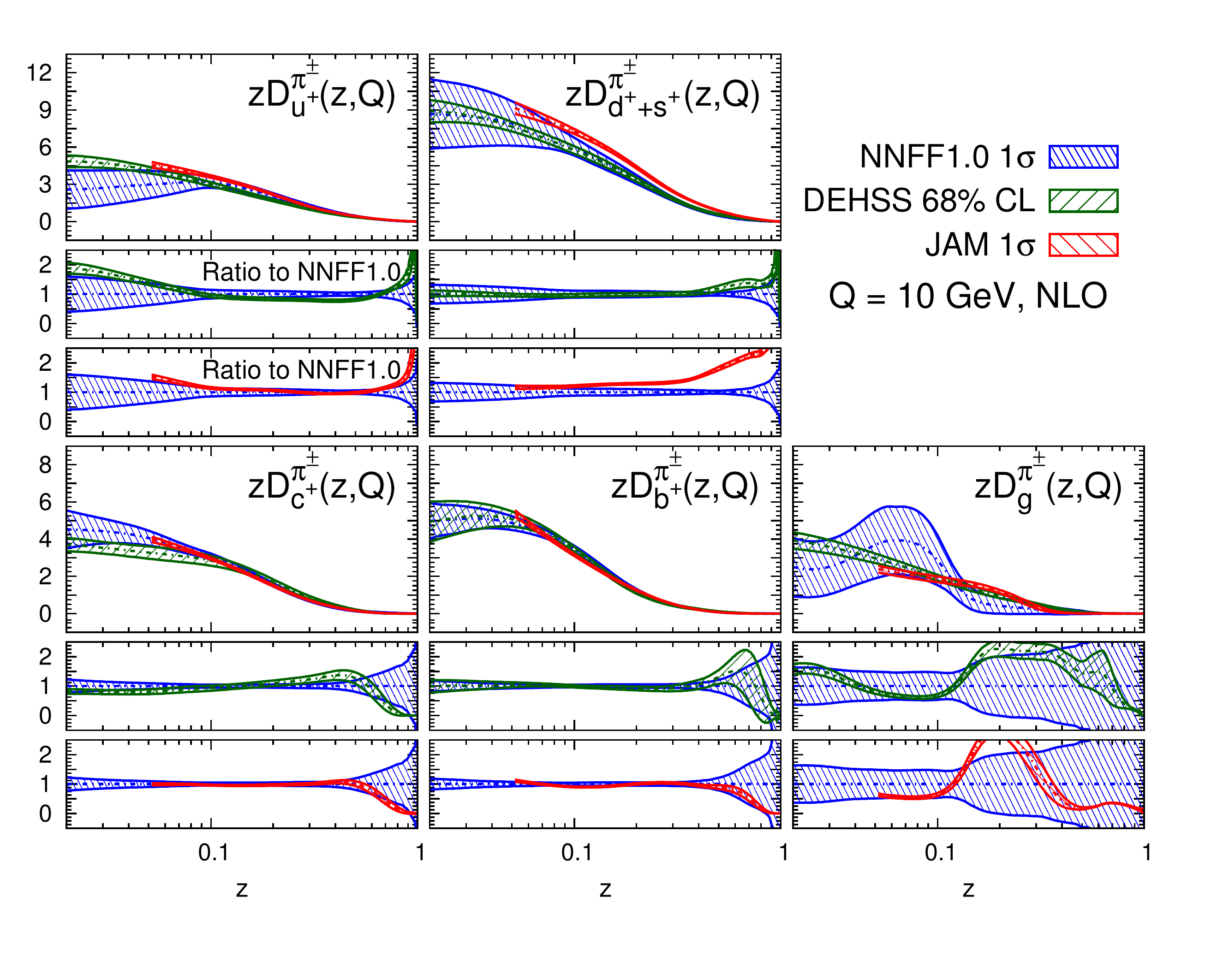}\\
\caption{\small Comparison among the NLO NNFF1.0, DEHSS and JAM FF sets
  for the sum of charged pions, $\pi^\pm$.
  The FFs in the parametrisation basis, Eq.~(\ref{eq:parambasis}), are shown at 
  $Q=10$ GeV as a function of $z$, together with their corresponding
  one-$\sigma$ uncertainties.
  The ratios of NNFF1.0 to DEHSS and JAM are displayed respectively in
  the two insets below each FF.}
\label{fig:compPI}
\end{figure}

\begin{figure}[!t]
\centering
\includegraphics[scale=0.6,angle=270,clip=true,trim=0 0 1.5cm 0]
{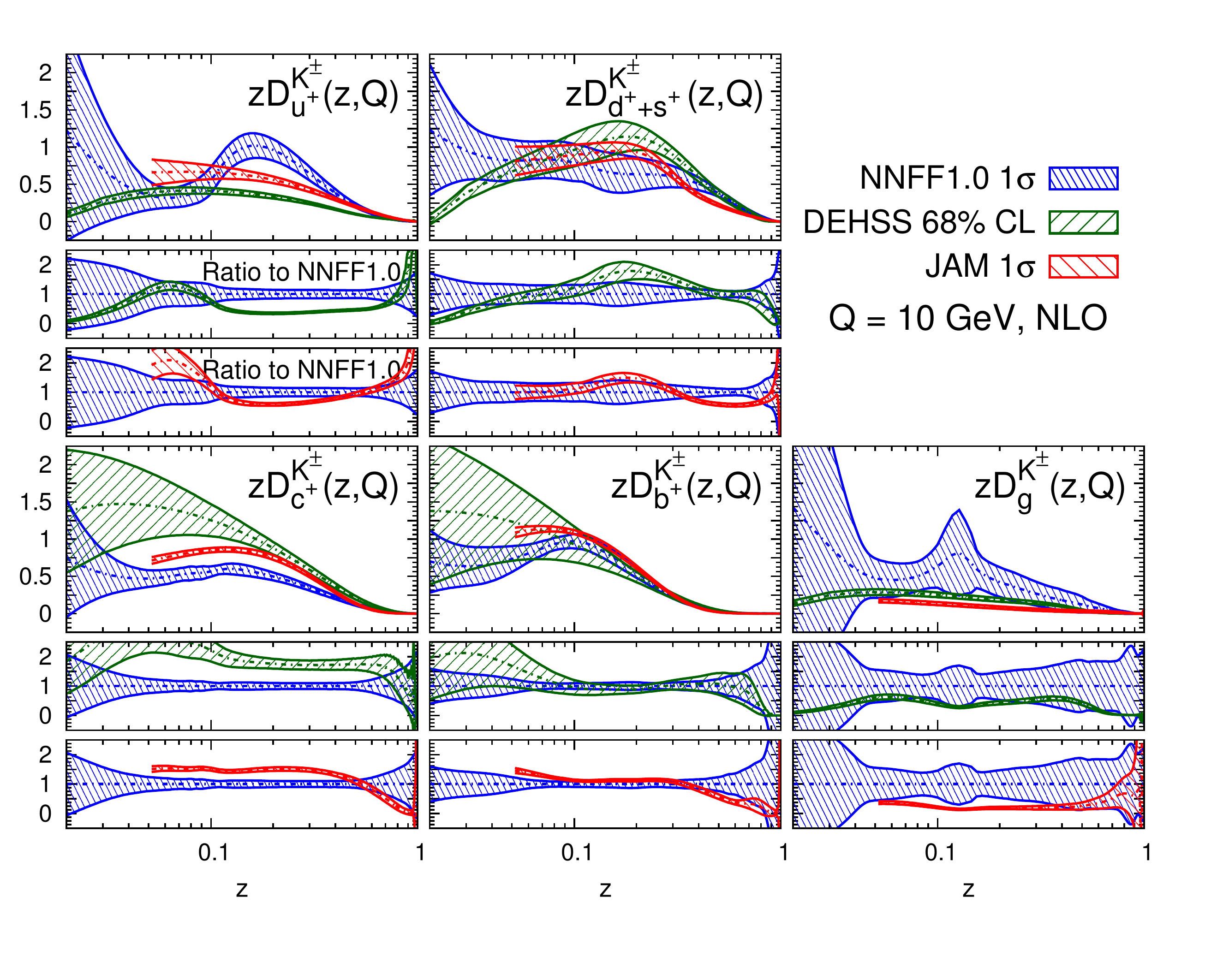}\\
\caption{\small Same as Fig.~\ref{fig:compPI}, but for the sum of
  charged kaons, $K^\pm$.}
\label{fig:compKA}
\end{figure}

We now compare our FFs to the most recent determinations available in the 
literature, namely the DEHSS~\cite{deFlorian:2014xna,deFlorian:2017lwf}
and the JAM~\cite{Sato:2016wqj} sets.
The HKKS sets~\cite{Hirai:2016loo} mentioned in
Sect.~\ref{sec:dataset} were also recently presented but were not
intended to be publicly released~\cite{Kumano:2016pcm}.
Since these analyses were performed only for pions and kaons at NLO, we limit 
the comparison to these hadronic species and this perturbative order.
Such a comparison is shown in Figs.~\ref{fig:compPI} and~\ref{fig:compKA} 
at $Q=10$ GeV in the basis of Eq.~(\ref{eq:parambasis}).
The upper panel of each plot shows the absolute distributions, while
the central and the lower insets show the ratio to NNFF1.0 of DEHSS
and JAM respectively.
Note that the JAM FFs are not extrapolated below the lowest kinematic
cut used in their fits ($z=0.05$), hence the truncated curves in
Figs.~\ref{fig:compPI} and~\ref{fig:compKA}.


Concerning the shapes of the FFs, a number of interesting differences
between the three sets can be seen from the comparisons in
Figs.~\ref{fig:compPI} and~\ref{fig:compKA}.

For the charged pion FFs, the NNFF1.0 and DEHSS results are
in fairly good agreement, despite differences in the dataset (see
Sect.~\ref{sec:dataset}).
Moderate differences are observed only for $D_{u^+}^{\pi^\pm}$ at
$0.2\lesssim z \lesssim 0.5$, for $D_{u^+}^{\pi^\pm}$ and
$D_{c^+}^{\pi^\pm}$ below $z\sim 0.1$, and for all quark combinations
of FFs above $z\sim 0.7$.
A more pronounced difference in shape is observed for the gluon FF,
$D_g^{\pi^\pm}$, for which the NNFF1.0 distribution is more suppressed
at large $z$. 
However, the two sets still agree at the one-$\sigma$ level.
The NNFF1.0 and JAM results are also in fair agreement, except for
$D_{d^++s^+}^{\pi^\pm}$ above $z\sim 0.1$ and for $D_g^{\pi^\pm}$
around $z\sim 0.2$, where the discrepancy exceeds two standard deviations.
Again, the gluon FF from NNFF1.0 is more suppressed at large $z$
than that from JAM.
Differences are also seen for all quark FF combinations at large $z$,
although they are always compatible within uncertainties.

For charged kaons, the differences in shape among the three FF sets
are more marked than in the case of charged pions.
A fair agreement is observed only in the case of $D_{b^+}^{K^\pm}$.
The discrepancies are typically within a couple of standard deviations
for the $D_{d^++s^+}^{K^\pm}$ and $D_g^{K^\pm}$ and more than
three/four standard deviations for $D_{u^+}^{K^\pm}$ and $D_{c^+}^{K^\pm}$.

The origin of the differences among the three sets, at low $z$ for
most of the quark FFs and over the whole $z$ range for the gluon FF,
is likely to be mostly due to the hadron-mass corrections.
These are included in NNFF1.0 but not in the other two sets.
Using a more conservative small-$z$ cut in the NNFF1.0 analysis,
similar to that adopted in the DEHSS and JAM analyses, can exclude the region
where hadron-mass corrections are sizeable (see Fig.~\ref{fig:corrections}).
If a fit is performed with such a conservative cut, most of the
differences among the three FF sets are reconciled, as we will explicitly 
show in Sect.~\ref{sec:kincutdep}.
The differences at large $z$ might arise from our choice of the
kinematic cut too.
Indeed, we exclude all data above $z=0.9$, which are instead retained
in the DEHSS and JAM analyses.

Concerning the FF uncertainties, we observe that for the quark
distributions the three FF sets are in good agreement in the region
covered by the common data, roughly $0.1\lesssim z \lesssim 0.7$.
Conversely, in the regions where a different amount of experimental
information is included or where such information is more sparse,
differences are more significant.
Typically, the uncertainties of the NNFF1.0 FFs are larger than those
of their DEHSS and JAM counterparts at small and large values of $z$,
where data are less abundant or even absent.
Exceptions to this trend are the uncertainties of the
$D_{c^+}^{K^\pm}$ and $D_{b^+}^{K^\pm}$ DEHSS distributions below
$z\sim0.1$. 
They are larger than the NNFF1.0 ones, again because of
their more conservative small-$z$ cuts.

The uncertainty of the gluon FFs, for both pions and kaons, deserves a
separate comment.
As already mentioned, SIA cross-sections are directly sensitive to the
gluon FF only beyond LO.
As a consequence, one would expect that the gluon FF is
determined with larger uncertainties than the quark FFs.
This is clearly shown in Figs.~\ref{fig:compPI}-\ref{fig:compKA}
for the NNFF1.0 sets.
The gluon FFs of the DEHSS and JAM sets, instead, have uncertainties
comparable to those of the quark FFs.
While the smaller uncertainties of the DEHSS gluon FFs may be due to
the larger dataset used in their analysis (which also includes
$pp$ measurements sensitive to the gluon FF already at LO),
this is not the case for the JAM sets, whose dataset mostly coincides
with that of NNFF1.0 (see Sect.~\ref{sec:dataset}).
We ascribe this difference to the more restrictive functional form
used in the JAM analysis to parametrise their FFs.
An underestimate of the gluon FF uncertainty due to the functional
form might also affect the DEHSS determinations.

\subsubsection{The momentum sum rule}

We conclude this section with a brief discussion on the momentum sum rule
\begin{equation}
\sum_h\int_0^1dz\, z D_i^h(z,Q) = 1
\qquad
i=q,\bar{q},g
\,\mbox{,}
\label{eq:sumrules}
\end{equation}
which must be satisfied by FFs irrespective of the value of $Q$. 
Note that the sum in Eq.~(\ref{eq:sumrules}) runs over all
possible hadrons $h$ produced in the fragmentation of the parton $i$, not
only over those determined in the present analysis.
The physical interpretation of Eq.~(\ref{eq:sumrules}) is very simple:
it ensures that the momentum carried by all hadrons produced in the
fragmentation of a given parton $i$ is the same as that carried by the
parton itself.
If Eq.~(\ref{eq:sumrules}) is true at some scale $Q$, it
must remain true at all scales. 
This is guaranteed by DGLAP evolution as a direct consequence of
the energy conservation.

In principle Eq.~(\ref{eq:sumrules}) could be used in a fit to constrain 
simultaneously the behaviour of the FFs for different hadrons, especially in 
the small-$z$ region where no experimental information is available.
In practice we determine the FFs of pions, kaons, and
protons/antiprotons separately and we do not impose the momentum sum
rule.
The momentum sum rule cannot be enforced in our fits for two
reasons. 
First, it requires the knowledge of the FFs of all hadronic
species $h$, while we consider only a subset of them. 
Second, it requires to integrate FFs down to $z=0$, while our FFs are 
determined only down to $z=10^{-2}$.

This said, Eq.~(\ref{eq:sumrules}) can still be used as an {\it a posteriori} 
check of the consistency of the fitted FFs. 
In particular, one expects that
\begin{equation}
  M_i(Q)=\sum_{h=\pi^\pm,K^\pm,p/\bar{p}}\int_{z_{\rm min}}^1dz\, z D_i^h(z,Q) < 1
  \,\mbox{,}
\label{eq:redsumrules}
\end{equation}
with $z_{\rm min}=10^{-2}$.
Since the lower bound of the integral in
Eq.~(\ref{eq:redsumrules}) is not zero, the quantity $M_i$ depends on
$Q$.
We have computed the central value and the uncertainty of $M_i$ for
the gluon FFs ($i=g$) using the NNLO NNFF1.0 FF sets for three
different values of $Q$, obtaining
\begin{equation}
\begin{array}{lcl}
M_g(Q=5\mbox{ GeV}) &=&  0.82 \pm 0.18 \,,\\ 
M_g(Q=10\mbox{ GeV}) &=&  0.79 \pm 0.16\,,\\ 
M_g(Q=M_Z) &=&  0.70\pm 0.12\,.
\end{array}
\end{equation}
The uncertainties from each hadronic species have been added in
quadrature.
Remarkably, we find that Eq.~(\ref{eq:redsumrules}) is fulfilled in the 
range of energies covered by the data included in the fits.

We cannot unambiguously check Eq.~(\ref{eq:redsumrules}) for
the individual quark and antiquark FFs, unless some {\it ad hoc}
assumptions are imposed to separate them from the fitted FF
combinations in Eq.~(\ref{eq:parambasis}).
Nevertheless, we can check a modified, less restrictive, 
version of Eq.~(\ref{eq:redsumrules})
\begin{equation}
  M_i(Q) = \displaystyle\sum_{h=\pi^\pm,K^\pm,p/\bar{p}}\int_{z_{\rm min}}^1dz\, z D_{i}^h(z,Q) < N
\,\mbox{,}
\label{eq:redsumrules2}
\end{equation}
where $N=2$ for $i=u^+,c^+,b^+$ and $N=4$ for $i=d^++s^+$.
We have verified that also Eq.~(\ref{eq:redsumrules2}) is satisfied by
the NNFF1.0 FFs.
For instance, at $Q=5$ GeV we find
\begin{equation}
\begin{array}{lcl}
M_{u^+}(Q=5\mbox{ GeV}) &=&  1.41 \pm 0.13 \,,\\ 
M_{d^++s^+}(Q=5\mbox{ GeV}) &=&  2.12 \pm 0.25\,,\\ 
M_{c^+} (Q=5\mbox{ GeV}) &=&  1.04\pm 0.06\,,\\ 
M_{b^+} (Q=5\mbox{ GeV}) &=&  1.01\pm 0.06\,,
\end{array}
\end{equation}
in agreement with the expectations.

\section{Fit stability}
\label{sec:stability}

In this section we study the stability of the results presented in
Sects.~\ref{sec:fitquality}-\ref{sec:ffs} upon variations of the
small-$z$ kinematic cuts and of the dataset defined in
Sect.~\ref{sec:dataset}.

\subsection{Dependence on the small-$z$ kinematic cuts}
\label{sec:kincutdep}

We first study the dependence of our results upon the small-$z$
kinematic cuts applied to the data included in the NNFF1.0 fits.
Our aim is to assess the interplay between higher-order QCD corrections and 
the description of small-$z$ data, in order to motivate the choice of 
kinematic cuts made in Sect.~\ref{sec:observables}.
To this purpose, we perform some additional fits, all to the same baseline 
dataset described in Sect.~\ref{sec:dataset} but with different small-$z$ cuts.
For each value of the kinematic cut, the additional fits are performed
at LO, NLO, and NNLO.

The various small-$z$ kinematic cuts considered here are summarised in
Tab.~\ref{tab:varkincuts}, together with our baseline choice (denoted
as {\tt BL} henceforth).
For each hadronic species we consider the two limiting cases in which
the small-$z$ kinematic cuts are either completely removed or set to
{\it conservative} values. 
The latter are defined in such a way that only data in the
kinematic region where both hadron-mass and NNLO QCD corrections are
expected to be negligible are included in the fit.
Conservative cuts are similar to those adopted in other analyses of FFs.
Additional choices of small-$z$ kinematic cuts between the two cases
above are also investigated.
As the data for charged pions extend down to smaller values of $z$
than data for charged kaons and protons/antiprotons, a more dense
scanning is adopted in the first case.
\begin{table}[!t]
\renewcommand{\arraystretch}{1.3}
\centering
\scriptsize
\begin{tabular}{lcccccccccc}
\toprule
Hadron & \multicolumn{2}{c}{{\tt BL}} 
       & \multicolumn{2}{c}{no cuts} 
       & \multicolumn{2}{c}{con. cut}
       & \multicolumn{2}{c}{cut1}
       & \multicolumn{2}{c}{cut2} \\
       & $z_{\rm min}^{(M_Z)}$ & $z_{\rm min}$ 
       & $z_{\rm min}^{(M_Z)}$ & $z_{\rm min}$ 
       & $z_{\rm min}^{(M_Z)}$ & $z_{\rm min}$ 
       & $z_{\rm min}^{(M_Z)}$ & $z_{\rm min}$
       & $z_{\rm min}^{(M_Z)}$ & $z_{\rm min}$ \\
\midrule
$\pi^\pm$    
       & 0.02 & 0.075 & 0.00 & 0.00 & 0.05 & 0.10 & 0.01 & 0.05 & 0.01 & 0.075\\
$K^\pm$      
       & 0.02 & 0.075 & 0.00 & 0.00 & 0.10 & 0.20 & 0.05 & 0.10 & ---  & --- \\
$p/\bar{p}$ 
       & 0.02 & 0.075 & 0.00 & 0.00 & 0.10 & 0.20 & ---  & ---  & ---  & ---  \\
\bottomrule
\end{tabular}
\caption{\small Summary of the various choices for the small-$z$ kinematic 
  cuts for each hadronic species that are investigated here.
  For experiments taken at $\sqrt{s}=M_Z$ ($\sqrt{s}<M_Z$), data points
  with  $z<z_{\rm min}^{(M_Z)}\,(z<z_{\rm min})$ are excluded from the fit.
  }
\label{tab:varkincuts}
\end{table}

\begin{figure}[!t]
\centering
\includegraphics[scale=0.21,angle=270]{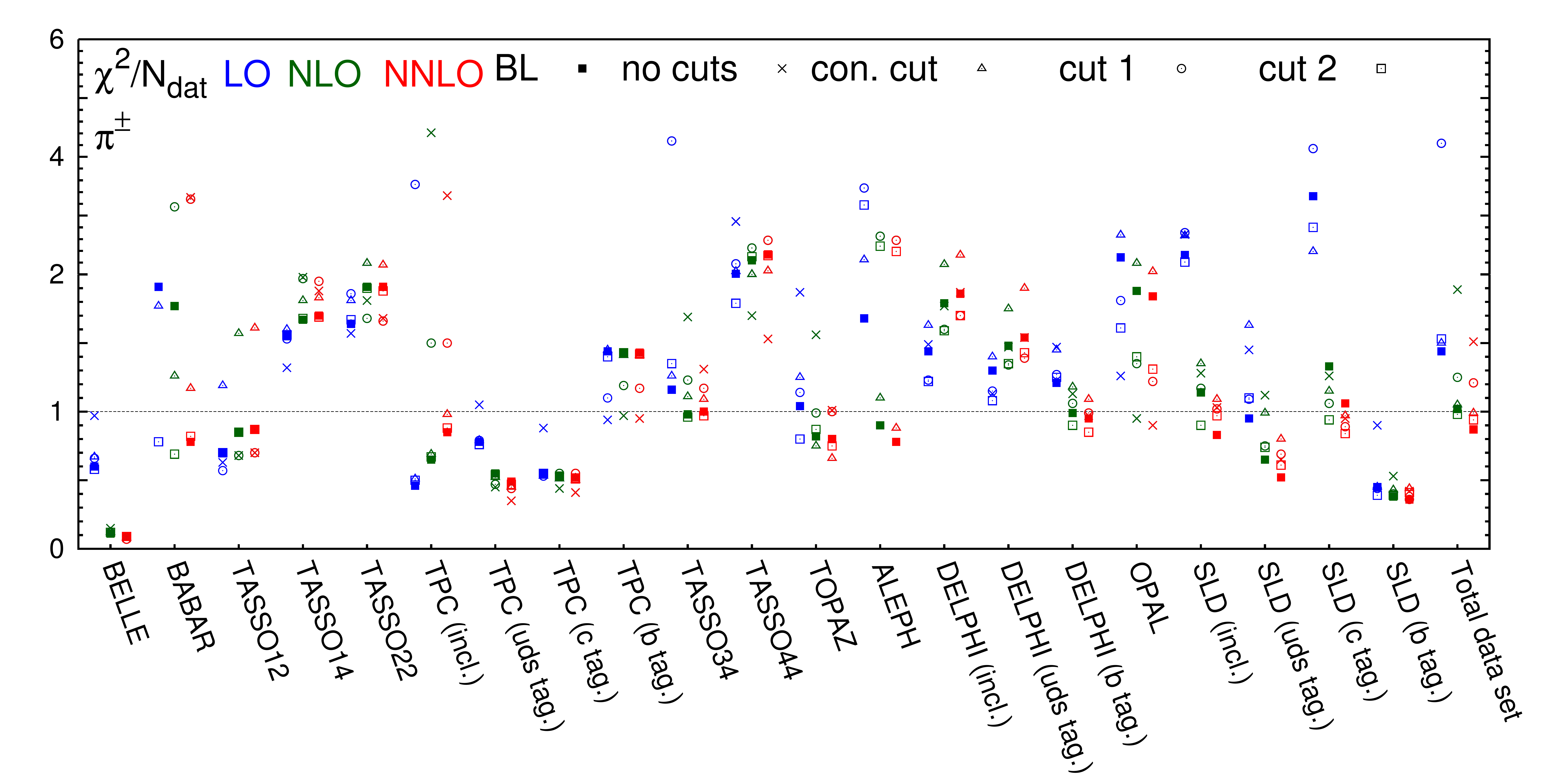}\\
\caption{\small The values of $\chi^2/N_{\rm dat}$ for each
fit to $\pi^\pm$ data with the choices of small-$z$ kinematic cuts summarised
in Tab.~\ref{tab:varkincuts}, at LO, NLO, and NNLO.}
\label{fig:cutsPIbd}
\end{figure}
\begin{figure}[!t]
\centering
\includegraphics[scale=0.21,angle=270]{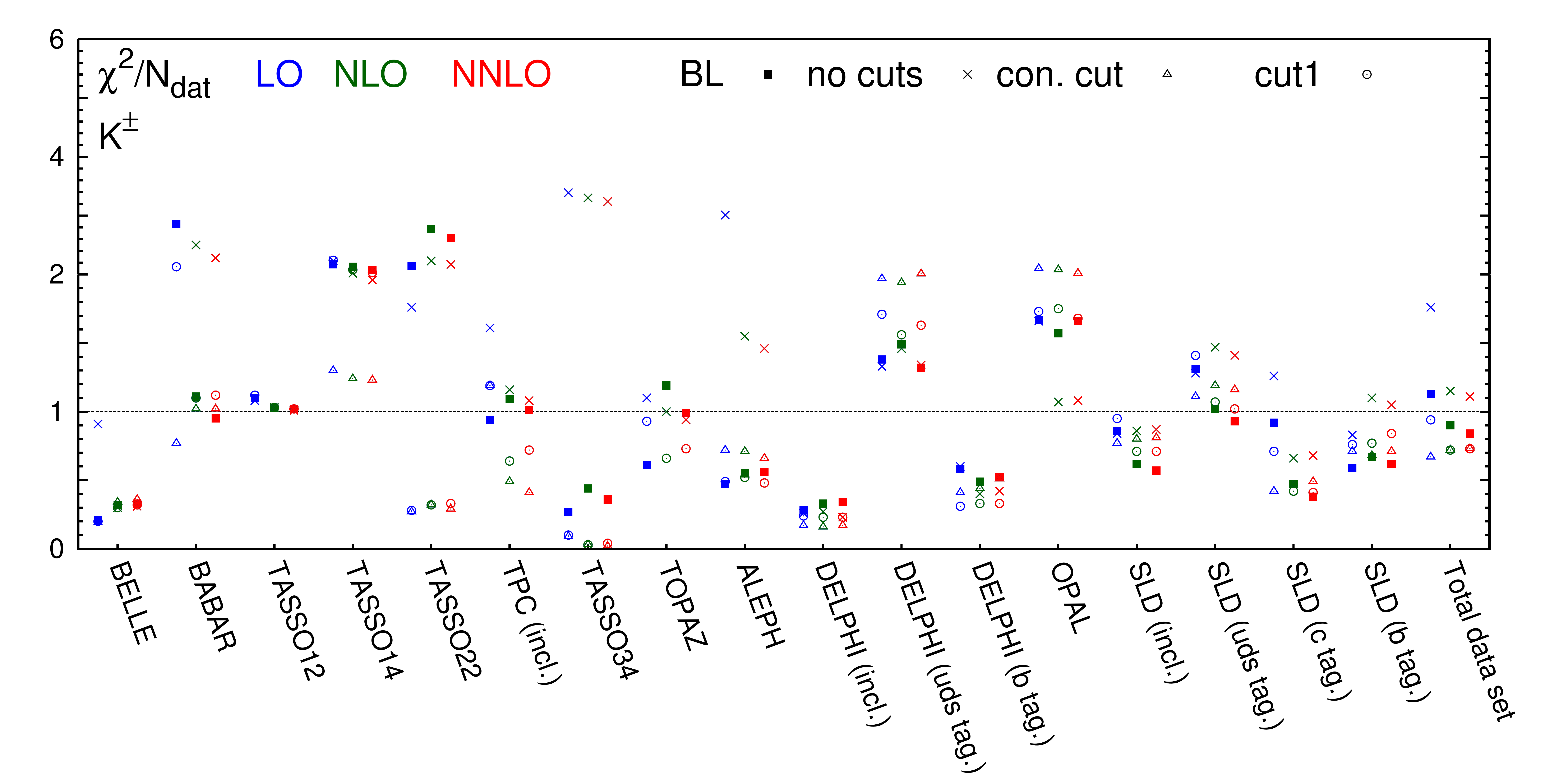}\\
\caption{\small Same as Fig.~\ref{fig:cutsPIbd}, but for $K^\pm$.}
\label{fig:cutsKAbd}
\end{figure}
\begin{figure}[!t]
\centering
\includegraphics[scale=0.21,angle=270]{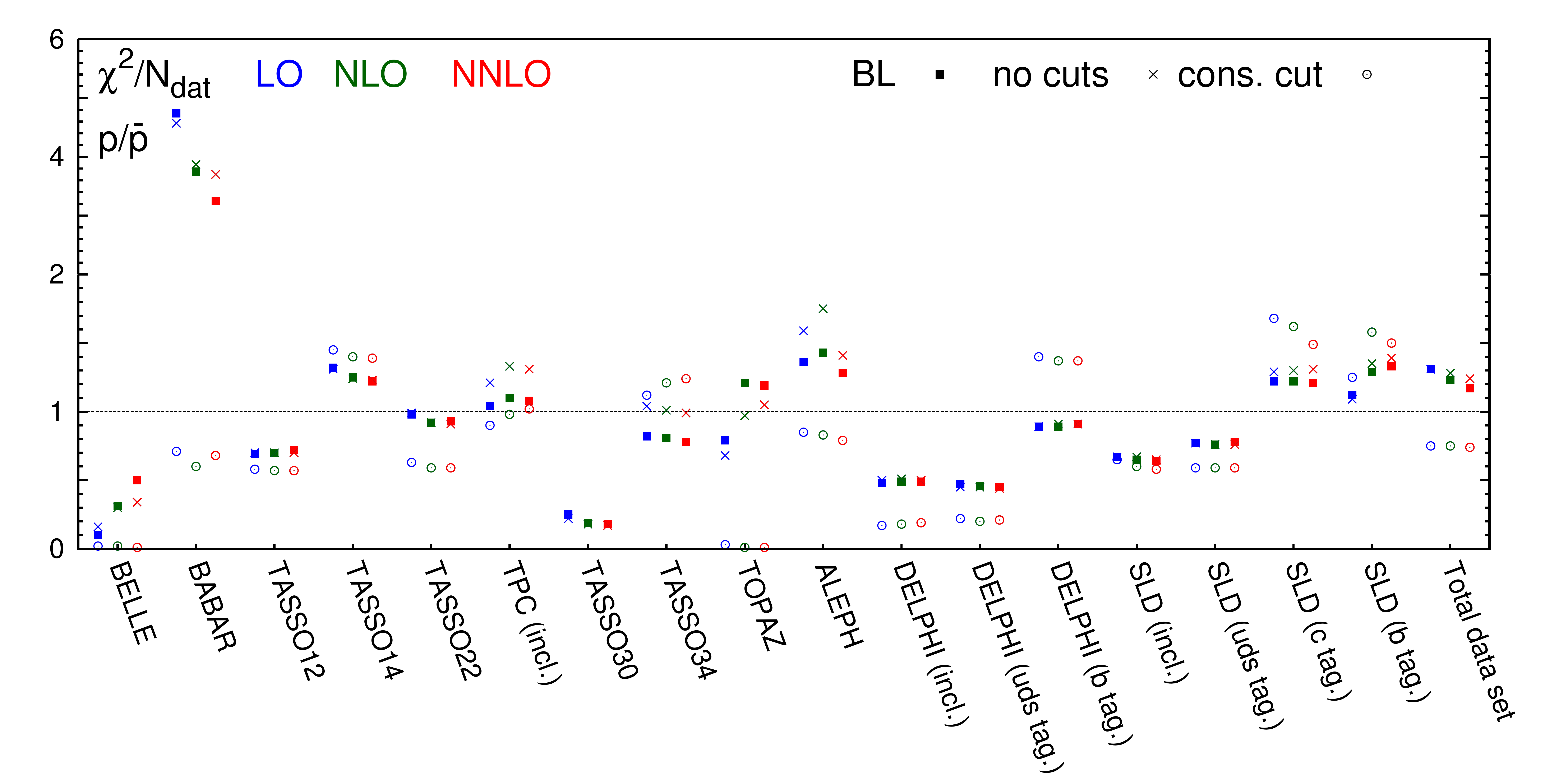}\\
\caption{\small Same as Fig.~\ref{fig:cutsPIbd}, but for $p/\bar{p}$.}
\label{fig:cutsPRbd}
\end{figure}

In Figs.~\ref{fig:cutsPIbd}-\ref{fig:cutsPRbd} we show the values of
$\chi^2/N_{\rm dat}$ for the LO, NLO, and NNLO fits of charged pions,
charged kaons, and protons/antiprotons FFs performed with the
kinematic cuts in Tab.~\ref{tab:varkincuts}.
Inspection of the $\chi^2/N_{\rm dat}$ values for the total dataset in
Figs.~\ref{fig:cutsPIbd}-\ref{fig:cutsPRbd} allows us to draw three remarks.

First, there is clear evidence of perturbative convergence:
irrespective of the specific choice of the small-$z$ cuts, the
$\chi^2/N_{\rm dat}$ values at NNLO are always lower than at NLO,
which are in turn always lower than at LO.

Second, the spread of the $\chi^2/N_{\rm dat}$ values for different cuts at a 
fixed perturbative order is reduced as the perturbative order is increased.
The value of the $\chi^2/N_{\rm dat}$ for the less restrictive cuts moves closer 
to the corresponding value for the conservative cuts.
This confirms that the inclusion of higher-order QCD corrections
significantly improves the description of the data at small $z$ and
that the results become accordingly less dependent on the choice of small-$z$
cuts.
These results are consistent with what was reported in
Ref.~\cite{Anderle:2016czy} where, at least for charged pions, it was
found that a fixed-order NNLO fit is able to describe data down to
$z_{\rm min}=0.02$ with the same accuracy as a small-$z$ resummed
NNLO+NNLL fit.

Third, at any perturbative order, the $\chi^2/N_{\rm dat}$ of the fit
with baseline kinematic cuts is always very close to the lowest
$\chi^2/N_{\rm dat}$, usually associated to the fit with conservative cuts.
The only exception is the proton/antiproton case, where the fit with
the baseline cuts has a $\chi^2/N_{\rm dat}$ significantly larger than
the fit with conservative cuts.
This behaviour is mostly driven by the high value of the
$\chi^2/N_{\rm dat}$ for the BABAR data.
However, as already mentioned in Sect.~\ref{sec:fitquality}, this is
due to a genuine tension between the BABAR and TPC/TASSO34 data below
$z=0.2$.
This will be explicitly demonstrated in Sect.~\ref{sec:datdep} by
studying fits to reduced datasets.

Similar conclusions as those for the total dataset can be drawn for
the individual datasets, based on
Figs.~\ref{fig:cutsPIbd}-\ref{fig:cutsPRbd}.
However, the baseline cuts do not always minimise the
$\chi^2/N_{\rm dat}$ of the individual experiments, especially of
those with a limited number of data points.
This is a feature of any global analysis, where the total $\chi^2$ always 
represents a compromise among the different pulls from individual experiments.

\begin{figure}[!t]
\centering
\includegraphics[scale=0.6,angle=270,clip=true,trim=0 0 1.5cm 0]
{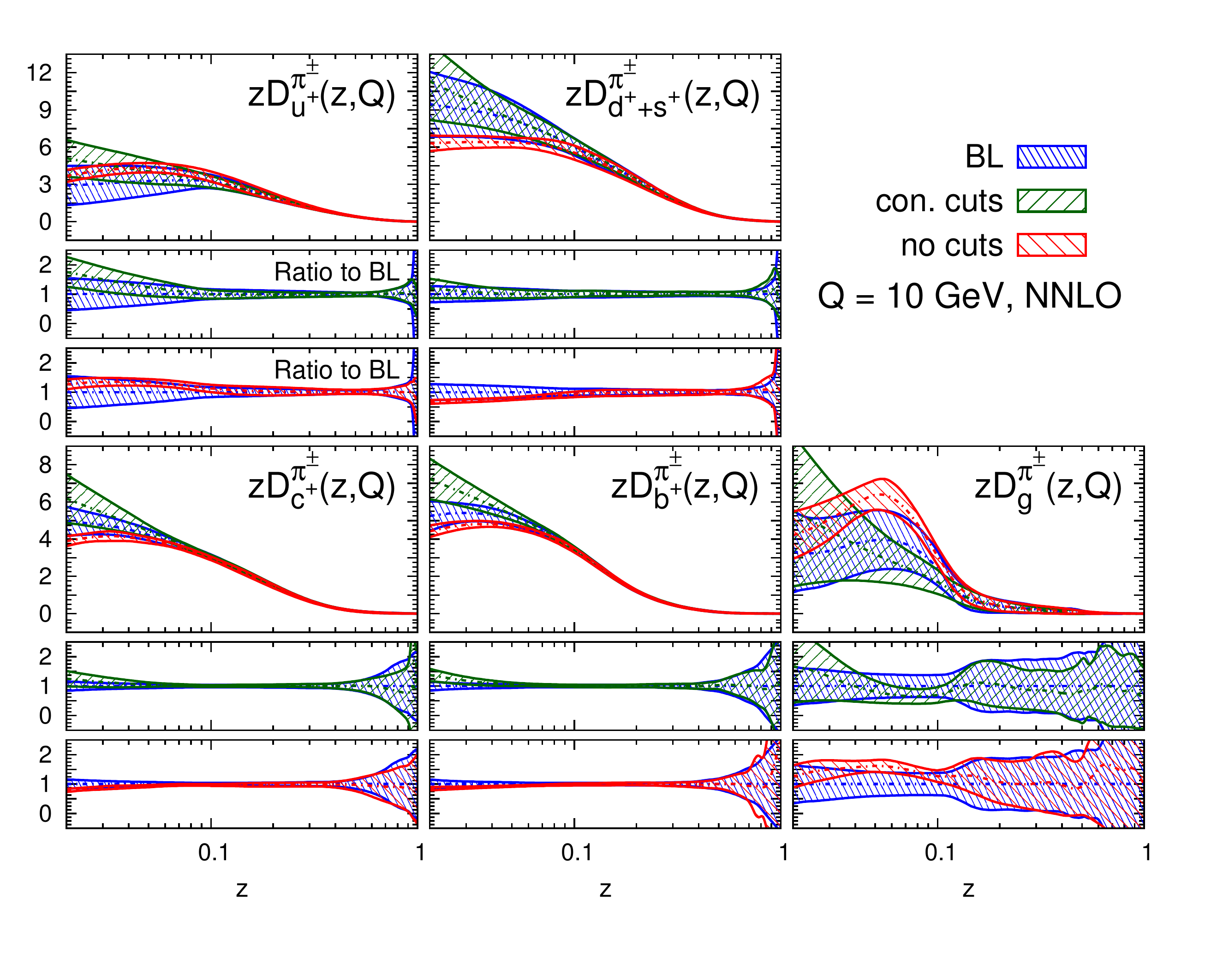}\\
\caption{\small Comparison among the NNLO charged pion FFs at $Q=10$ GeV
  for three different small-$z$ kinematic cuts: baseline,
  conservative cuts, and no cuts (see Tab.~\ref{tab:varkincuts}).
  The two insets below each distribution show the ratios of the fits with
  conservative cuts and without cuts to the baseline fit.}
\label{fig:cutsPI}
\end{figure}

\begin{figure}[!t]
\centering
\includegraphics[scale=0.6,angle=270,clip=true,trim=0 0 1.5cm 0]
{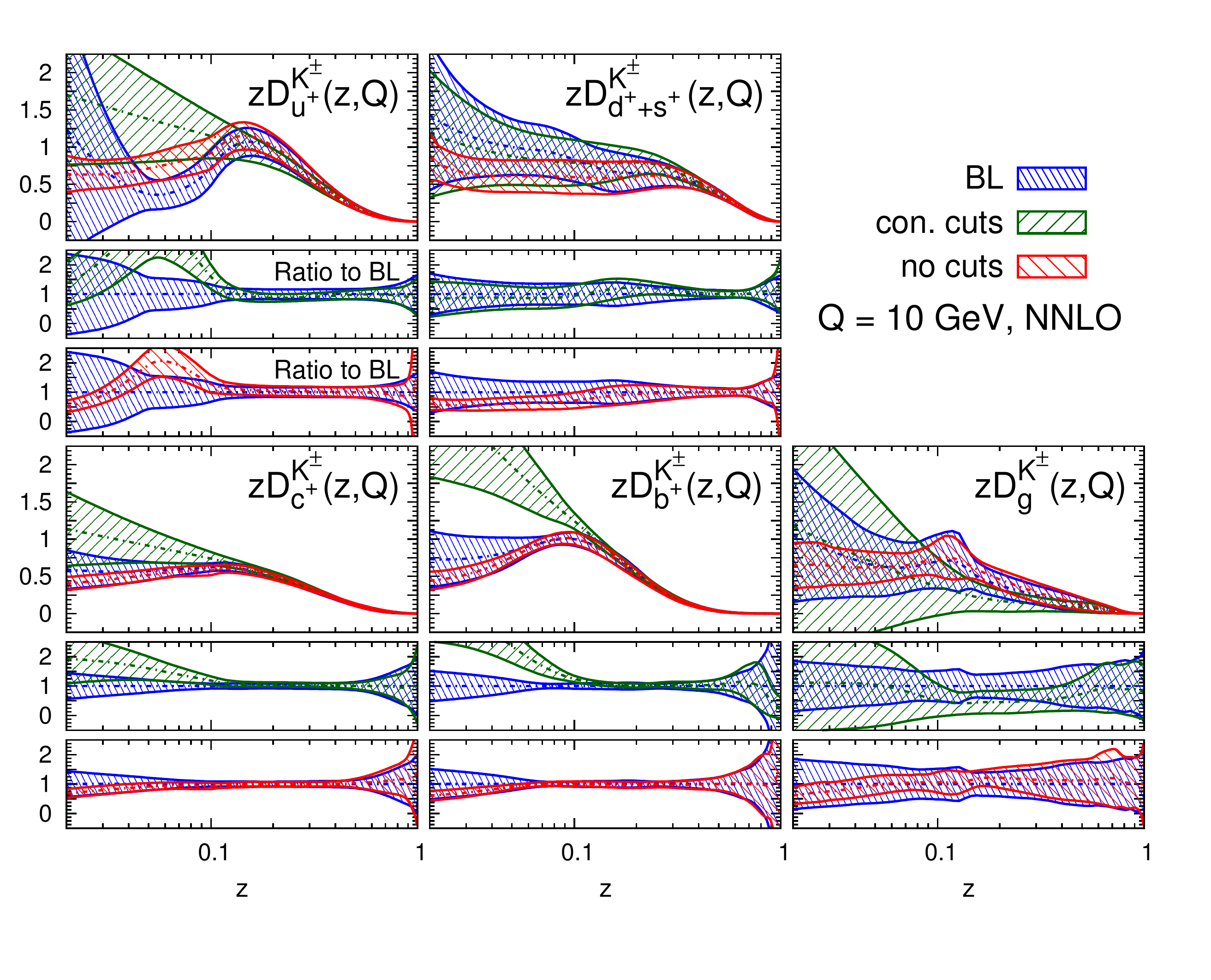}\\
\caption{\small Same as Fig.~\ref{fig:cutsPI}, but for the sum of charged 
  kaons, $K^\pm$.}
\label{fig:cutsKA}
\end{figure}

\begin{figure}[!t]
\centering
\includegraphics[scale=0.6,angle=270,clip=true,trim=0 0 1.5cm 0]
{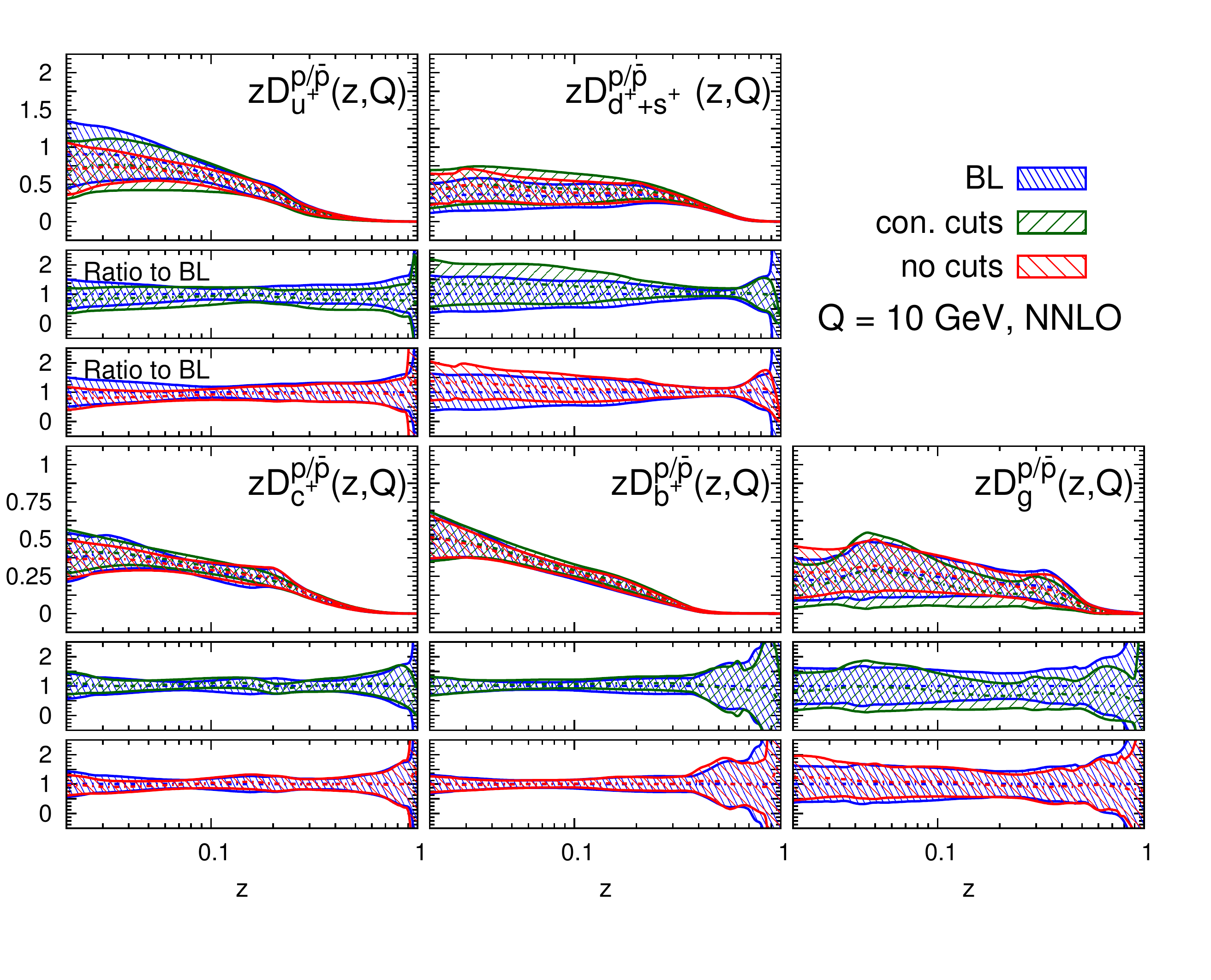}\\
\caption{\small Same as Fig.~\ref{fig:cutsPI}, but for the sum of protons and 
  antiprotons, $p/\bar{p}$.}
\label{fig:cutsPR}
\end{figure}

In order to gauge the impact of the different choices of small-$z$
kinematic cuts on FFs, in Figs.~\ref{fig:cutsPI}-\ref{fig:cutsPR} 
we compare the NNLO results from the two extreme choices (no kinematic cuts 
and conservative cuts) with those from the baseline fit.
By comparing the fit with no cuts to the baseline results, one can
infer by how much uncertainties would decrease if the small-$z$ data
excluded by the baseline cuts were included.
This is relevant in view of a possible fit with small-$z$ resummation,
which is expected to provide a better description of the small-$z$ data
below our baseline choice of $z_{\rm min}$.


We find that varying the small-$z$ kinematic cuts does not affect the FFs for 
any hadronic species in the region above $z=0.1$.
Conversely, significant differences in shape emerge at small $z$,
where the typical effect of the data is to suppress FFs.
This behaviour is observed for all FFs and all hadronic species,
particularly when moving from the conservative to the baseline cuts.
The effect is milder, especially for protons/antiprotons, when the
cuts are completely removed because the amount of the additional data
included in this case is more limited.

Importantly, the gluon FFs for charged pions and kaons are
particularly affected by the choice of the small-$z$ cuts.
In the fit with conservative cuts the shape of the gluon FFs becomes
similar to that of their counterparts in the JAM and DEHSS sets, and
they are always compatible with them within uncertainties.
This is not unexpected because our conservative cuts are similar to
those adopted in these analyses.

Concerning the FF uncertainties in the small-$z$ region, they decrease
approximately by a factor two for charged pions and kaons when moving
from the conservative to the baseline cuts, while they remain almost
unchanged for protons/antiprotons.
They decrease by a further factor of two for charged pions and kaons
in comparison to the baseline when cuts are removed, while they remain
mostly unchanged for protons/antiprotons.
This reduction highlights the importance of including small-$z$ data
to tame the FF uncertainties in the small-$z$ region, provided that the
theoretical calculations used are accurate enough to describe the 
corresponding measurements.

\subsection{Dependence on the fitted dataset}
\label{sec:datdep}

We now study the dependence of the NNFF1.0 NNLO FFs upon two variations of the 
fitted dataset.
In comparison to the baseline dataset listed in Tab.~\ref{tab:datasets}, 
we consider first a dataset from which the BELLE and BABAR experiments are 
removed, and second a dataset in which only the BELLE, 
BABAR, ALEPH, DELPHI, OPAL, and SLD experiments are retained.
The first dataset is denoted as {\tt noBB}, the second dataset is denoted
as {\tt BBMZ}.

For each hadronic species, we perform an additional NNLO fit to each of these
reduced datasets.
All fit settings, including the kinematic cuts, are identical to the baseline 
fits.
With the first fit we intend to assess the impact on the light-quark
FF flavour separation and on the gluon FF of the $B$-factory data, the
most recent and precise piece of experimental information.
The motivation for the second fit is instead to assess the impact on the FFs 
and their uncertainties of the older and less accurate SIA measurements.

\begin{figure}[!t]
\centering
\includegraphics[scale=0.6,angle=270,clip=true,trim=0 0 1.5cm 0]
{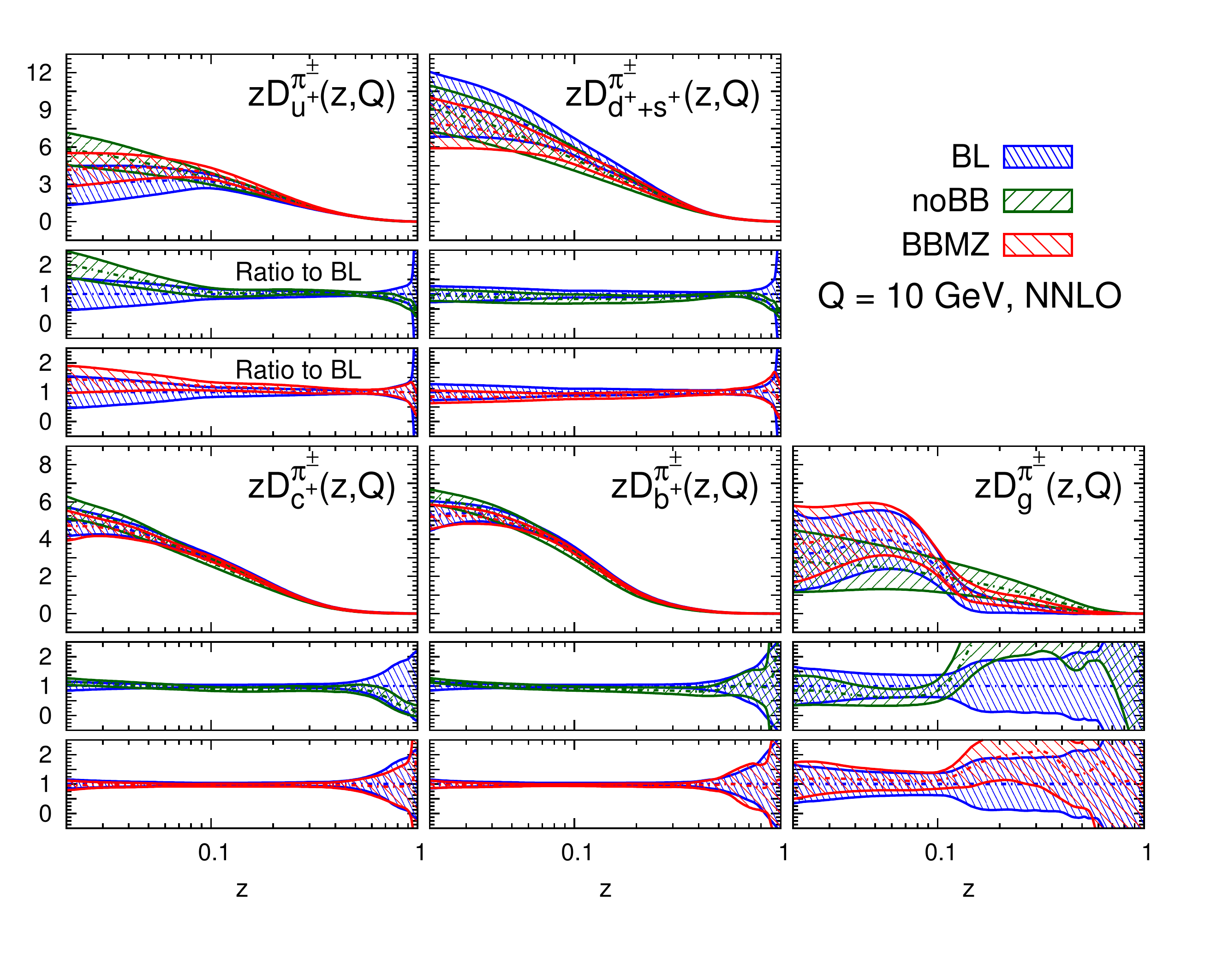}\\
\caption{\small Comparison among the NNFF1.0 NNLO  FFs for charged
  pions for three variations of the 
  fitted dataset: {\tt BL}, {\tt noBB} and {\tt BBMZ} 
  (see the text for details).
  The bands indicate their one-$\sigma$ uncertainties.
  The ratios of the {\tt noBB} and {\tt BBMZ} fits to the {\tt BL} fit 
  are displayed in the lower insets.}
\label{fig:dataPI}
\end{figure}

\begin{figure}[!t]
\centering
\includegraphics[scale=0.6,angle=270,clip=true,trim=0 0 1.5cm 0]
{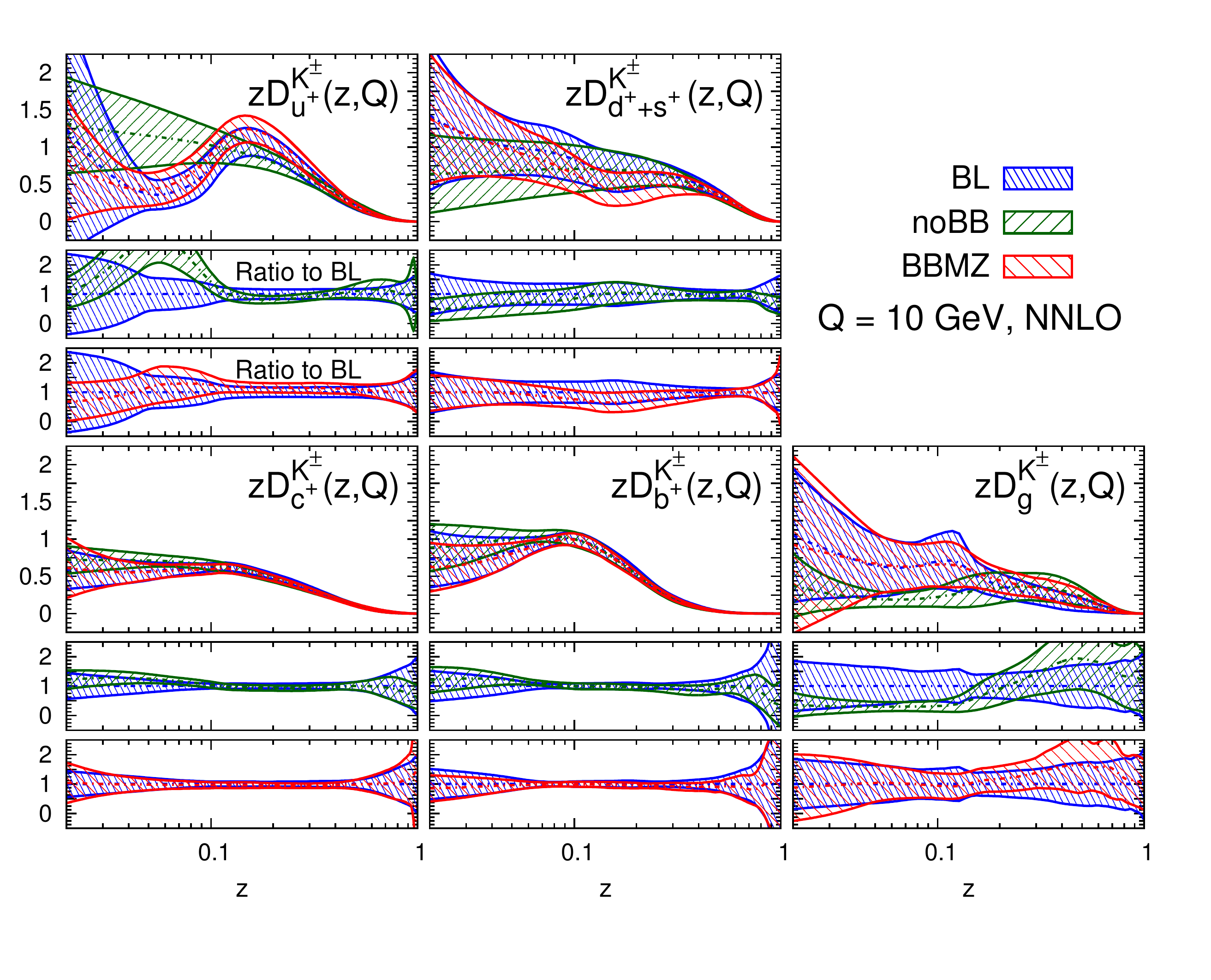}\\
\caption{\small Same as Fig.~\ref{fig:dataPI}, but for the sum of charged 
kaons, $K^\pm$.}
\label{fig:dataKA}
\end{figure}

\begin{figure}[!t]
\centering
\includegraphics[scale=0.6,angle=270,clip=true,trim=0 0 1.5cm 0]
{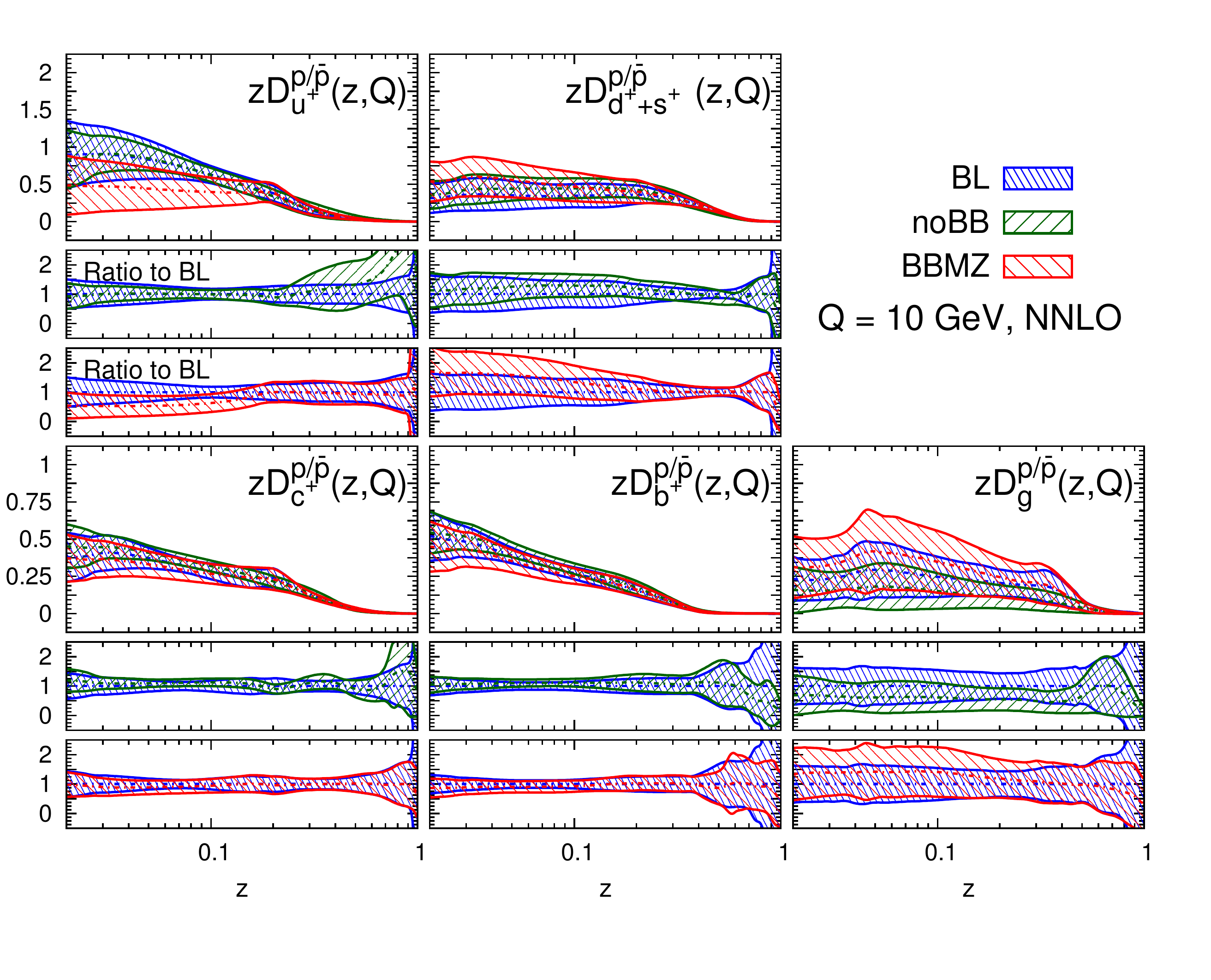}\\
\caption{\small Same as Fig.~\ref{fig:dataPI}, but for the sum of protons and 
antiprotons, $p/\bar{p}$.}
\label{fig:dataPR}
\end{figure}

In Tab.~\ref{tab:chi2sdata} we show the values of $\chi^2/N_{\rm dat}$ for the 
fits to the {\tt noBB} and {\tt BBMZ} datasets.
For ease of comparison, we also report the $\chi^2/N_{\rm dat}$ values for the
fit to the baseline dataset (denoted as {\tt BL}) from Tab.~\ref{tab:chi2s}.
The numbers in squared brackets refer to the experiments not included
in the corresponding fits.
We display the resulting FFs at $Q=10$ GeV in Fig.~\ref{fig:dataPI}
for charged pions, in Fig.~\ref{fig:dataKA} for charged kaons, and in
Fig.~\ref{fig:dataPR} for protons/antiprotons.

\begin{table}[!t]
\renewcommand{\arraystretch}{1.4}
\centering
\scriptsize
\begin{tabular}{lccccccccc}
\toprule
     & \multicolumn{3}{c}{$\chi^2/N_{\rm dat}$ ($h=\pi^\pm$)}
     & \multicolumn{3}{c}{$\chi^2/N_{\rm dat}$ ($h=K^\pm$)}
     & \multicolumn{3}{c}{$\chi^2/N_{\rm dat}$ ($h=p/\bar{p}$)}\\
Exp. & {\tt BL} & {\tt noBB} & {\tt BBMZ}  
     & {\tt BL} & {\tt noBB} & {\tt BBMZ} 
     & {\tt BL} & {\tt noBB} & {\tt BBMZ} \\
\midrule
BELLE               & 0.09 & [4.92] & 0.09 
                    & 0.33 & [13.0] & 0.32
                    & 0.50 & [25.9] & 0.74   \\
BABAR               & 0.78 & [144]  & 0.88 
                    & 0.95 & [208]  & 1.21 
                    & 3.25 & [32.8] & 0.84   \\
TASSO12             & 0.87 & 0.52   & [0.87]
                    & 1.02 & 1.07   & [1.02] 
                    & 0.72 & 0.78   & [0.74] \\
TASSO14             & 1.70 & 1.38   & [1.71]
                    & 2.07 & 1.50   & [2.22] 
                    & 1.22 & 1.41   & [1.13] \\
TASSO22             & 1.91 & 1.29   & [2.15]  
                    & 2.62 & 1.10   & [2.87] 
                    & 0.93 & 0.88   & [1.25] \\
TPC (incl.)         & 0.85 & 2.12   & [0.81]  
                    & 1.01 & 0.59   & [1.66] 
                    & 1.08 & 0.88   & [3.86] \\
TPC ($uds$ tag.)    & 0.49 & 0.54   & [0.77] 
                    & ---  & ---    & ---  
                    & ---  & ---    & ---    \\
TPC ($c$ tag.)      & 0.52 & 0.74   & [0.58] 
                    & ---  & ---    & ---  
                    & ---  & ---    & ---    \\
TPC ($b$ tag.)      & 1.43 & 1.60   & [1.48]  
                    & ---  & ---    & ---  
                    & ---  & ---    & ---    \\
TASSO30             & ---  & ---    & ---  
                    & ---  & ---    & ---   
                    & 0.18 & 0.11   & [0.64] \\
TASSO34             & 1.00 & 1.17   & [1.38]  
                    & 0.36 & 0.10   & [0.47] 
                    & 0.78 & 0.48   & [2.37] \\
TASSO44             & 2.34 & 2.52   & [2.97]  
                    & ---  & ---    & ---  
                    & ---  & ---    & ---    \\
TOPAZ               & 0.80 & 0.92   & [1.72]  
                    & 0.99 & 0.39   & [1.60]
                    & 1.19 & 1.08   & [0.87] \\
ALEPH               & 0.78 & 0.57   & 0.74 
                    & 0.56 & 0.51   & 0.58
                    & 1.28 & 1.38   & 1.23   \\
DELPHI (incl.)      & 1.86 & 1.97   & 1.82 
                    & 0.34 & 0.34   & 0.36 
                    & 0.49 & 0.53   & 0.46   \\
DELPHI ($uds$ tag.) & 1.54 & 1.56   & 1.42  
                    & 1.32 & 1.29   & 1.41 
                    & 0.45 & 0.49   & 0.44   \\
DELPHI ($b$ tag.)   & 0.95 & 1.01   & 0.95  
                    & 0.52 & 0.54   & 0.53 
                    & 0.91 & 0.98   & 0.90   \\
OPAL                & 1.84 & 1.75   & 1.92  
                    & 1.66 & 1.60   & 1.66 
                    & ---  & ---    & ---    \\
SLD (incl.)         & 0.83 & 0.87   & 0.95  
                    & 0.57 & 0.52   & 0.57 
                    & 0.64 & 0.51   & 0.63   \\
SLD ($uds$ tag.)    & 0.52 & 0.53   & 0.63  
                    & 0.93 & 0.77   & 0.91 
                    & 0.78 & 0.73   & 0.80   \\
SLD ($c$ tag)       & 1.06 & 0.69   & 0.96  
                    & 0.38 & 0.38   & 0.37 
                    & 1.21 & 0.98   & 1.21   \\
SLD ($b$ tag)       & 0.36 & 0.49   & 0.37  
                    & 0.62 & 0.80   & 0.66 
                    & 1.33 & 1.21   & 1.21   \\
\midrule
Total dataset       & \bf 0.87 & \bf 1.06   & \bf 0.82 
                    & \bf 0.73 & \bf 0.70   & \bf 0.69 
                    & \bf 1.17 & \bf 0.87   & \bf 0.87   \\
\bottomrule
\end{tabular}
\caption{\small The values of $\chi^2/N_{\rm dat}$ for the baseline ({\tt BL}) 
  NNLO fit and for the fits with dataset variations,
  {\tt noBB} (excluding BELLE and BABAR), and {\tt BBMZ} (including only
  $B$-factories, LEP and SLD experiments), for each hadronic species.
  The values for
  datasets not included in a fit are indicated in squared brackets.}
\label{tab:chi2sdata}
\end{table}

We now discuss the main features of these two fits based on reduced datasets.

\paragraph{The {\tt noBB} fit.}

In comparison to the baseline, the overall quality of the {\tt noBB} fit, as
quantified by its total $\chi^2/N_{\rm dat}$ value, slightly deteriorates for 
charged pions, while it slightly improves for charged kaons and 
protons/antiprotons.
For pions, this effect is due to a significant deterioration in the
description of the TPC measurements, in particular the inclusive
multiplicities, for which the $\chi^2/N_{\rm dat}$ grows from 0.85 in
the baseline fit to 2.12 in the fit without the BELLE and BABAR data.
For kaons, the improvement is driven by a better description of the
TPC, TOPAZ, and all the TASSO datasets, except TASSO12.
For protons/antiprotons, the improvement is determined by the
exclusion of the BABAR dataset, whose rather high $\chi^2/N_{\rm dat}$
raises the total $\chi^2/N_{\rm dat}$ in the baseline fit.

Apart from these small differences, the overall quality of the fit to
the reduced dataset is comparable to that of the baseline fit.
We note however that the BELLE and BABAR datasets are poorly described
if they are not included in the fit. 
In this case their $\chi^2/N_{\rm dat}$ value is indeed significantly higher,
particularly for the latter experiment.
This indicates that these two experiments carry a significant amount
of information.

The effect of this information on FFs, is apparent from  
Figs.~\ref{fig:dataPI}-\ref{fig:dataPR}.
Flavour separation between the light-quark FF
combinations $D_{u^+}^h$ and $D_{d^++s^+}^h$ is moderately affected.
For pions, $D_{u^+}^{\pi^\pm}$ is slightly more suppressed below
$z=0.1$ in the {\tt BL} fit than in the {\tt noBB} fit, while
$D_{d^++s^+}^{\pi^\pm}$ is slightly larger, especially in the region
$0.05\lesssim z \lesssim 0.5$.
For kaons, differences in the FF shapes are more marked, especially
in the small-$z$ region, where hadron-mass and higher-order QCD
corrections are more important for the BELLE and BABAR data than for
the data at higher energies.
For protons/antiprotons, the shape of $D_{u^+}^{p/\bar{p}}$ and
$D_{d^++s^+}^{p/\bar{p}}$ is almost unaffected by the BELLE and BABAR
data.
For all hadronic species, the uncertainties of the light-quark FF
combinations are slightly reduced when the measurements from BELLE and
BABAR are included in the fit.

The gluon FFs is also affected.
For pions and kaons, significant differences in shape are observed
over the whole $z$ range, while for protons/antiprotons only a small
enhancement is seen when the BELLE and BABAR data are included.
As expected, heavy-quark FFs for all hadronic species, $D_{c^+}^h$ and
$D_{b^+}^h$, are not affected by the BELLE and BABAR data to which
they are not directly sensitive.

The importance of the $B$-factory measurements is demonstrated also by
the fact that the uncertainty of the gluon FFs for all hadronic
species is reduced by up to a factor two above $z=0.4$ upon their
inclusion in the fit.
These results prove that the BELLE and BABAR data represent an important 
ingredient for any state-of-the-art determination of FFs.

\paragraph{The {\tt BBMZ} fit.}

In comparison to the baseline, the overall quality of the {\tt BBMZ} 
fit improves for all hadronic species, see Tab.~\ref{tab:chi2sdata}.
Individual experiments included in both fits are described with
similar accuracy in most cases.
The only exception is the BABAR experiment, for which the
$\chi^2/N_{\rm dat}$ increases from 0.78 to 0.88 for charged pions,
from 0.95 to 1.21 for charged kaons, and decreases from 3.25 to 0.84
for protons/antiprotons when moving from the baseline to the {\tt BBMZ} fit.
In the case of pions and kaons, the BABAR measurements stabilise the fit. 
In the case of protons/antiprotons instead they might
be in tension with the rest of the dataset.

This is confirmed by the $\chi^2/N_{\rm dat}$ values of the experiments
excluded from the {\tt BBMZ} fit.
In most cases, they are equally good or slightly worse than in the
baseline fit.
In the case of protons/antiprotons, instead, the $\chi^2/N_{\rm dat}$
value of the TPC and TASSO34 experiments is significantly worse in the
fit to the reduced dataset than in the baseline fit.
Since this deterioration is accompanied by an improvement in the
$\chi^2/N_{\rm dat}$ value of the BABAR experiment, we conclude that
there is some tension between this experiment and the TPC and TASSO34
data.
The origin of this tension is likely to be due to a limited number of
data points at small $z$, as this effect disappears if more
conservative kinematic cuts are applied (see Sect.~\ref{sec:kincutdep}).

At the level of FFs, results are remarkably stable as one can see by
comparing the FFs from the {\tt BL} fit to those from the {\tt BBMZ} fit 
in Figs.~\ref{fig:dataPI}-\ref{fig:dataPR}.
For all hadronic species, the FFs in the {\tt BL} and {\tt BBMZ} fits
are compatible within uncertainties and no significant differences in
shape are observed.
As far as flavour separation is concerned, $D_{u^+}^h$ is slightly
larger (smaller) in the fit to the reduced dataset than in the fit to
the baseline dataset for charged pions and kaons (protons/antiprotons).
This is accompanied by a slightly smaller (larger) $D_{d^++s^+}^h$, so
that the total singlet FF is almost equivalent in the two fits.
The gluon FF is slightly larger for all hadronic species in the {\tt BBMZ} fit, 
although this effect is mostly localised above $z=0.2$
for charged pions and kaons and below $z=0.2$ for protons/antiprotons.
As expected, heavy quark FFs $D_{c^+}^h$ and $D_{b^+}^h$ for all
hadronic species are unaffected.
The two fits do not differ by any relevant heavy-quark tagged measurements 
(except for the TPC tagged data for pions, which however carry a very 
small weight).
Uncertainties of FFs are slightly smaller for all hadronic species and 
flavours in the baseline fit as compared to the {\tt BBMZ} fit.

We conclude that the {\tt BBMZ} fit is competitive with the baseline fit.
Nonetheless, we also find that the measurements at $\sqrt{s}$ between
the $B$-factory scale and $M_Z$ still carry some amount of
experimental information that should be taken into account.

\section{Conclusions and outlook}
\label{sec:conclusions}

In this work we have presented NNFF1.0, a new determination of the FFs of
charged pions, charged kaons, and protons/antiprotons at LO, NLO, and
NNLO accuracy in perturbative QCD.
This analysis is based on a comprehensive set of SIA data, including
the recent and precise measurements from the $B$-factory experiments
BELLE and BABAR.
The well-established NNPDF fitting methodology, widely used to determine 
polarised and unpolarised PDFs, was extended to FFs here for the first time.
This methodology is specifically designed to provide a faithful representation 
of the experimental uncertainties and to minimise any bias related to the 
parametrisation of FFs and to the minimisation procedure.

In this analysis we have introduced some methodological improvements aimed at 
reducing even further any possible procedural bias.

As a first improvement, we have removed from the usual NNPDF
parametrisation the preprocessing function governing the FF behaviour
in the small- and large-$z$ extrapolation regions.
As a consequence, we do not need to iterate the fits anymore in order to 
determine the optimal ranges of the preprocessing exponents.
This came at the price of modifying the activation function in the neural 
network in order to avoid an unphysical behaviour of the FFs in the small- 
and large-$z$ extrapolation regions.

As a second improvement, we have used a minimisation procedure
based on the CMA-ES family of algorithms. 
This procedure allows for a more efficient exploration of the parameter space 
in comparison to the genetic algorithm used in previous NNPDF fits of PDFs.
The fitting framework has finally been validated by means of closure tests.

We have presented the NNFF1.0 sets of FFs.
We have discussed the quality of our fits and showed that the
inclusion of QCD corrections up to NNLO improves the
description of the data for all the hadronic species considered,
especially in the small-$z$ region.
We have then examined the FFs resulting from our fits.
We highlighted their perturbative stability and observed a reduction
of the FF uncertainties at NLO and NNLO with respect to LO.
We have then compared the NNFF1.0 FFs to the recent DEHSS and JAM FFs
for charged pions and kaons at NLO.
We found a general fair agreement among the three sets with some
noticeable differences mostly for the gluon FFs and their
uncertainties.

We concluded our discussion by studying the stability of our fits upon
variations of the small-$z$ kinematic cuts and of the fitted dataset.
The primary aim was that of justifying our particular choices for the
default kinematic cuts and dataset.
However, these studies have also clarified the role of the
higher-order QCD corrections on the description of the low-$z$ data
and shed light on the tension among some of the datasets included
in our analysis.

The analysis presented in this paper represents the first step of a
broader program. 
A number of updates and improvements are foreseen.

The most important limitation of the NNFF1.0 analysis is the fact that
it is based only on SIA measurements.
Despite SIA is the cleanest process for the
determination of FFs, it carries little information on flavour
separation, is scarcely sensitive to the gluon FF, and is completely
blind to the separation between quark and antiquark FFs.
To improve on this, future updates of the NNFF fits will include
measurements from other processes that provide a handle on these
aspects.
This will be achieved by including in our future analyses SIDIS data 
($e.g.$ from the COMPASS and HERMES experiments) and $pp$ collision data 
($e.g.$ from the LHC and RHIC experiments).
This will require an efficient numerical implementation of the
corresponding observables which are more involved than SIA
observables.

A further improvement for future NNFF analyses is the inclusion of
heavy-quark mass corrections.
Such corrections are expected to improve the description of the data
at the lowest center-of-mass energy.

Finally, as a long-term project, we aim at carrying out a simultaneous
determination of FFs and (un)polarised PDFs.
We will take advantage of the fact that unpolarised and polarised PDFs,
and now FFs, are already separately available from the common, mutually 
consistent NNPDF fitting framework.

The NNFF1.0 FF sets presented in this work are available via the 
{\tt LHAPDF6} interface~\cite{Buckley:2014ana}
\begin{center}
 {\bf \url{http://lhapdf.hepforge.org/}~}
\end{center}
The list of the available FF sets is the following.
\begin{itemize}
\item The FF sets for $\pi^{\pm}=\pi^++\pi^-$ ({\tt PIsum}), $\pi^+$
  ({\tt PIp}), and $\pi^-$ ({\tt PIm}) at LO, NLO, and NNLO:
  
  {\tt NNFF10\_PIsum\_lo},     {\tt NNFF10\_PIp\_lo},     {\tt NNFF10\_PIm\_lo}, \\
  {\tt NNFF10\_PIsum\_nlo},   {\tt NNFF10\_PIp\_nlo},   {\tt NNFF10\_PIm\_nlo},\\
  {\tt NNFF10\_PIsum\_nnlo}, {\tt NNFF10\_PIp\_nnlo}, {\tt NNFF10\_PIm\_nnlo}.

\item The FF sets for $K^{\pm}=K^++K^-$ ({\tt KAsum}), $K^+$ ({\tt
    KAp}), and $K^-$ ({\tt KAm}) at LO, NLO, and NNLO:

  {\tt NNFF10\_KAsum\_lo},     {\tt NNFF10\_KAp\_lo},     {\tt NNFF10\_KAm\_lo}, \\
  {\tt NNFF10\_KAsum\_nlo},   {\tt NNFF10\_KAp\_nlo},   {\tt NNFF10\_KAm\_nlo},\\
  {\tt NNFF10\_KAsum\_nnlo}, {\tt NNFF10\_KAp\_nnlo}, {\tt NNFF10\_KAm\_nnlo}.

\item The FF sets for $p/\bar{p}=p+\bar{p}$ ({\tt PRsum}), $p$ ({\tt
    PRp}), and $\bar{p}$ ({\tt PRm}) at LO, NLO, and NNLO:

  {\tt NNFF10\_PRsum\_lo},     {\tt NNFF10\_PRp\_lo},     {\tt NNFF10\_PRm\_lo}, \\
  {\tt NNFF10\_PRsum\_nlo},   {\tt NNFF10\_PRp\_nlo},   {\tt NNFF10\_PRm\_nlo},\\
  {\tt NNFF10\_PRsum\_nnlo}, {\tt NNFF10\_PRp\_nnlo}, {\tt NNFF10\_PRm\_nnlo}.
\end{itemize}
We refer the reader to Appendix~\ref{sec:delivery} for the details on
the assumptions used to construct these sets of FFs and on
the features of the $(x,Q)$ tabulation.

\subsection*{Acknowledgements}
We thank A.~Accardi, D.~P.~Anderle, R.~Sassot and N.~Sato for useful 
discussions.
We thank I.~Garzia for assistance with the BABAR measurements,
and R.~Seidl for his help with the BELLE measurements, in particular
for providing us with the proton/antiproton cross-section measurements
of Ref.~\cite{Seidl:2015lla}.
We thank N.~Sato for providing us with the TPC tagged data of
Ref.~\cite{Lu:1986mc}.
We thank D.~P.~Anderle for a benchmark of the NNLO computation of SIA
observables.
We thank M.~Stratmann for providing us with the charged pion FF sets
determined in Ref.~\cite{deFlorian:2014xna}.
We thank R.~Sassot for providing us with the charged kaon FF sets
determined in Ref.~\cite{deFlorian:2017lwf}, and for his help with the
corresponding interpolating routines.
We are finally grateful to our colleagues of the NNPDF Collaboration
for their support.
In particular, we thank S.~Forte and M.~Ubiali for a critical reading of the 
manuscript.

V.~B., N.~P.~H. and J.~R. are supported by an European Research
Council Starting Grant ``PDF4BSM''; S.~C. is supported by the HICCUP
ERC Consolidator grant (614577); E.~R.~N. is supported by the UK STFC
grant ST/M003787/1.

\appendix
\section{Delivery of the NNFF1.0 sets}
\label{sec:delivery}

The determination of FFs presented in this work is based on SIA data
whose measurements are provided for the sum of the charged hadrons of
a given species (see Sect.~\ref{sec:dataset}).
Specifically, cross-sections are measured for $\pi^\pm=\pi^++\pi^-$,
$K^\pm=K^++K^-$, and $p/\bar{p} =p+\bar{p}$ production.
As a consequence, for each partonic species, what we actually extract
is the sum of the distributions belonging to the positive and negative
hadrons.
However, many phenomenological applications require sets of FFs for
positive and negative hadrons separately.
In this appendix, we discuss the assumptions adopted to perform this
separation.

Considering that opposite-charge hadrons are related by charge
conjugation
\begin{equation}\label{eq:chargesymmetryq}
D_{q(\bar{q})}^{h^+} = D_{\bar{q}(q)}^{h^-}\,,
\end{equation}
and that the relevant SIA observables are sensitive only to the sum of
quark and antiquark distributions ($i.e.$
$D_{q^+}^h=D_{q}^h+D_{\bar{q}}^h$ with $h=\pi^\pm,K^\pm,p/\bar{p} $
and $q=u,d,s,c,b$), it is possible to separate quark and antiquark
contributions as follows:
\begin{equation}\label{eq:quarkantiquarkseparation}
\small
D_{q^+}^h  =
D_{q}^{h^+}+D_{\bar{q}}^{h^+}+D_{q}^{h^-}+D_{\bar{q}}^{h^-} =
\left\{
\begin{array}{l}
D_{q}^{h^+}+D_{q}^{h^-}+D_{q}^{h^-}+D_{q}^{h^+} = 2 D_{q}^{h}\\
D_{\bar{q}}^{h^-}+D_{\bar{q}}^{h^+}+D_{\bar{q}}^{h^+}+D_{\bar{q}}^{h^-} = 2 D_{\bar{q}}^{h}
\end{array}
\right.\Longrightarrow
D_{q}^h = D_{\bar{q}}^h = \frac{D_{q^+}^h}{2}\,,
\end{equation}
where we have omitted the dependence of FFs on the momentum fraction
$z$ and the factorisation scale $Q$. 
In fact, charge conjugation is an \textit{exact} symmetry that
connects particles and antiparticles. Therefore, the rightmost
relation in Eq.~(\ref{eq:quarkantiquarkseparation}) has to be true for
any value of $z$ and $Q$.

In order to disentangle the up- and down-quark contributions to the
pion FFs and the up- and strange-quark contributions to the kaon FFs,
we assume SU(2) and SU(3) isospin symmetry respectively.
This assumption leads to the equalities
\begin{equation}\label{eq:isospinsymmetry}
  D_{u^+}^{\pi^\pm} = D_{d^+}^{\pi^\pm}\quad\mbox{and}\quad D_{u^+}^{K^\pm} = D_{s^+}^{K^\pm}\, ,
\end{equation}
which we take to be valid for all values of $z$ and $Q$.
However, SU(2) and SU(3) isospin symmetries are only approximate, in
that they are broken by terms proportional to the difference of quark
masses $m_u-m_d$ and $m_u-m_s$, respectively.
Given the spread between the light-quark masses, this implies that
SU(2) is expected to be more accurate than SU(3).
Indeed, the amount of SU(2) violation in a fit of pion FFs was found
to be negligible in Ref.~\cite{deFlorian:2014xna}.

For the protons fragmentation functions it is not possible to write
similar equalities based on isospin symmetry because the SU(2) isospin
transformation turns protons into neutrons rather than connecting
protons with antiprotons.
However, in what follows we will derive a relation of the same kind of
Eq.~(\ref{eq:isospinsymmetry}) that will allow us to separate up and
down contributions also for protons.
As we will see, this will require the introduction of further
assumptions.

A further step is that of separating the distributions of positive and
negative hadrons.
This step is necessarily artificial because the experimental
information included in our fits is not sensitive to such a
separation.
Consequently, we need to make some assumptions that will allow us to
isolate the positive contributions from the negative ones.
However, it should be kept in mind that the resulting distributions
might be affected by a bias that only the inclusion of processes
sensitive to this separation, such as SIDIS and hadron-collider
measurements, might possibly resolve.

The assumption that we make in order to separate the opposite-charge
contributions is based on the concept of {\it favoured}, {\it
  unfavoured}, and {\it sea} components. The (anti)flavours that
preferably fragment into a particular hadron are said favourite.
For the species considered in this work, such contributions are the
following:
\begin{equation}
\begin{array}{rcrc}
\pi^+: & \left(\frac12u,\frac12\bar{d}\right), & \pi^-: & \left (\frac12\bar{u},\frac12d
                                            \right),\\
\\
K^+: & \left (\frac12u,\frac12\bar{s}\right), & K^-: & \left (\frac12\bar{u},\frac12s \right),\\
\\
p: & \left (\frac23u,\frac13d \right), & \bar{p}: & \left (\frac23\bar{u},\frac13\bar{d}\right),
\end{array}
\end{equation}
where for each quark we have also indicated the ``preference''
factors, which is the relative pro\-ba\-bi\-lity with which it
fragments into the child hadron.
We assume that the favoured distributions of a given
hadron are equal to one another up to a factor equal to the inverse of
the preference factor.
This requirement, together with charge conjugation symmetry, leads to
the following relations
\begin{equation}\label{eq:favourite}
\begin{array}{lllll}
  D_u^{\pi^+} &= D_{\bar{d}}^{\pi^+} &= D_{\bar{u}}^{\pi^-} &= D_d^{\pi^-}  &\equiv D_{\rm fav}^\pi, \\[0.1cm]
  D_u^{K+}     &= D_{\bar{s}}^{K^+}   &= D_{\bar{u}}^{K^-} &=D_s^{K^-}  &\equiv   D_{\rm fav}^K, \\[0.1cm]
  D_u^{p}/2   &= D_d^{p}               &= D_{\bar{u}}^{\bar{p}}/2 &= D_{\bar{d}}^{\bar{p}} &\equiv   D_{\rm fav}^p.
\end{array}
\end{equation}
Next, we assume that all unfavoured distributions, defined as the
antiparticle counterparts of the favoured ones, and the light sea
distributions, defined as the remaining distributions associated to
light flavours, are equal. Again using charge conjugation symmetry,
this translates into the following equalities:
\begin{equation}\label{eq:unfavourite}
\begin{array}{lllllllll}
D_{\bar{u}}^{\pi^+} &= D_{d}^{\pi^+}&=D_s^{\pi^+} &= D_{\bar{s}}^{\pi^+}&= D_{u}^{\pi^-} &= D_{\bar{d}}^{\pi^-}&=D_s^{\pi^-} &=D_{\bar{s}}^{\pi^-} &\equiv D_{\rm unf}^\pi\,,\\[0.1cm]
D_{\bar{u}}^{K^+} &= D_{s}^{K^+}&=D_d^{K^+} &= D_{\bar{d}}^{K^+}&= D_{u}^{K^-} &= D_{\bar{s}}^{K^-}&=D_d^{K^-} &= D_{\bar{d}}^{K^-} &\equiv D_{\rm unf}^K\,,\\[0.1cm]
D_{\bar{u}}^{p} &= D_{\bar{d}}^{p}&=D_s^{p} &= D_{\bar{s}}^{p}&= D_{u}^{\bar{p}} &= D_{d}^{\bar{p}}&=D_s^{\bar{p}} &=D_{\bar{s}}^{\bar{p}} &\equiv D_{\rm unf}^p\,.
\end{array}
\end{equation}
We emphasise again that the assumptions that lead to
Eqs.~(\ref{eq:favourite}) and~(\ref{eq:unfavourite}) are not based on
any exact or approximated physical symmetry.
Rather, they are instrumental in separating distributions that
otherwise could not be disentangled neither on the base of the
experimental data nor considering a physical symmetry.
Therefore, we limit ourselves to impose these equalities only at the
initial scale $Q_0=5$ GeV and let the perturbative evolution generate
any possible breaking at higher scales.

We can now come back to the question of separating up and down
contributions of the proton/antiproton FFs. Using
Eqs.~(\ref{eq:favourite}) and~(\ref{eq:unfavourite}), we can
establish the following relation:
\begin{equation}\label{eq:isospinsymmetry2}
  D_{u^+}^{p/\bar{p}} = 2 D_{d^+}^{p/\bar{p}}-\frac12 D_{s^+}^{p/\bar{p}}\,,
\end{equation}
that, on the same footing as Eq.~(\ref{eq:isospinsymmetry}), provides
a further constrain to separate the sum
$D_{d^++u^+}^{p/\bar{p}}=D_{d^+}^{p/\bar{p}}+D_{s^+}^{p/\bar{p}}$
which is the quantity determined in our fits.

The remaining step to separate quark FFs of the opposite-charge
hadrons is to relate the favoured and unfavoured combinations to the
fitted FFs.
This is achieved by using Eqs.~(\ref{eq:quarkantiquarkseparation}),
(\ref{eq:isospinsymmetry}), (\ref{eq:isospinsymmetry2}),
(\ref{eq:favourite}), and~(\ref{eq:unfavourite}) to relate the
unfavoured distributions $D_{\rm fav}^\pi$, $D_{\rm fav}^K$, and
$D_{\rm fav}^p$ and the unfavoured distributions $D_{\rm unf}^\pi$,
$D_{\rm unf}^K$, and $D_{\rm unf}^p$ to the distributions $D_{u^+}^h$
and $D_{d^++s^+}^h$ that are extracted from the fits, finding the
following relations:
\begin{equation}
\begin{array}{ll}
  \displaystyle 
  D_{\rm fav}^\pi = \frac34 D_{u^+}^{\pi^\pm}-\frac14
  D_{d^++s^+}^{\pi^\pm}\,,&  \displaystyle D_{\rm unf}^\pi = -\frac14 D_{u^+}^{\pi^\pm}+\frac14 D_{d^++s^+}^{\pi^\pm}\,,\\
  \\
  \displaystyle 
  D_{\rm fav}^K = \frac34 D_{u^+}^{K^\pm}-\frac14
  D_{d^++s^+}^{K^\pm}\,,&   \displaystyle 
                          D_{\rm unf}^K = -\frac14 D_{u^+}^{K^\pm}+\frac14 D_{d^++s^+}^{K^\pm}\,,\\
  \\
  \displaystyle 
  D_{\rm fav}^p = \frac3{10} D_{u^+}^{p/\bar{p}}-\frac1{10}
  D_{d^++s^+}^{p/\bar{p}} \,,  & \displaystyle 
  D_{\rm unf}^p = -\frac1{10} D_{u^+}^{p/\bar{p}}+\frac1{5} D_{d^++s^+}^{p/\bar{p}}\,.
\end{array}
\end{equation}
Finally, in order to determine the gluon and the heavy-quark
distributions of the single hadrons, we simply assume that
\begin{equation}\label{eq:gluonheavyquarks}
D_g^{h^+} = D_g^{h^-} = \frac12 D_g^h\quad\mbox{and}\quad D_i^{h^+} =
D_{\bar{i}}^{h^-} = \frac14  D_{i^+}^h\,,\quad i=c,b\,.
\end{equation}
Eqs.~(\ref{eq:favourite})-(\ref{eq:gluonheavyquarks}) are used to
tabulate the FFs determined in our fits in the {\tt LHAPDF}
format~\cite{Buckley:2014ana}.
For each hadronic species and each perturbative order considered in
this work, we deliver an {\tt LHAPDF} grid for the sum of the charged
hadrons and two additional grids for the positive and the negative
hadrons.
It should be stressed that for kaons and pions, the grids associated
to the sum of the opposite-charge hadrons reflect very closely the
information extracted from the fit because they only rely on the exact
charge conjugation symmetry, Eq.~(\ref{eq:quarkantiquarkseparation}),
and the SU(2) and SU(3) isospin symmetries,
Eq.~(\ref{eq:isospinsymmetry}).
Conversely, all the grids for proton/antiproton FFs and the grids for
the separate charged pions and kaons have been produced by making
empirical assumptions on the relation between favoured, unfavoured,
and sea distributions, Eqs.~(\ref{eq:favourite})
and~(\ref{eq:unfavourite}), which are not based on any fundamental
physical symmetry.

Another important remark concerns the tabulation range in $z$ and $Q$
of the {\tt LHAPDF} grids produced in this work.
We have chosen to deliver our FF set in the range $[10^{-2}:1]$ in $z$
and $[1:10000]$ GeV in $Q$.
The choice of the range in $z$ is motivated by the fact that our
lowest default kinematic cut is $z_{\rm min} = 0.02$.
Therefore, in order to avoid unreliable extrapolations far below
$z_{\rm min}$, our grids extend only slightly below this value. 
As far as the range in $Q$ is concerned, despite our FFs are
parametrised at $Q_0=5$ GeV, our grids extend down to 1 GeV in order
to make them usable also for low-energy predictions.
As discussed in Sect.~\ref{sec:numericalimplementation}, in this
analysis $Q_0$ has been chosen to be larger of the bottom threshold in
such a way to avoid any crossing of heavy-quark thresholds during the
fit.
However, 1 GeV is below the charm threshold $m_c$ and thus we need to
evolve our FFs backward from $Q_0$ to 1 GeV crossing both the bottom
and the charm thresholds.
Such crossings are delicate for two reasons. The first is related to
the fact that we fit charm and bottom FFs that thus contain a
non-perturbative contribution that is not accounted by the
perturbative matching conditions at the thresholds.
To overcome this problem, we set to zero the bottom and charm FFs
below the respective thresholds.
The second reason has to do with the fact that time-like matching
conditions are currently known to $\mathcal{O}(\alpha_s)$, $i.e.$ NLO.
Therefore, when evolving backward our NNLO determinations, we still
assume NLO matching conditions at the heavy-quark thresholds.

\bibliography{nnff10}

\end{document}